\documentclass[fleqn,usenatbib,useAMS]{mnras}

\usepackage{graphicx}	
\usepackage{amsmath}	
\usepackage{multicol}        
\usepackage{bm}		
\usepackage{pdflscape}	
\usepackage{ulem} 
\usepackage{comment} 

\usepackage[T1]{fontenc}
\usepackage{ae,aecompl}

\usepackage{newtxtext,newtxmath}

\title[Systematic opacity calculations for kilonovae -- II.]{Systematic opacity calculations for kilonovae -- II. Improved atomic data for singly ionized lanthanides}

\author[Kato et al.]{
  Daiji Kato$^{1,2}$\thanks{Contact e-mail: \href{kato.daiji@nifs.ac.jp}{kato.daiji@nifs.ac.jp}},
  Masaomi Tanaka$^{3,4}$,
  Gediminas Gaigalas$^5$, 
  Laima Kitovien\.{e}$^5$, 
  Pavel Rynkun$^5$
\\
$^1$National Institute for Fusion Science, 322-6 Oroshi-cho, Toki 509-5292, Japan\\
$^2$Interdisciplinary Graduate School of Engineering Sciences, Kyushu University, Kasuga, Fukuoka 816-8580, Japan\\
$^3$Astronomical Institute, Tohoku University, Sendai 980-8578, Japan
\\	
$^4$Division for the Establishment of Frontier Sciences, Organization for Advanced Studies, Tohoku University, Sendai 980-8577, Japan\\
$^5$Institute of Theoretical Physics and Astronomy, Vilnius University, Saul\.{e}tekio Ave. 3, LT-10257 Vilnius, Lithuania
}

\date{}

\pubyear{2024}

\begin{document}
\label{firstpage}
\pagerange{\pageref{firstpage}--\pageref{lastpage}}
\maketitle

\begin{abstract}
Lanthanides play most important roles in the opacities for kilonova,
ultraviolet-optical-infrared emission from neutron star mergers.
Although several efforts have been made to construct atomic data,
the accuracy in the opacity is not fully assessed and understood.
In this paper, we perform atomic calculations for
singly ionized lanthanides with improved strategies, 
aiming at understanding the physics of the lanthanide opacities in kilonova
ejecta and necessary accuracy in atomic data.
Our results show systematically lower energy level distributions
as compared with our previous study (Paper I).
As a result, the opacities evaluated with our new results are
higher by a factor of up to $3 - 10$,
depending on the element and wavelength range.
For a lanthanide-rich element mixture,
our results give a higher opacity than that 
in Paper I
by a factor of about 1.5.
We also present opacities by using the results of {\it ab-initio} atomic calculations by using {\sc Grasp2K} code.
In general, our new opacities show good agreements
with those with {\it ab-initio} calculations.
We identify that structure of the lanthanide opacities
are controlled by transition arrays among several configurations,
for which derivation of accurate energy level distribution
is important to obtain reliable opacities.
\end{abstract}





\section{Introduction}
\label{sec:introduction}

Neutron star (NS) mergers have been expected to be
a promising site for rapid neutron capture process
\citep[$r$-process, e.g.,][]{lattimer74,eichler89,goriely11,korobkin12,bauswein13,wanajo14}.
The ejected material (or ejecta) can emit thermal
electromagnetic radiation, so called ``kilonova'',
which is powered by radioactive decays of newly synthesized $r$-process
nuclei \citep[e.g.,][]{li98,metzger10}.
By reflecting the temperature and opacities in the ejecta,
kilonova emission is expected to be mainly in ultraviolet (UV),
optical, and infrared (IR) wavelengths for a timescale of about $1-10$ days
after the merger
\citep[e.g.,][]{metzger10,kasen13,barnes13,tanaka13,metzger14}.

In 2017, by following the detection of gravitational waves from
a NS merger \citep[GW170817][]{abbott17},
an electromagnetic counterpart has been observed \citep{abbott17MMA}.
In UV, optical, and IR wavelengths, the counterpart (AT2017gfo)
shows thermal emission.
The properties of AT2017gfo is broadly consistent with
expected properties of kilonova
\citep[e.g.,][]{kasen17,perego17,tanaka17,rosswog18,kawaguchi18},
confirming $r$-process nucleosynthesis in NS merger.

Properties of kilonova, i.e., luminosity, timescale,
and color or spectral shapes, are mainly determined by
the mass and velocity of the ejecta and elemental compositions
in the ejecta.
In particular, elemental compositions play important roles
as they control the opacity in the ejecta.
In the NS merger ejecta, with a typical temperature of $T \sim 10^3 - 10^5$ K,
the main opacity source is the bound-bound transitions of heavy elements.
In particular, lanthanides (atomic number $Z=57-71$)
have high opacities by reflecting
their dense energy levels \citep{kasen13,tanaka13,fontes20,tanaka20}.
Thus, the presence or absence of lanthanides largely
affects the light curves of kilonova.
Thanks to these properties, we can infer the nucleosynthesis
in NS mergers through observational data of kilonovae.

To reliably connect the nucleosynthesis information with observed properties of kilonovae,
accurate understanding of the opacities in NS merger ejecta
is crucial.
The opacities are determined by a large number of transitions
including those from excited states.
Thus, complete energy levels and transition probabilities
are necessary to evaluate the opacity,
even under the simplest assumption of local thermodynamic equilibrium (LTE).
Since it is not yet practical to derive such complete information from
experimental data, current understanding of the kilonova opacities
relies on theoretical atomic calculations.
In fact, there has been significant progress in the atomic calculations
for application to kilonovae in the past decade
\citep[e.g.,][]{kasen17,tanaka18,gaigalas19,wollaeger18,tanaka20,fontes20,banerjee20,fontes23,carvajalgallego23,carvajalgallego24,banerjee24}.
Thanks to these efforts, atomic opacities for essentially all the elements
relevant to kilonova have been constructed (up to about 10th ionization).

However, the accuracy of the opacities is not entirely assessed.
Due to the complexity, theoretical atomic calculations covering
many elements and ionization stages often involve simplifications
in the calculations, such as a parameterized effective potential.
Some studies have studied the accuracy of the results,
but such works only focused on one or a few elements \citep{tanaka18,gaigalas19,floers23}.
It is, thus, not yet clear in general how good the accuracy of the currently available opacities is.
In particular, since lanthanides give the dominant opacities in kilonovae, it is important to understand which configurations play important roles to the lanthanide opacities and how good accuracy is necessary to derive the reliable opacities.

Recently, we have performed 
{\it ab-initio} calculations for singly ionized lanthanides covering
twelve elements with $Z=59-70$
(\citealt{gaigalas19}, \citealt{radziute20} and \citealt{radziute21}, hereafter denoted as G19, R20 and R21, respectively) by using {\sc Grasp2K} code \citep{jonsson13}.
The calculated energy levels and transition probabilities are 
intensively compared with available atomic data,
and thus, the atomic data serve as benchmark results for singly ionized lanthanides.
Due to the computational cost, it is not practical to perform such detailed calculations covering all the elements and ionization states which are important in kilonova. 
Thus, it is also important to provide accurate atomic data with more approximated calculations, which can cover many elements and ionization states as demonstrated in our previous work (\citealt{tanaka20}, hereafter denoted as Paper I).

In this paper, by using the privilege of G19, R20 and R21,
we aim at obtaining deeper understanding of the lanthanide opacities in kilonova ejecta, 
and at finding a pathway to provide accurate atomic data with approximated calculations.
In Section \ref{sec:atomic}, we perform atomic calculations using {\sc Hullac} code~\citep{bar-shalom01} with improved strategies as compared with those in Paper I.
In Section \ref{sec:opacity}, we calculate the opacities using our new results and results from G19, R20 and R21.
In Section \ref{sec:discussions}, we discuss properties of lanthanide opacities and implication to kilonova.
Finally, we give summary of the paper in Section \ref{sec:summary}.

\begin{table*}
\caption{Strategy for effective potentials. The second column shows occupied orbitals of Equation~(\ref{eq:potential}). Inner shells:  $n=1-3$, $4spd$, and $5sp$ are fully occupied. The potentials for orbitals with (g) were weighted-averaged as in Equation~(\ref{eq:g-function}). The third column shows electronic configurations included in the first-order energy minimization.
$\Delta_{\rm median}$ is the median of absolute values of deviation from the reference values, i.e.,~$\Delta = |E - E^{\rm (ref)}|/E^{\rm (ref)}$, in \% for the calculated lowest levels (see text for more details).
Note that the strategy of Yb II is the same as in Paper I.
}
\label{tab:strategy}
\begin{tabular}{lllc}
\hline\hline
Ion & Potential & First-order energy & $\Delta_{\rm median}$  \\
\hline
Pr II & $4f^3$ & $4f^3~6s$ & 12 
\\ \\

Nd II & $4f^4$ & $4f^4~6s$  & 25 
\\ \\

Pm II & $4f^4~6s$ (g) & $4f^5~6s$ & 17 \\ \\

Sm II & $4f^5~5d$ & $4f^6~6s$ & 13  \\ \\
      
Eu II & $4f^7$ & $4f^7~6s$ & 13  \\ \\

Gd II & $4f^7~5d$ & $4f^8~6s$ & 42 \\ \\

Tb II & $4f^8~5d$ & $4f^9~6s$ & 23  \\ \\

Dy II & $4f^{10}$ & $4f^{10}~6s$ & 16 \\ \\

Ho II & $4f^{10}~5d$ & $4f^{11}~6s$ & 18 \\ \\

Er II & $4f^{11}~5d$ & $4f^{12}~\left( 6s, 5d, 6p \right), 4f^{11}~6s^2$ & 11 \\ \\

Tm II & $4f^{12}~6s$ & $4f^{13}~6s$ & 14 \\ \\
      
Yb II & $4f^{14}$ (g) & $4f^{14}~\left(6s, 5d, 6p, 7s\right), 4f^{13}~6s^2$ & 14 \\

\hline
\hline
\end{tabular}
\end{table*}

\begin{table*}
\caption{Comparison of the number of levels for each parity in 6 eV from the ground level. The first and second rows are {\sc Hullac} results of the present calculation and Paper I, respectively, and the third {\sc Grasp} results from G19 for Nd, R20 for Pr and Pm - Gd, and R21 for Tb - Yb, respectively. For Yb II, the present results and those of Paper I are identical as the same strategy was used.
The total number of levels by the present RCI calculation is also shown for each configuration in the following columns.}
\label{tab:levels}
    \begin{tabular}{ccccccccccc}
    \hline\hline
Ion & \multicolumn{2}{c}{$N_{\rm level}$ ($\le$ 6 eV)} & $4f^q~6s$ & $4f^q~5d$ & $4f^{q-1}~5d^2$ & $4f^{q-1}~5d~6s$ & $4f^{q-1}~6s^2$ & $4f^q~6p$ & $4f^{q-1}~5d~6p$ & $4f^{q-1}~6s~6p$ \\ \cline{2-3}
& Even & Odd & & & & & & & & \\
\hline
Pr II ($q=3$) & 700 & 523& 87 & 358 & 392 & 242 & 22 & 270 & 601 & 186  \\
& 679 & 486& & & & & & & &  \\
& 758 & 731& & & & & & & &  \\ \\

Nd II ($q=4$) & 690 & 1213 & 256 & 896 & 1438 & 801 & 113 & 687 & 2048 & 690 \\
& 647 & 970 & & & & & & & & \\
& 929 & 1337 & & & & & & & & \\ \\

Pm II ($q=5$) & 1874 & 895 & 598 & 1793 & 3575 & 2072 & 340 & 1465 & 5165 & 1694 \\
& 576 & 792  & & & & & & & & \\
& 1466 & 1113 & & & & & & & & \\ \\

Sm II ($q=6$) & 446 & 611 & 1002 & 2949 & 6141 & 4225 & 517 & 2521 & 9456 & 3133 \\
& 532 & 159 & & & & & & & & \\
& 544 & 750 & & & & & & & & \\ \\
      
Eu II ($q=7$) & 140 & 50 & 1387 & 4231 & 8977 & 6051 & 936 & 3544 & 13513 & 4109 \\
& 103 & 48 & & & & & & & & \\
& 137 & 85 & & & & & & & & \\ \\

Gd II ($q=8$) & 301 & 327 & 1256 & 3537 & 9169 & 6882 & 1204 & 4238 & 15298 & 5149 \\
& 68 & 235 & & & & & & & & \\
& 270 & 222 & & & & & & & & \\ \\

Tb II ($q=9$) & 820 & 538 & 841 & 3042 & 8552 & 5987 & 942 & 3267 & 13437 & 4434 \\
& 862 & 352 & & & & & & & & \\
& 925 & 851 & & & & & & & & \\ \\

Dy II ($q=10$) & 273 & 698 & 370 & 1667 & 5986 & 4345 & 555 & 1658 & 9182 & 3205 \\
& 231 & 535 & & & & & & & & \\
& 458 & 951 & & & & & & & & \\ \\

Ho II ($q=11$) & 478 & 178 & 108 & 559 & 3351 & 2279 & 251 & 645 & 2279 & 1814 \\
& 113 & 147 & & & & & & & & \\
& 701 & 295 & & & & & & & & \\ \\

Er II ($q=12$) & 162 & 628 & 25 & 123 & 1411 & 852 & 75 & 159 & 2028 & 660 \\
& 151 & 448 & & & & & & & & \\
& 188 & 534 & & & & & & & & \\ \\

Tm II ($q=13$) & 188 & 24 & 4 & 20 & 453 & 213 & 17 & 13 & 602 & 162 \\
& 223 & 34 & & & & & & & & \\
& 238 & 44 & & & & & & & &  \\ \\
      
Yb II ($q=14$) & 3 & 27 & 1 & 2 & 81 & 39 & 2 & 2 & 113 & 24 \\
& 3 & 27 & & & & & & & & \\
& 3 & 42 & & & & & & & & \\

\hline
\hline
    \end{tabular}
\end{table*}

\begin{figure*}
\centering
\begin{tabular}{cc}
    \includegraphics[width=8.5cm]{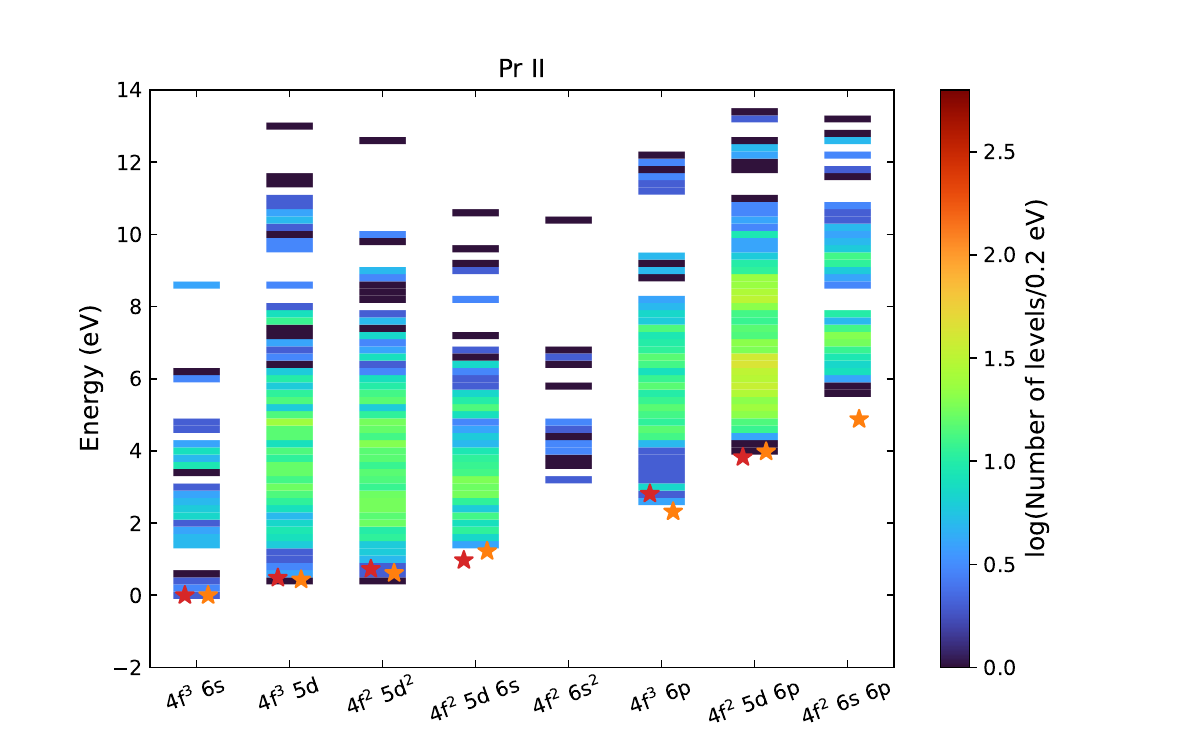}&
  \includegraphics[width=8.5cm]{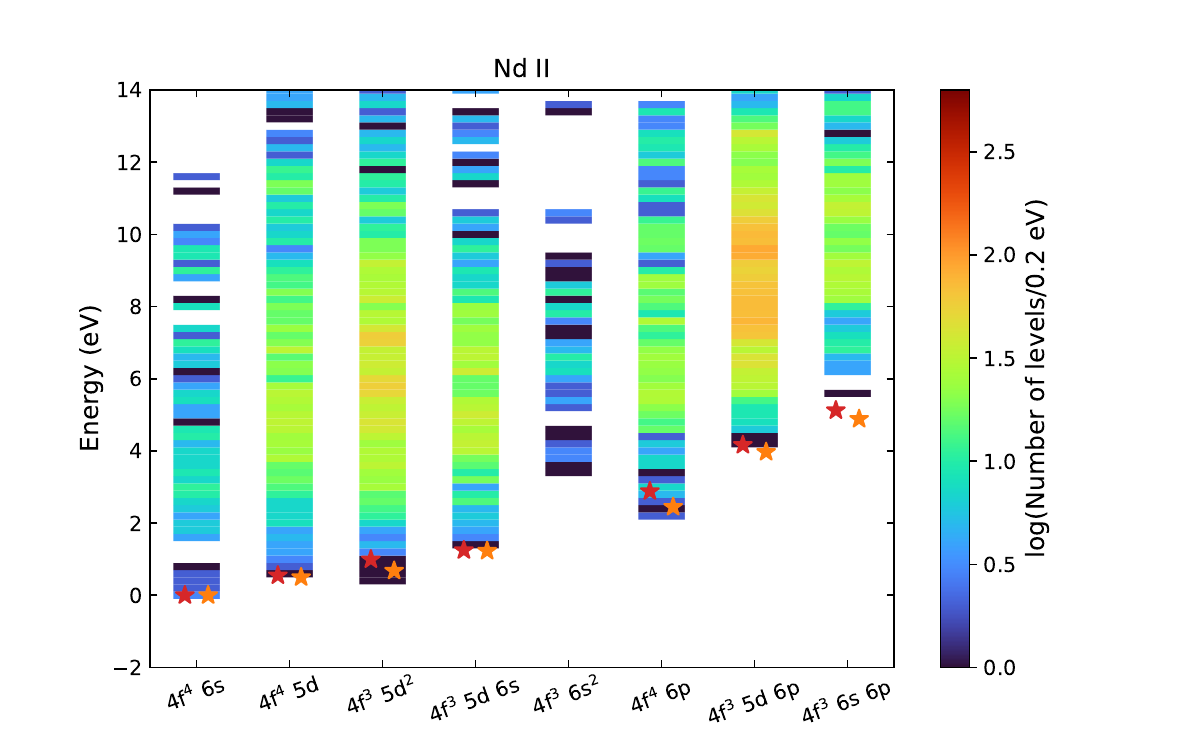}\\
    \includegraphics[width=8.5cm]{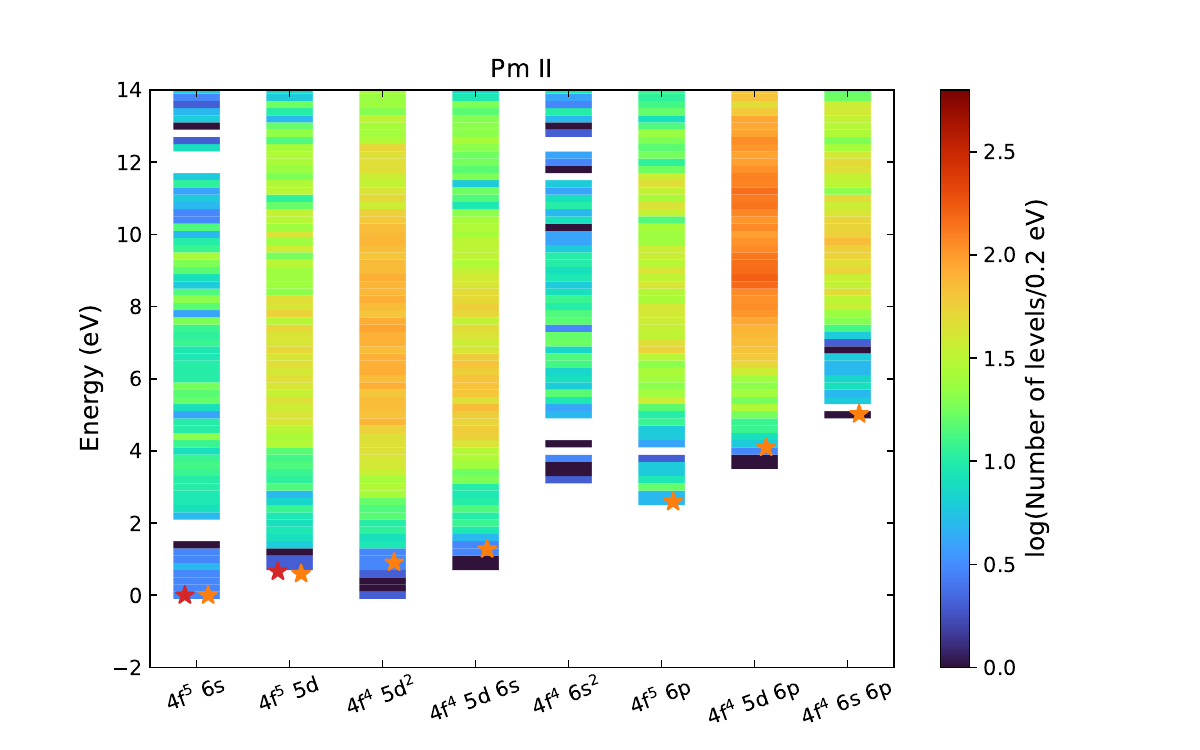}&
  \includegraphics[width=8.5cm]{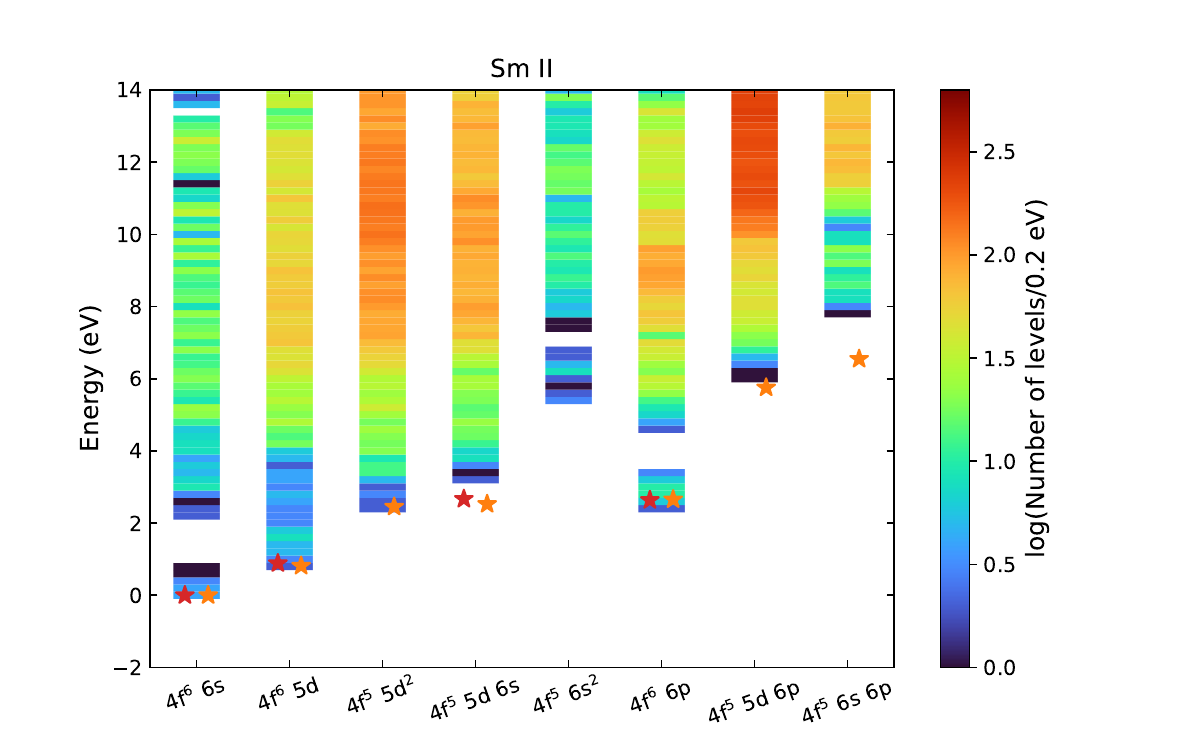}\\
    \includegraphics[width=8.5cm]{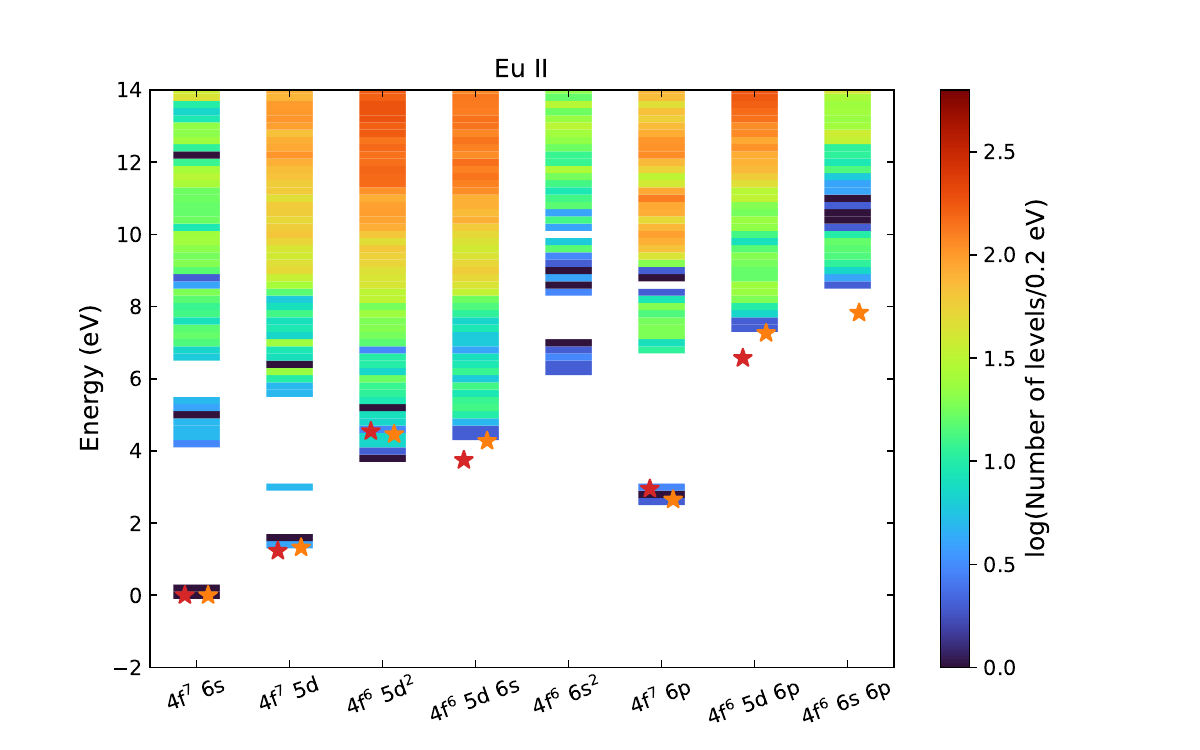}&
  \includegraphics[width=8.5cm]{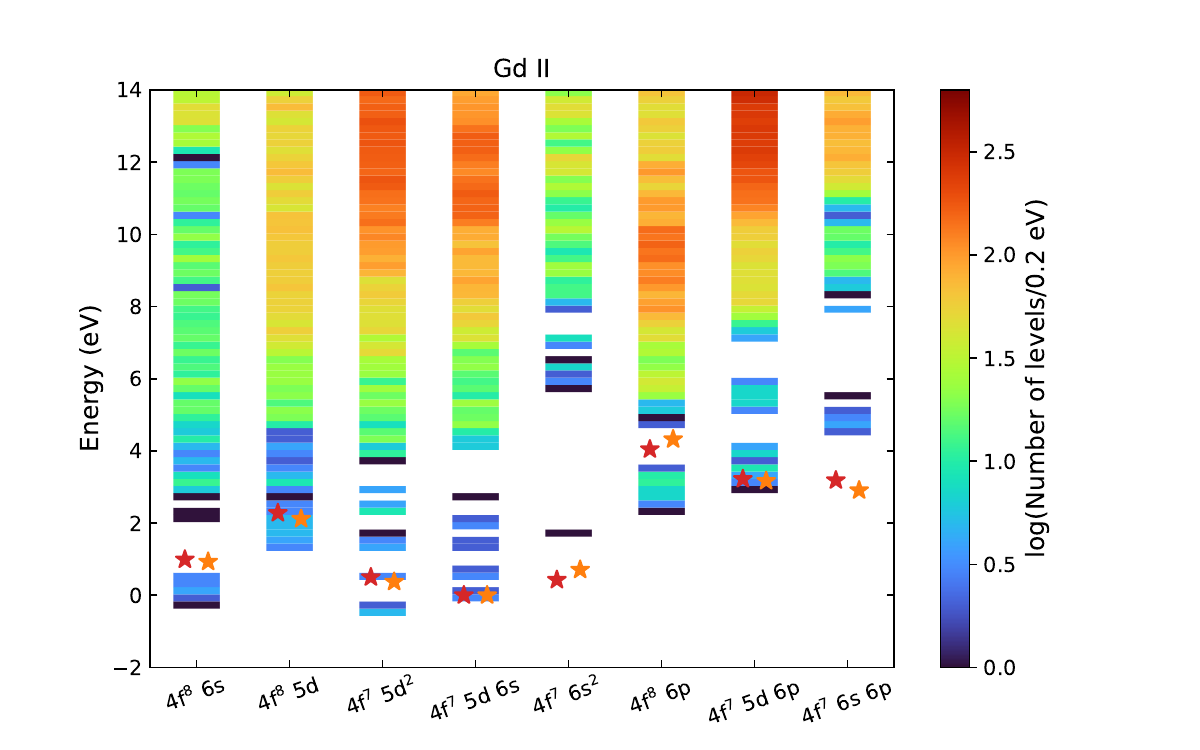}\\
\end{tabular}
\caption{
Calculated energy levels distribution for singly ionized ions ($Z=59-64$). The energy levels are shown for each configuration.
The colors represent the number of energy level in 0.2 eV bin. 
The red and orange star symbols represent the lowest energy for each configuration from the NIST ASD and {\sc GRASP} calculations (G19, R20, and R21), respectively.
The energy is measured from the lowest level of the correct ground state in the NIST ASD, i.e.,~$4f^7~5d~6s$ for Gd II and $4f^q~6s$ for the others.
Energy levels above $E=14$ eV are also calculated, but they are not shown in this figure to highlight the lower energy levels.
}
\label{fig:elevel_config1}
\end{figure*}

\begin{figure*}
\centering
\begin{tabular}{cc}
    \includegraphics[width=8.5cm]{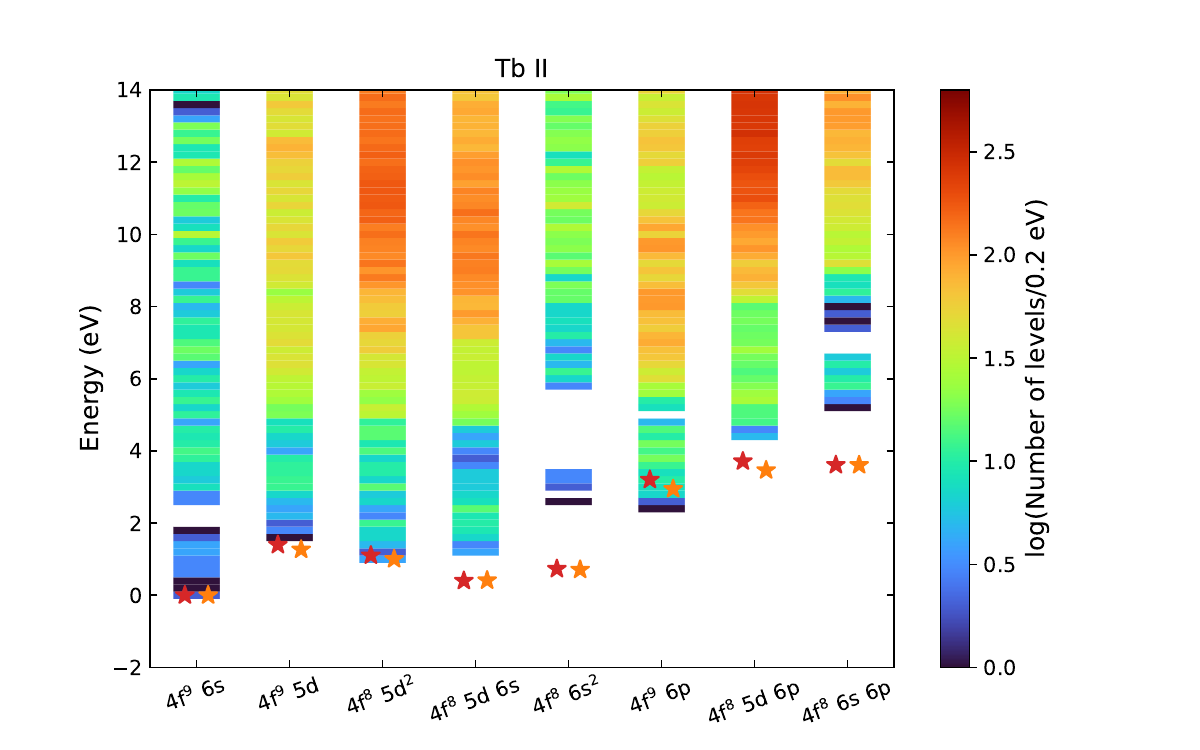}&
  \includegraphics[width=8.5cm]{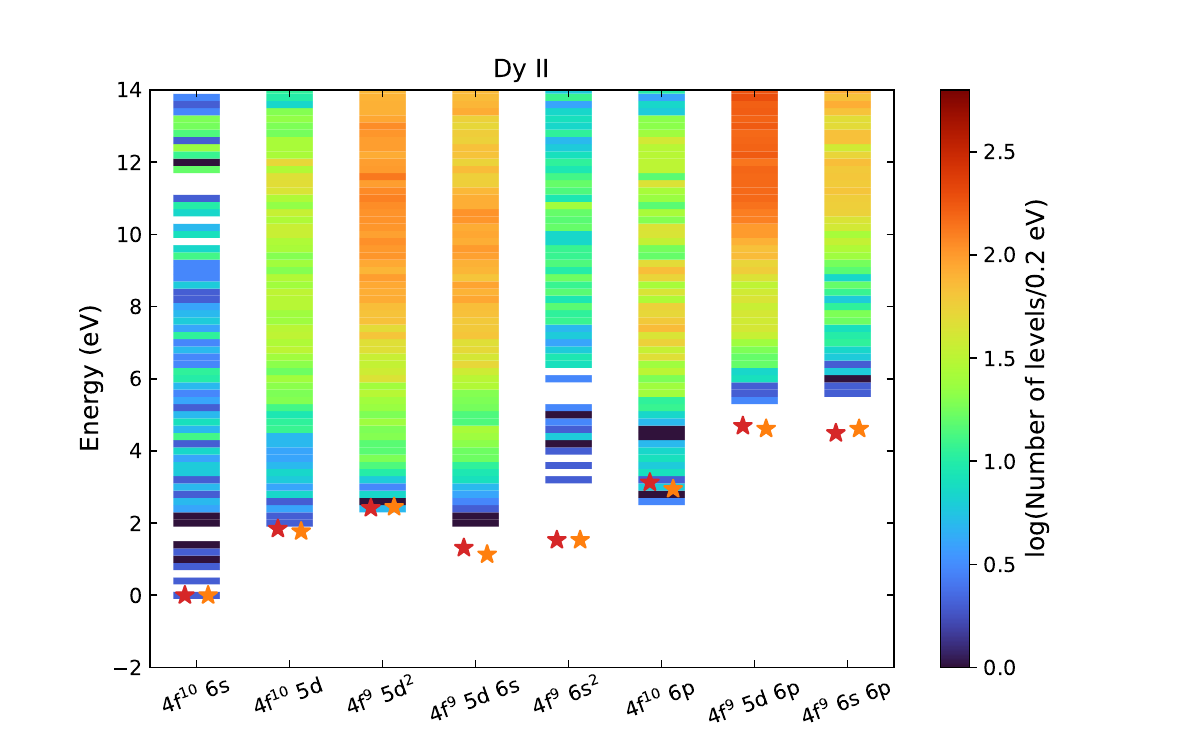}\\
    \includegraphics[width=8.5cm]{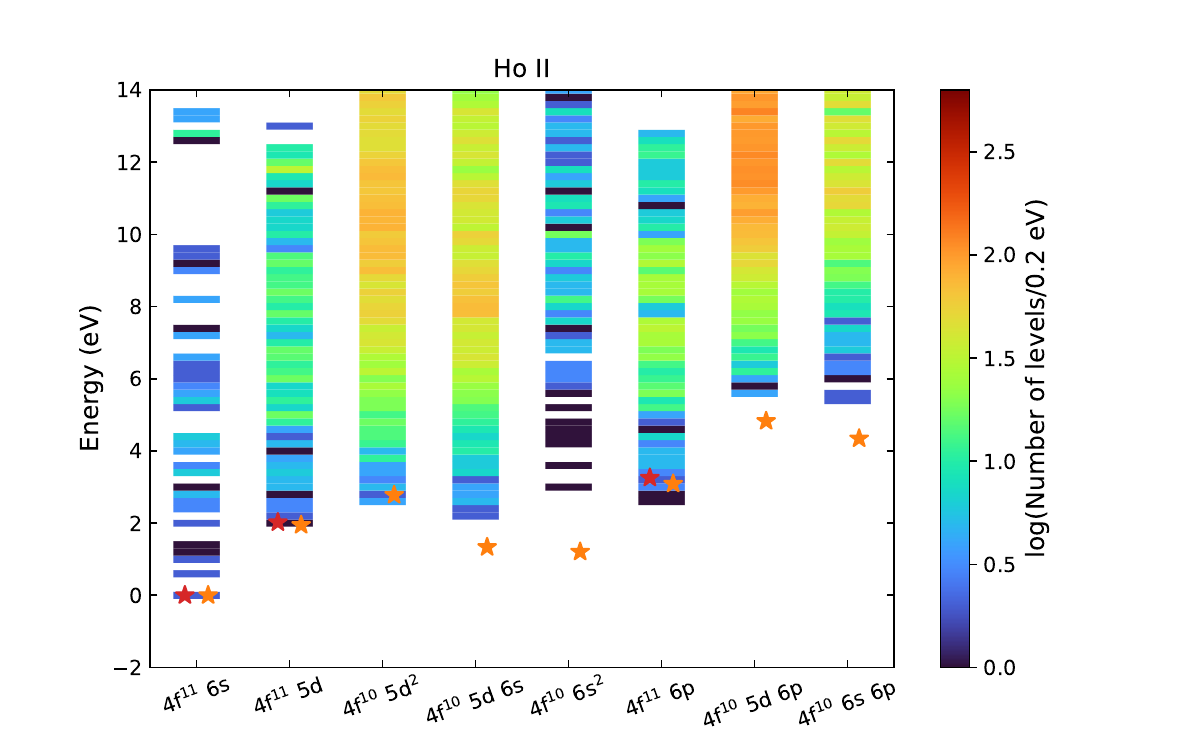}&
  \includegraphics[width=8.5cm]{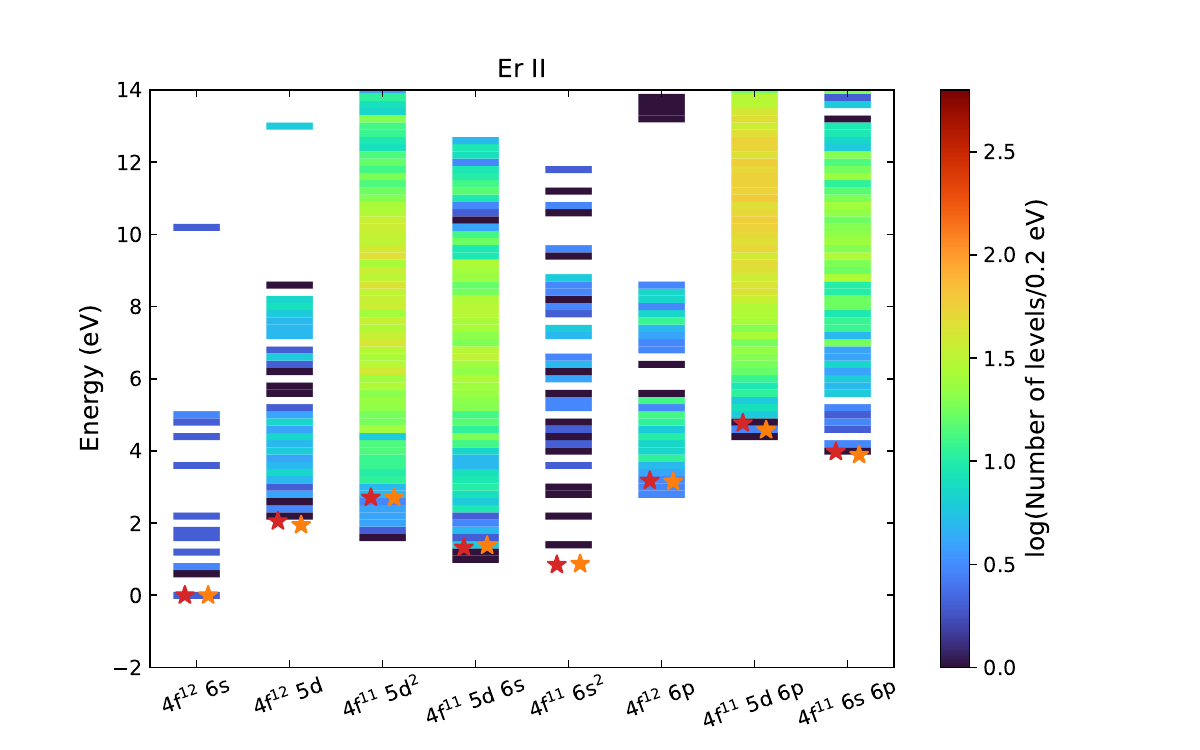}\\
    \includegraphics[width=8.5cm]{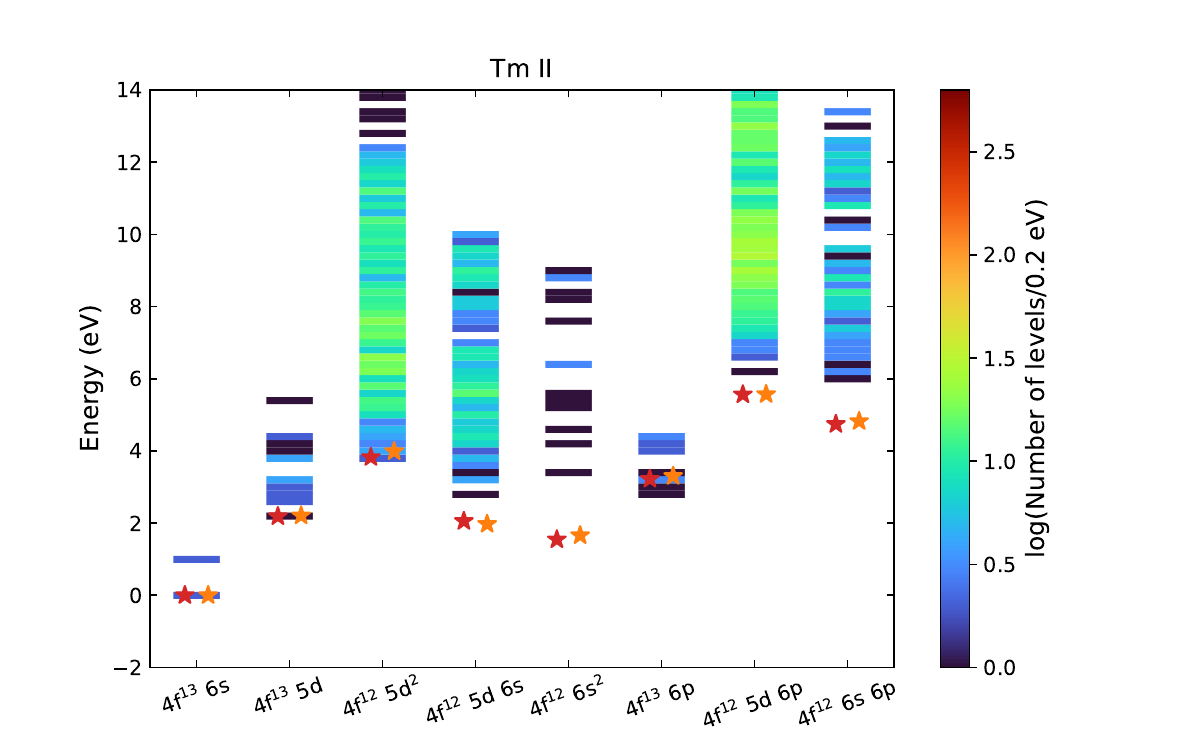}&
  \includegraphics[width=8.5cm]{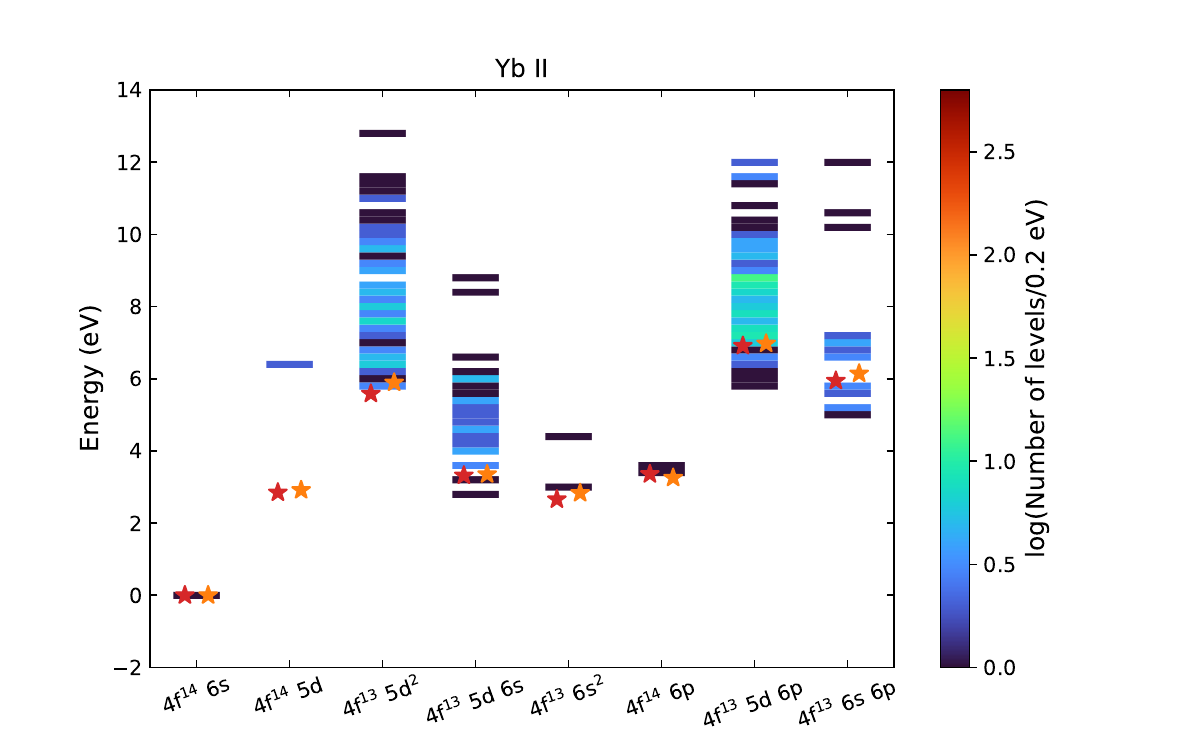}\\
\end{tabular}
\caption{Same with Figure \ref{fig:elevel_config1} but for the elements with $Z=65-70$.
}
\label{fig:elevel_config2}
\end{figure*}

\begin{figure*}
\centering
\begin{tabular}{cc}
\includegraphics[width=8.5cm]{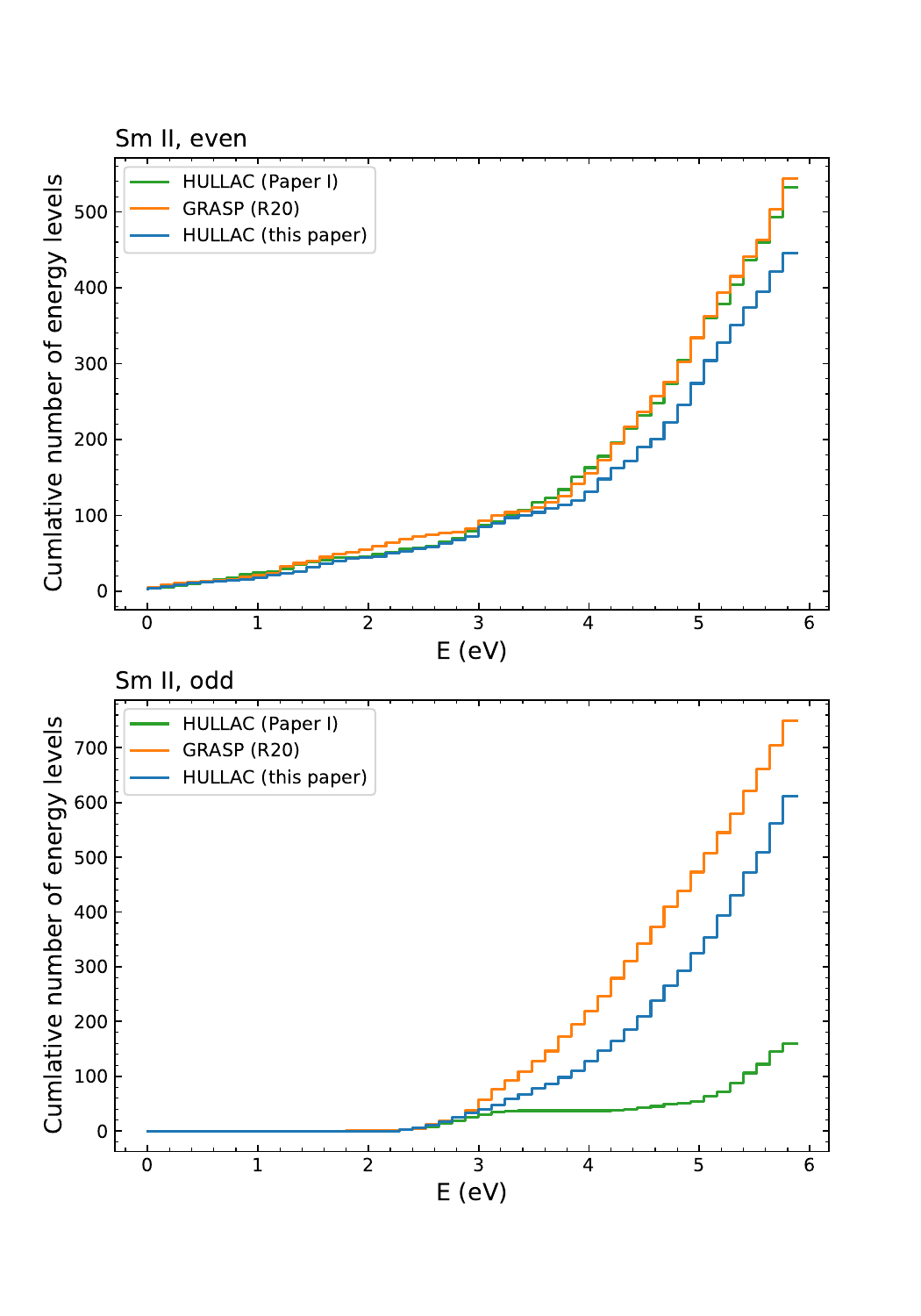} &
\includegraphics[width=8.5cm]{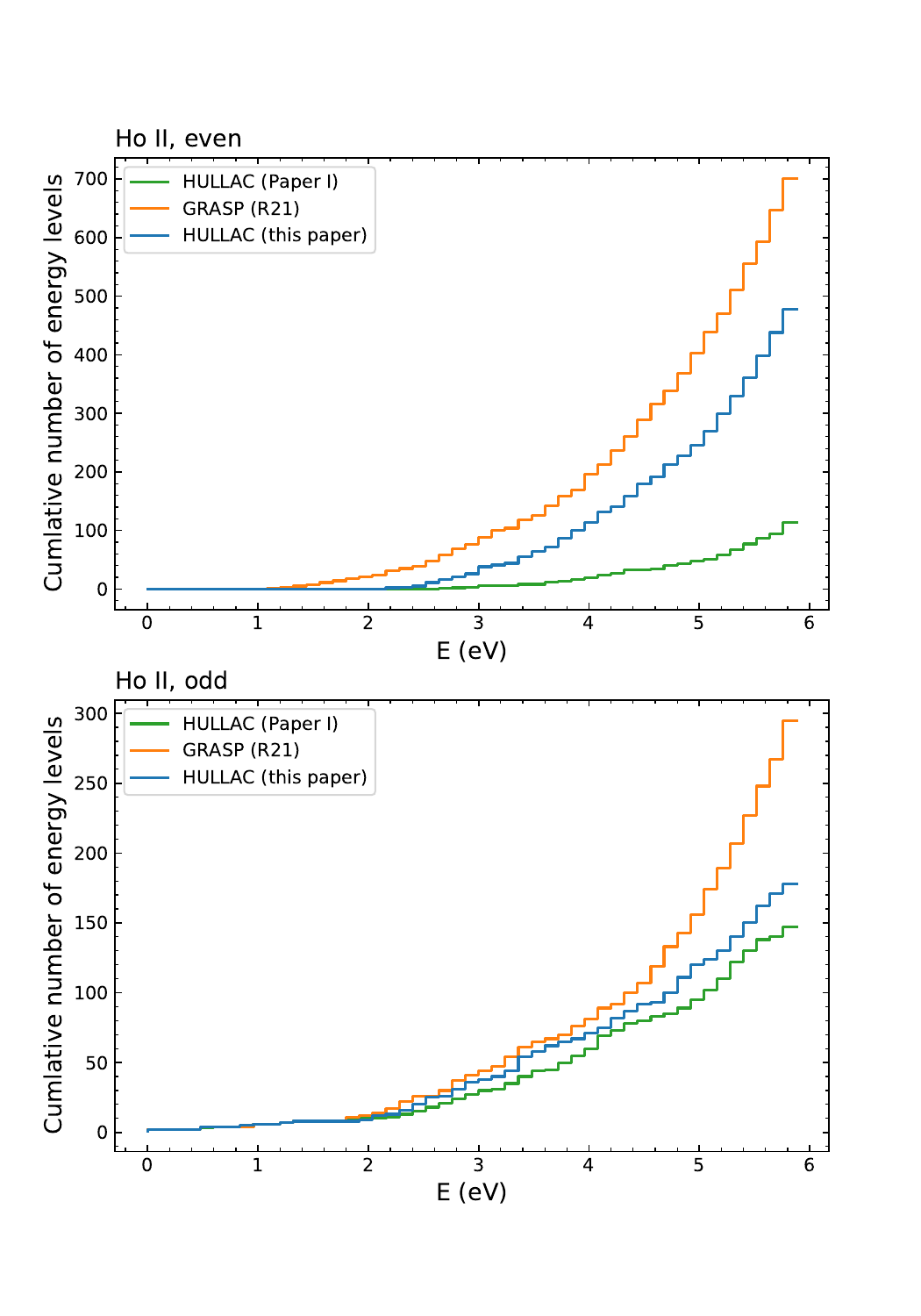}
\end{tabular}
\caption{Cumulative number distributions of the energy levels of Sm II and Ho II for each parity.}

\label{fig:Sm+Ho}
\end{figure*}


\section{Atomic calculations}
\label{sec:atomic}

\subsection{{\sc Hullac} calculations}

In {\sc Hullac}, the relativistic configuration interaction (RCI) method is performed using solutions of the single-electron Dirac equation with an effective central-field potential.
Accuracy of the RCI calculations is improved basically by increasing number of configurations. However, the size of the configurations can be exceedingly large for lanthanide due to the existence of open $4f$-shell. To reduce computational costs, we restricted the RCI to the minimal set of configurations of low energies that are most relevant to transitions of the opacity, i.e., $4f^q \left(6s, 5d, 6p\right)$ and $4f^{q-1} \left(5d^2, 5d~6s, 6s^2, 6s~6p, 5d~6p \right)$, $q=3-14$ for each element, respectively
(more details will be discussed in Section~\ref{sec:discussions}).
For Sm II and Yb II, $4f^7$ and $4f^{14}~7s$ were also added, respectively.
In the present calculations, therefore, we can improve the accuracy of the results by optimizing the effective potential.

In {\sc Hullac}, the effective potential for $N$-electron ions of the nuclear charge $Z$ is expressed as, 
\begin{equation}
U(r)=-\frac{1}{r}
\left[
\left(Z-N+1\right)+
\sum_i q_i f_{l_i,\alpha_i}(r)
\right],
\label{eq:potential}
\end{equation}
where $q_i$ is occupation numbers for the orbitals $(nl)_i$, and the total occupation number, $\sum q_i = N-1$.
$f_{l,\alpha}(r)$ is obtained from
the Slater-type charge distribution of an electron of the azimuthal quantum number $l$, which is expressed as, 
\begin{equation}
f_{l,\alpha}(r)=e^{-\alpha r}
\sum_{k=0}^{2l+1} \left(
1-\frac{k}{2l+2} \right)
\frac{(\alpha r)^k}{k!},
\label{eq:f-function}
\end{equation}
where $\alpha$ is related to the mean radius of the charge distribution by $\alpha = (2l+3)/\left<r\right>$. For closed shells, the weighted average of $f_{l,\alpha}(r)$ is used,
\begin{equation}
g_{L,\alpha}(r) = \frac{1}{2(L+1)^2}
\sum_{l=0}^{L}(4l+2)f_{l,\alpha^{(l)}}(r),
\label{eq:g-function}
\end{equation}
where $L \le n-1$, and $\alpha$ and $\alpha^{(l)}$ in the average are dependent by
\begin{equation}
\alpha^{(l)} = \alpha \times \frac{l+1}{1-\eta(l+1)},\ \ \ \eta=0.05.
\label{eq:alpha}    
\end{equation}
Note that the potential of Equation~(\ref{eq:potential}) satisfies the correct asymptotic conditions,
\begin{equation}
\lim_{r \to 0} U(r) = -\frac{Z}{r}, \\
\lim_{r \to \infty} U(r) = -\frac{Z-N+1}{r}.
\label{eq:asymptot}
\end{equation}

With a given set of the occupation numbers $q_i$, values of the $\alpha_i$ were varied until the expectation value of the energy (the first-order energy) for the ground state and low-lying excited states became minimum by the Nelder-Mead method.
The energy minimization was performed for several sets of the occupation numbers for the potential.

Then, we compared the calculated lowest energy level for each configuration with the value in the NIST Atomic Spectral Database~ \citep{kramida18}. For Pm II and Ho II, however, we compared also with the {\sc Grasp} results (R20 for Pm II and R21 for Ho II) for higher excited states since the data available in the database are limited. 
The agreement was evaluated by the median of absolute values of normalized errors from the reference values, i.e.,~$\Delta = |E - E^{\rm (ref)}|/E^{\rm (ref)}$, accounting for the lowest levels of $4f^q \left(6s, 5d, 6p\right)$ and $4f^{q-1} \left(5d^2, 5d~6s, 6s^2, 6s~6p, 5d~6p \right)$, where $E$ is measured from the lowest level of the correct ground state in the NIST ASD.
The best strategies for the potentials in the present calculations are summarized in Table~\ref{tab:strategy}.

\subsection{Results}

Figures~\ref{fig:elevel_config1} and \ref{fig:elevel_config2} show the calculated energy level distributions for each configuration and element.
In the figures, the lowest levels of each configuration from the NIST ASD and {\sc Grasp} results (G19 for Nd, R20 for Pr and Pm - Gd, and R21 for Tb - Yb, respectively) are also plotted for comparison (marked by stars).
The differences are 10 - 25 \% in the median, except for Gd II (42 \%) (see Table~\ref{tab:strategy}).
This agreement is significant as compared with Paper I that remained much larger differences (20 - 100 \%).

We optimized the potential for Gd II to the excited state $4f^8~6s$ exceptionally because the potential optimized to the ground state as for the other elements, i.e., ~$4f^7~5d~6s$ for Gd II, gave the lowest levels of $4f^8~nl$ configurations far high from reference values of the NIST ASD.
As a result, the low-lying levels of $4f^7~5d^2$ and $4f^8~6s$ spread below the lowest level of $4f^7~5d~6s$ in the present calculations. 
Nevertheless, the median of the errors from the reference values is smaller than that obtained with the potential optimized to the correct ground state.

It may be noteworthy that the energy sequence of the $4f^{q-1} (5d^2,5d~6s, 6s^2)$ levels seems different for light and heavy lanthanides: the energy levels of the $5d^2$ become higher for heavy lanthanides, while those of the $6s^2$ become relatively lower. This can be ascribed to variation of the binding energies of the $5d$ and $6s$ orbitals along $Z$.
As $Z$ increases, the $5d$ orbital becomes more loosely bounded due to screening of the nuclear charge, while the $6s$ orbital has an almost constant binding energy for $Z=59-69$. Therefore, because substituting electrons from the inner $4f$ orbital to the $5d$ orbital becomes energetically more unfavorable, the $5d^2$ levels become higher, and the $6s^2$ levels become relatively lower for heavy lanthanides.

The present calculations give basically lower level distributions of excited states than those of Paper I. Figure \ref{fig:Sm+Ho} shows examples for Sm II and Ho II below 6 eV. It is clear that the present calculations of cumulative level distributions tend to be consistent with the {\sc Grasp} results.
The lowering of the excited states level distributions will in principle give more bound-bound transitions at longer wavelengths resulting in an increase of the opacity (see Section~\ref{sec:opacity}).
Comparison of the number of levels in 6 eV from the ground level is given for all the elements of $Z=59-70$ in Table~\ref{tab:levels}.
The table also shows the total number of levels obtained by the present RCI calculations for each configuration. Assignment of the configuration is done by the leading composition of the eigenvector.

\begin{figure*}
\centering
\begin{tabular}{cc}
  \includegraphics[width=8.0cm]{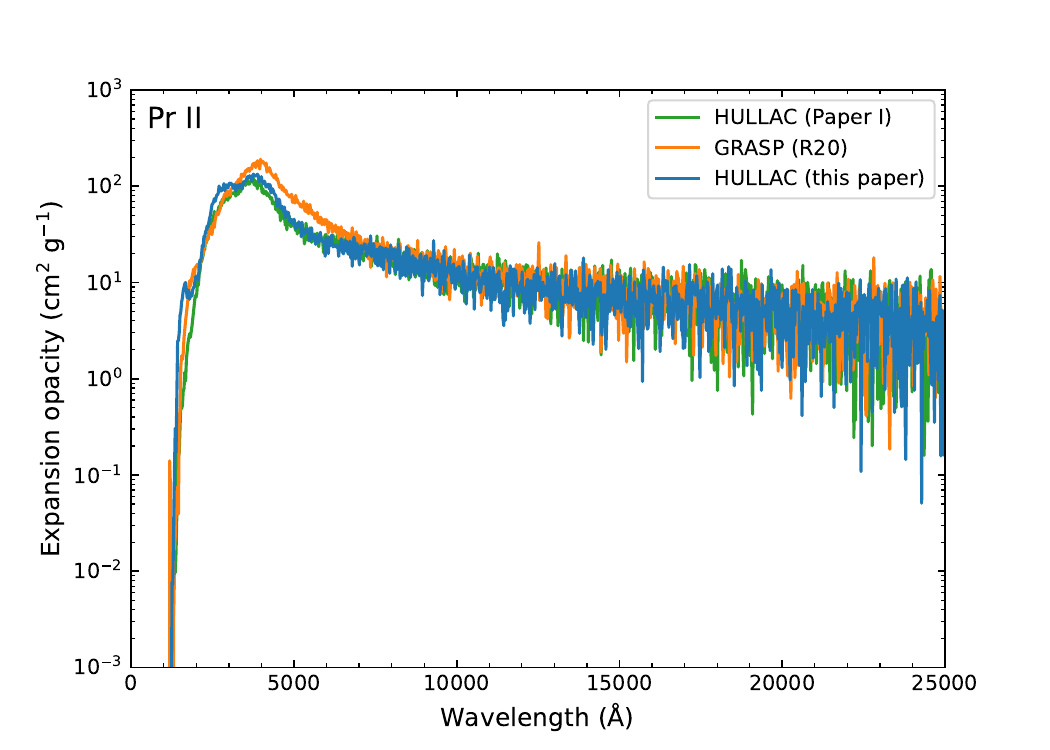}&
  \includegraphics[width=8.0cm]{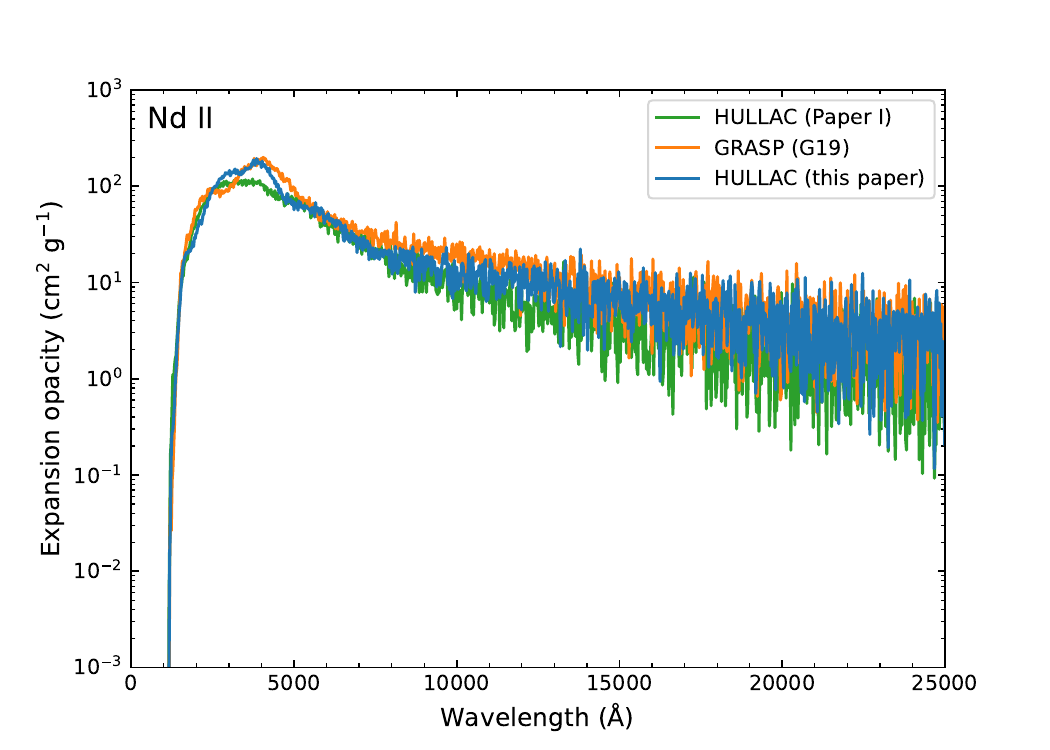}\\
  \includegraphics[width=8.0cm]{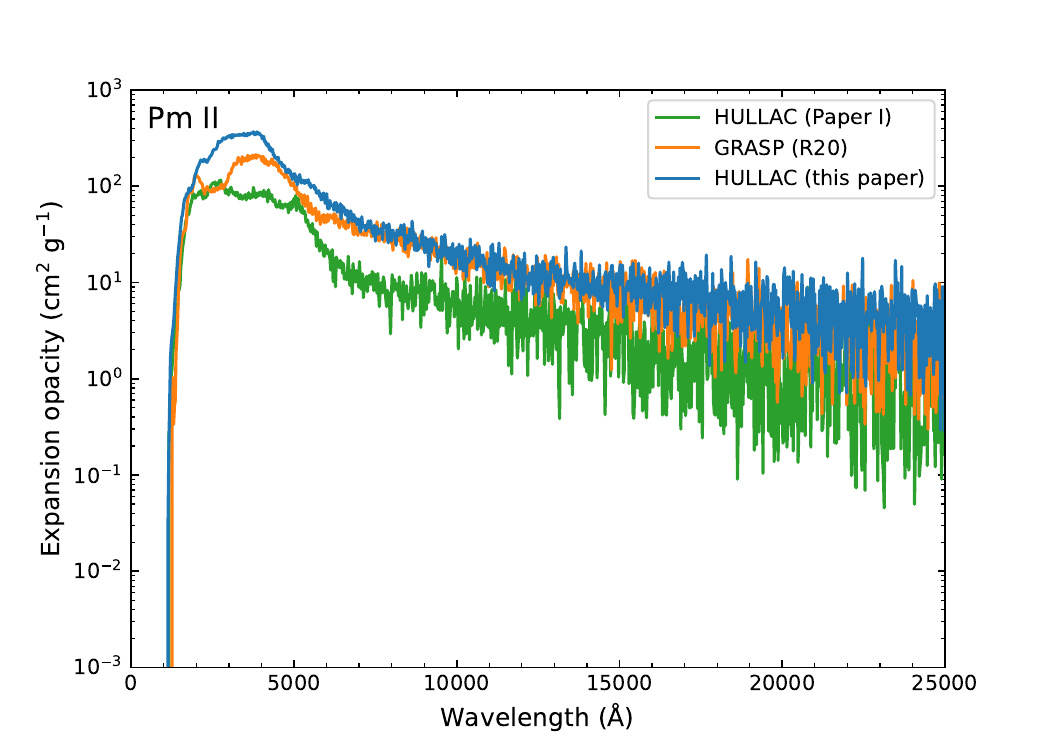}&
  \includegraphics[width=8.0cm]{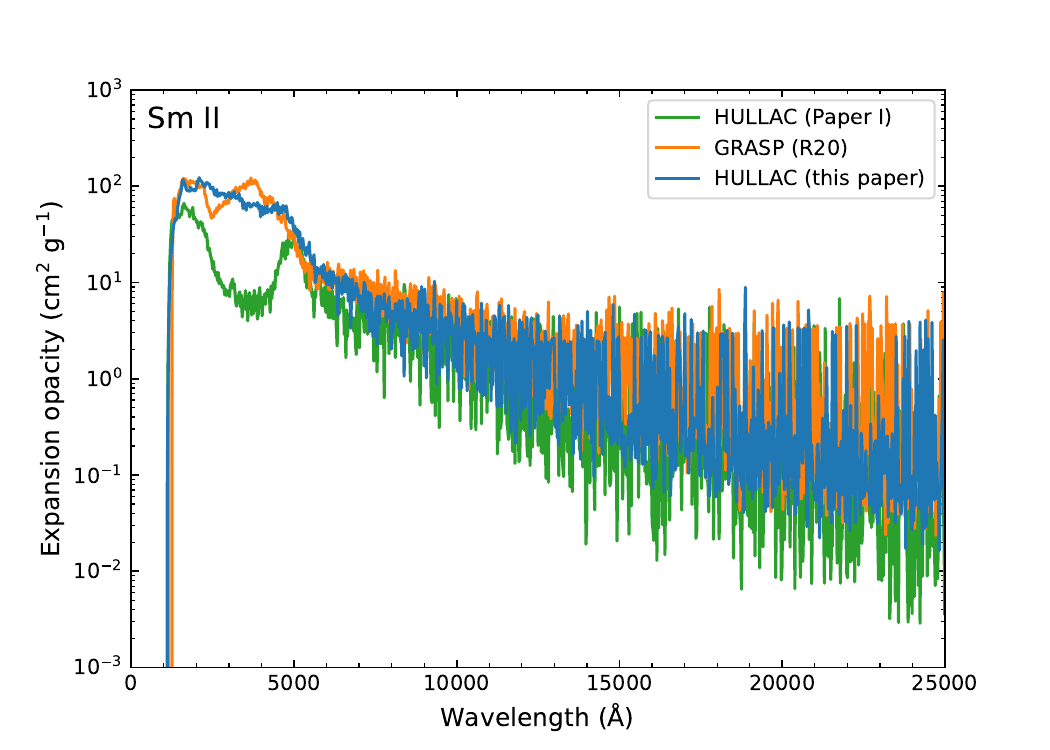}\\
  \includegraphics[width=8.0cm]{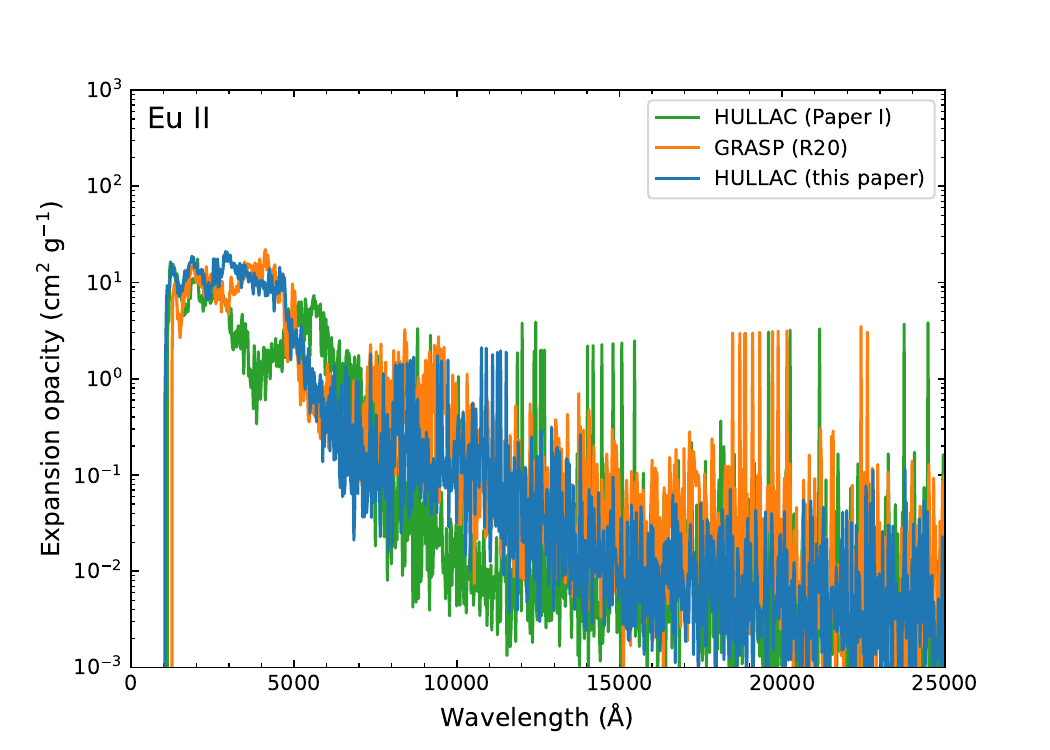}&
  \includegraphics[width=8.0cm]{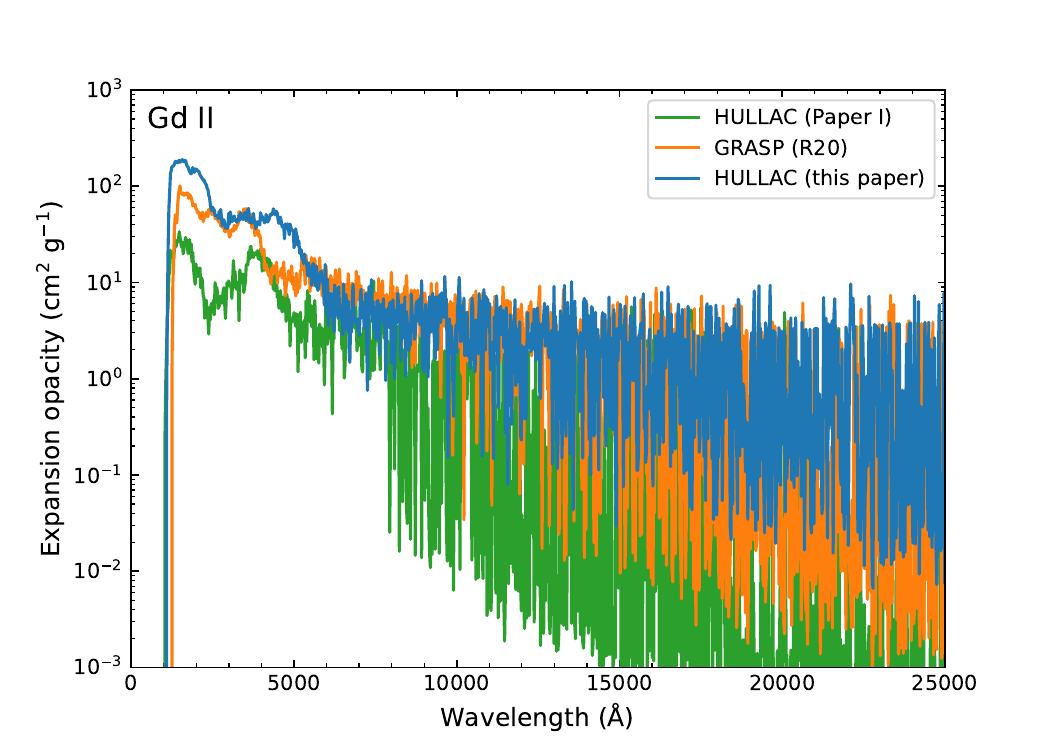}\\  
\end{tabular}
\caption{Expansion opacity as a function of wavelength for singly ionized lanthanides ($Z = 59-64$). The opacities are those for $\rho = 10^{-13} \ {\rm g\ cm^{-3}}$ and $T = 5000$ K at $t = 1$ day after the merger.}
\label{fig:kappa1}
\end{figure*}

\begin{figure*}
\centering
\begin{tabular}{cc}
  \includegraphics[width=8.0cm]{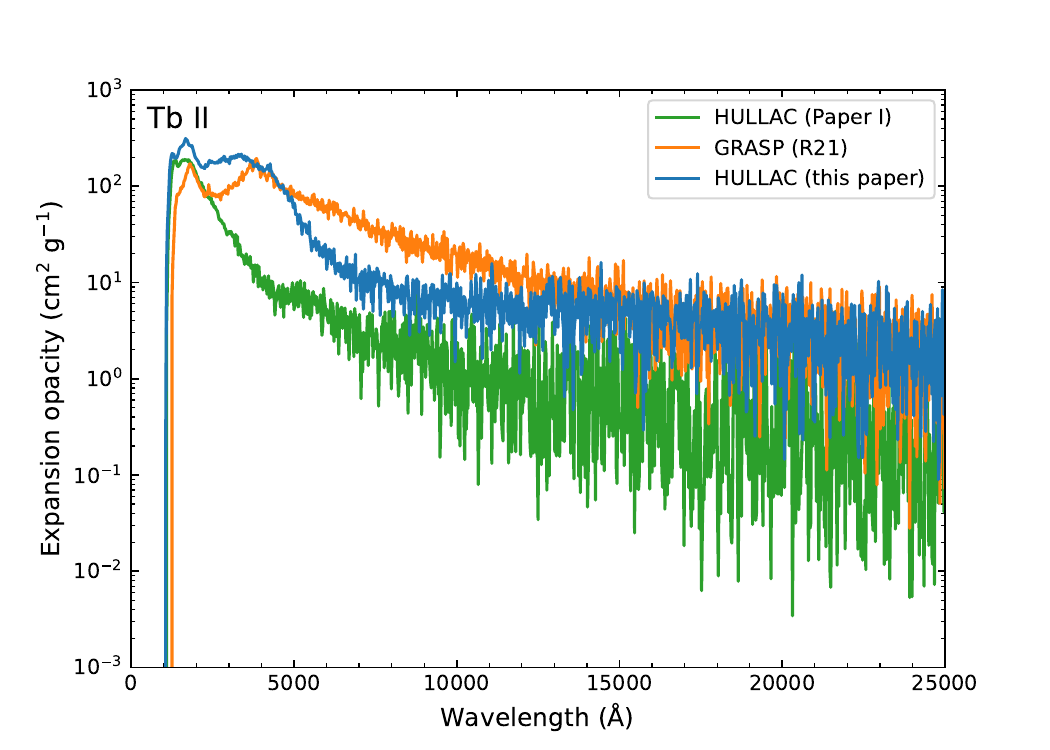}&
  \includegraphics[width=8.0cm]{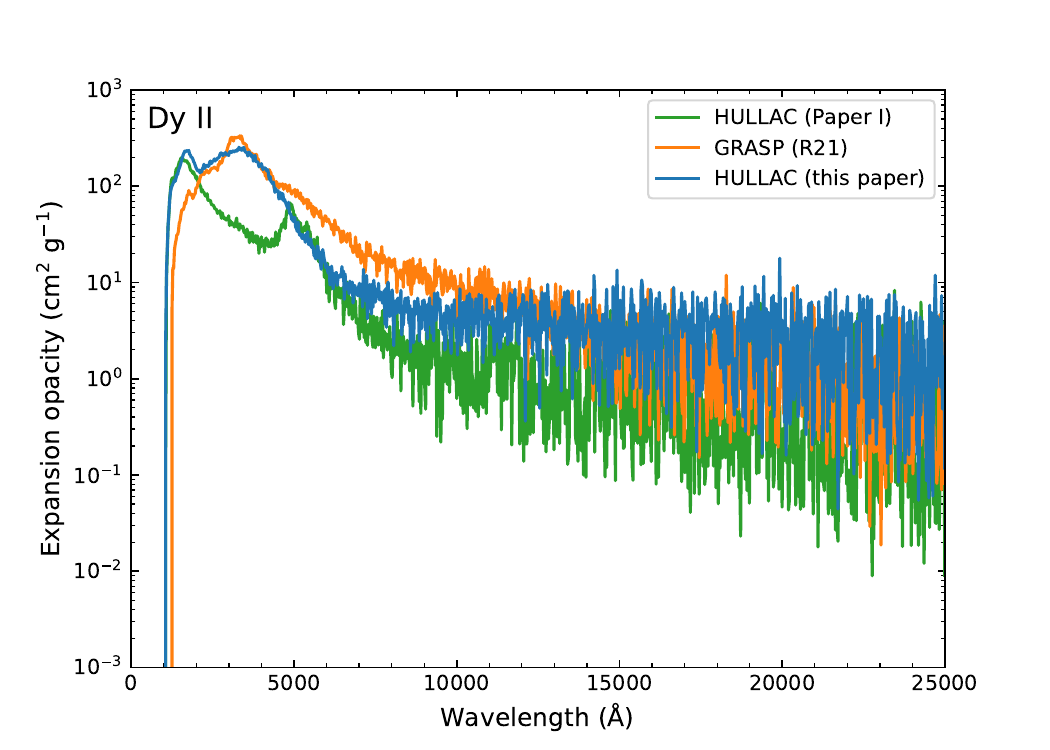}\\
  \includegraphics[width=8.0cm]{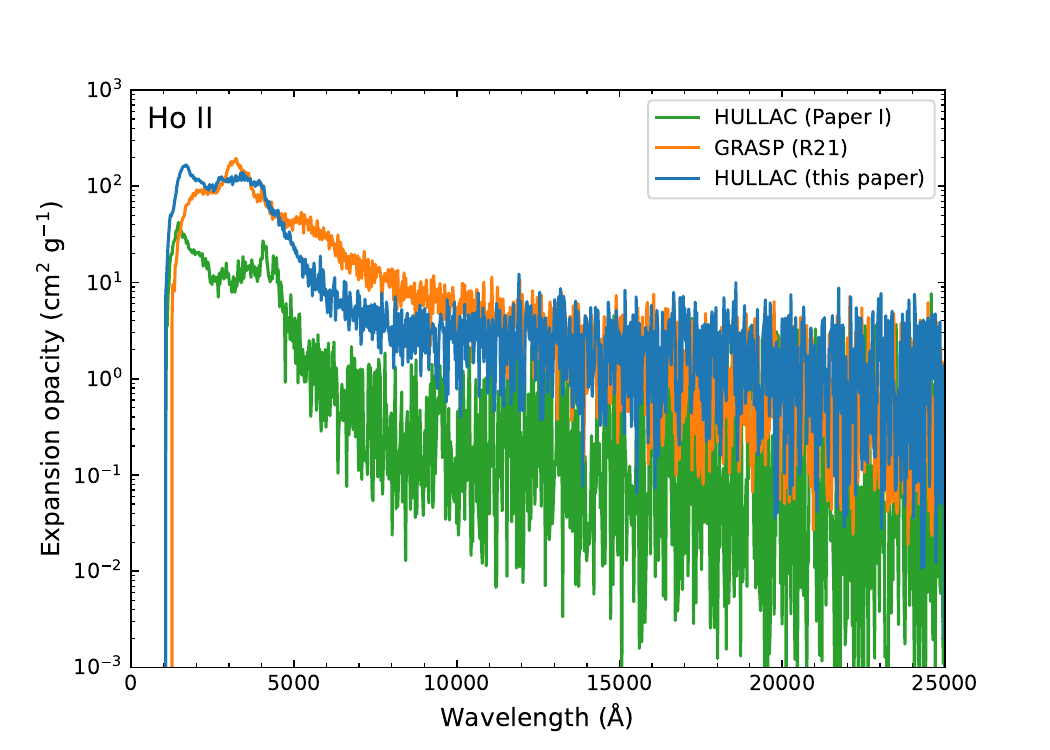}&
  \includegraphics[width=8.0cm]{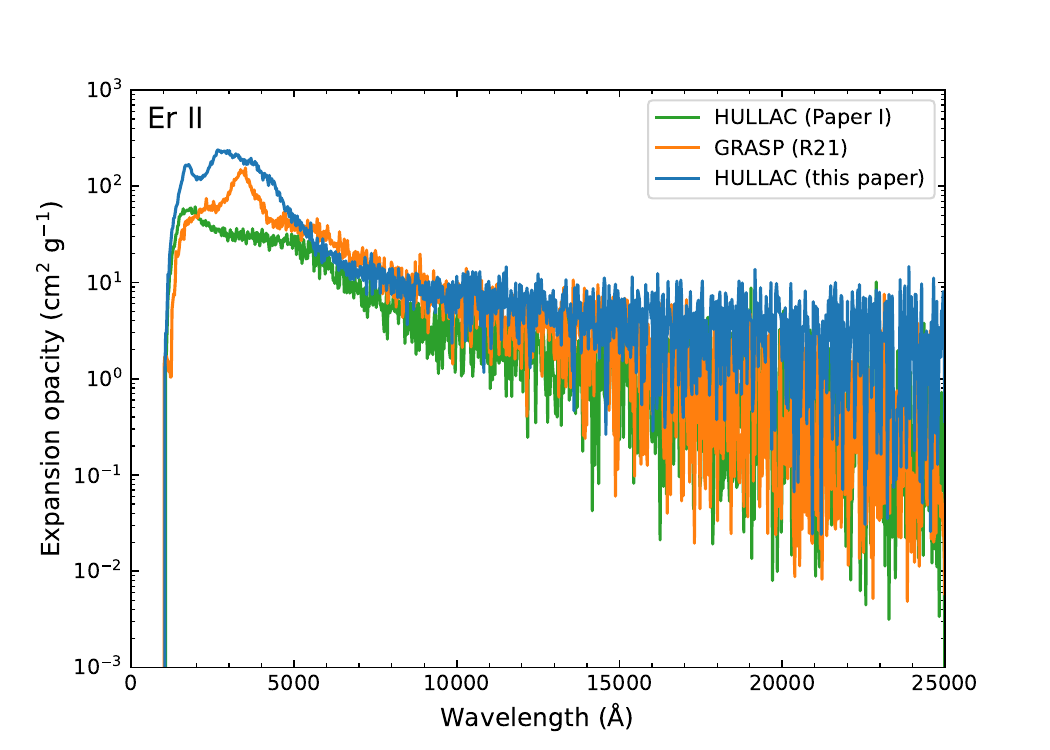}\\
  \includegraphics[width=8.0cm]{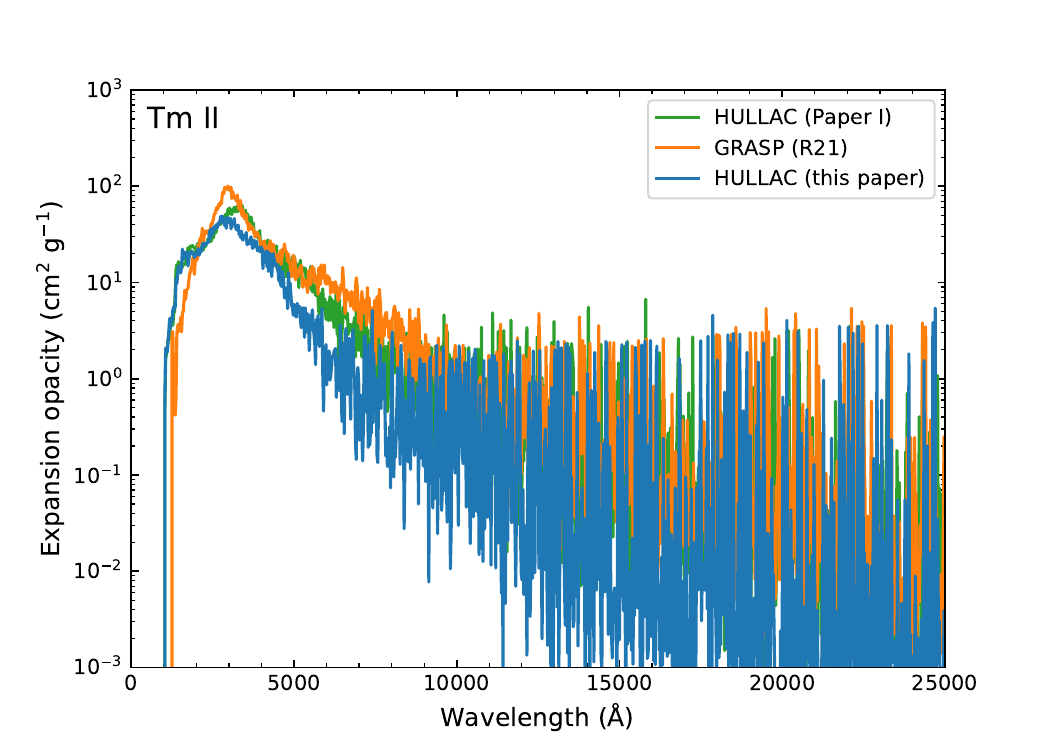}&
  \includegraphics[width=8.0cm]{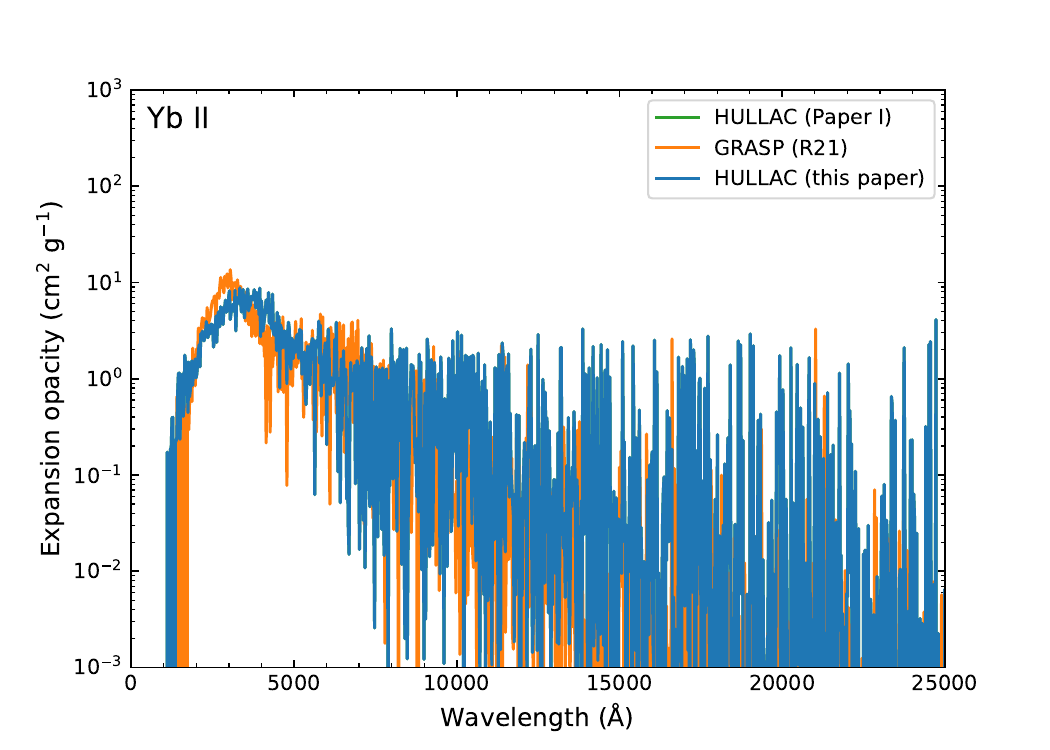}\\  
\end{tabular}
\caption{Same as Figure \ref{fig:kappa1} but for $Z=65-70$. Note that the {\sc HULLAC} results for Yb II in Paper I and this paper are identical.}
\label{fig:kappa2}
\end{figure*}


\section{Opacity calculations}
\label{sec:opacity}

\subsection{Methods}

By using the results of atomic calculations described in
Section \ref{sec:atomic}, we calculate the bound-bound opacities
in the ejecta of NS merger.
As we follow the same methods as in Paper I,
here we only give a brief overview.
In the rapidly expanding medium with a large velocity gradient,
such as ejecta of supernova or NS merger,
the bound-bound opacity for a certain wavelength grid
($\Delta \lambda$) can be evaluated by 
so called expansion opacity formalism
\citep[e.g.,][]{karp77,eastman93,kasen06}:
\begin{equation}
  \kappa_{\rm exp} (\lambda) 
= \frac{1}{ct \rho} 
\sum_l \frac{\lambda_l}{\Delta \lambda} \left(1 - e^{- \tau_l} \right).
\label{eq:kappa}
\end{equation}
Here $\lambda_l$ and $\tau_l$ are the transition wavelength
and the Sobolev optical depth for each transition, respectively, $\rho$ is mass density, and $t$ is time after the merger.
The summation in the equation is taken over all the bound-bound transitions
in a certain wavelength bin.
In the case of homologous expansion ($r=vt$),
which is a sound assumption in the ejecta at the epoch of interest
($t>$ a few hours),
the Sobolev optical depth is expressed as
\begin{equation}
  \tau_l = \frac{\pi e^2}{m_e c} f_{l} n_{i,j,k} t \lambda_{l},
\end{equation}
where $f_l$ is the oscillator strength of the transition
and $n_{i,j,k}$ is the number density of $n$-th element in
$m$-th ionization state and $k$-th excited state.

We calculate ionization and excitation under the assumption
of local thermodynamical equilibrium (LTE).
By Boltzman distribution, $n_{i,j,k}$ = $ n_{i,j} (g_k/\Sigma_{i,j} (T)) \exp (-E_k/kT)$,
where $g_k$ and $E_k$ are a statistical weight and energy of the
lower level of the transition, respectively.
Here $\Sigma_{i,j} (T)$ is the partition function for the $i-$th element
at $j-$th ionization state
\footnote{
In our previous works to calculate the opacities \citep{tanaka13,tanaka18},
we assumed $n_{i,j,k}$ = $ n_{i,j} (g_k/g_0) \exp (-E_k/kT)$,
and $g_0$ is evaluated as a sum of the statistical weight for 
the levels with the same $LS$ term with the ground level
as there were no atomic data covering the entire energy spectra are available.
In Paper I, we also used the same scheme.
In this paper (and also \citealt{banerjee24}),
we calculate a temperature dependent partition function.
We confirmed that the previous assumption has negligible impact to the opacity
in the temperature range of interest ($T < 25,000$ K).}.
The number density of the ion $n_{i,j}$ is evaluated by solving the Saha equation.
To derive the ionization degrees, we also need the partition functions for ionization states other than singly ionized states. For these, we used the results of Paper I.

\begin{table}
\begin{center}
\caption{Number transitions for each ion. $^a$ Total number calculated with {\sc HULLAC} in this paper. $^b$ The number of transitions that satisfy 
$gf \exp(-E_l/kT) > 10^{-5}$ at $T = 5000$ K.}
\label{tab:lines}
\begin{tabular}{crr}
\hline\hline
Ion & $N_{\rm total}$ $^a$ & $N_{\rm strong}$ $^b$\\
\hline
Pr II & 417~812      & 62~511   \\
Nd II & 4~001~851    &  67~934  \\   
Pm II & 21~472~279   & 111~540  \\
Sm II & 69~895~982   &  35~692 \\
Eu II & 132~942~648  & 2~330  \\
Gd II & 158~102~969  &  31~961  \\
Tb II & 119~471~719  & 98~484  \\
Dy II &  54~784~938  & 49~185  \\
Ho II &  15~301~547  &  29~399  \\
Er II & 2~432~667    &  30~397  \\
Tm II & 205~259      &  3~424  \\
Yb II & 8~110        &  274  \\
\hline
\end{tabular}
\end{center}
\end{table}

\subsection{Results}

In Figures \ref{fig:kappa1} and \ref{fig:kappa2},
we show the expansion opacity at $t=1$ day
as a function of wavelength for each element.
The opacities are calculated for
$\rho = 10^{-13} \ {\rm g\ cm^{-3}}$ and $T = 5000$ K,
which is a typical plasma condition for
the NS merger ejecta with an ejecta mass of an order of 0.01 $M_{\odot}$
and a typical velocity of about $v \sim 0.1 c$ at $t = 1$ day.
We choose this early time ($t=1$ day) as deviation from LTE
is known to be significant in particular in the outer ejecta
after several days after the merger \citep{hotokezaka21,pognan22}.

In Figures \ref{fig:kappa1} and \ref{fig:kappa2},
it is assumed that the ejecta consists of single element
(see Section \ref{sec:discussions} for more realistic elemental compositions).
To compare the opacity calculated from different atomic data,
we calculate the opacity only for the singly ionized states
as {\sc Grasp} data are available only for the singly ionized state.
Note that we still solve the ionization to derive the number density of each ion $n_{i,j}$.
At the adopted density and temperature, singly ionized states gives dominant contributions to the opacities. 

As shown in Figures \ref{fig:kappa1} and \ref{fig:kappa2},
the opacities evaluated with our new atomic data 
are generally higher than those in Paper I.
In particular, the opacities of Pm II, Sm II, Eu II, Gd II, Tb II, Dy II, Ho II, and Er II at $< 5000$ \AA\ show a large deviation up to by a factor of about 10.
These differences stem from the energy distribution as discussed in Section \ref{sec:atomic}. 
Our improved calculations tend to show lower energy level distributions as compared with Paper I.
As a result, the number of strong transitions increases by the higher population of excited states through the Boltzmann factor $\exp(-E_l /kT)$.

Our new opacities show reasonable agreements with 
those calculated with the results of {\sc Grasp} calculations (G19, R20 and R21).
In particular, for the elements with a large opacity increase with respect to Paper I, the agreement between our new opacities and {\sc Grasp} opacities is quite well in particular at $< 5000$ \AA.
However, there are still a few cases that show a large discrepancy at 5000-10000 \AA\ (Tb II, Dy II, and Ho II). This is discussed in Section \ref{sec:discussions} in more details.

\begin{figure}
\centering
\includegraphics[width=\hsize]{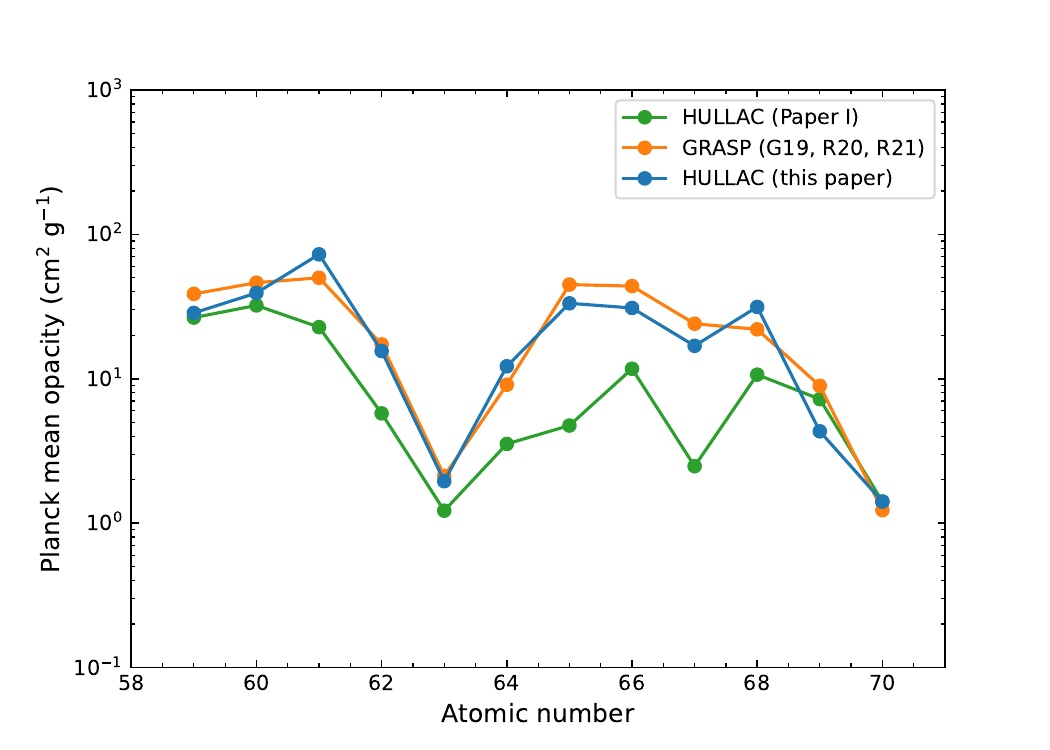}
\caption{Plank mean opacity (for $\rho = 10^{-13} \ {\rm g\ cm^{-3}}$ and $T = 5000$ K at $t = 1$ day after the merger) as a function of atomic number .
}
\label{fig:opacity_z}
\end{figure}

Figure \ref{fig:opacity_z} summarizes our results for all the elements with $Z=59-70$.
To define a characteristic opacity for each element, we evaluate Planck mean opacity with $T = 5000$ K.
As discussed above, the newly calculated opacities are in general higher than those in Paper I, giving a better agreement with {\sc Grasp} opacities.
The entire temperature dependences of the Planck mean opacities are shown in Appendix A (Figures \ref{fig:mean1} and \ref{fig:mean2}).

\begin{figure*}
\centering
\begin{tabular}{cc}
  \includegraphics[width=8.0cm]{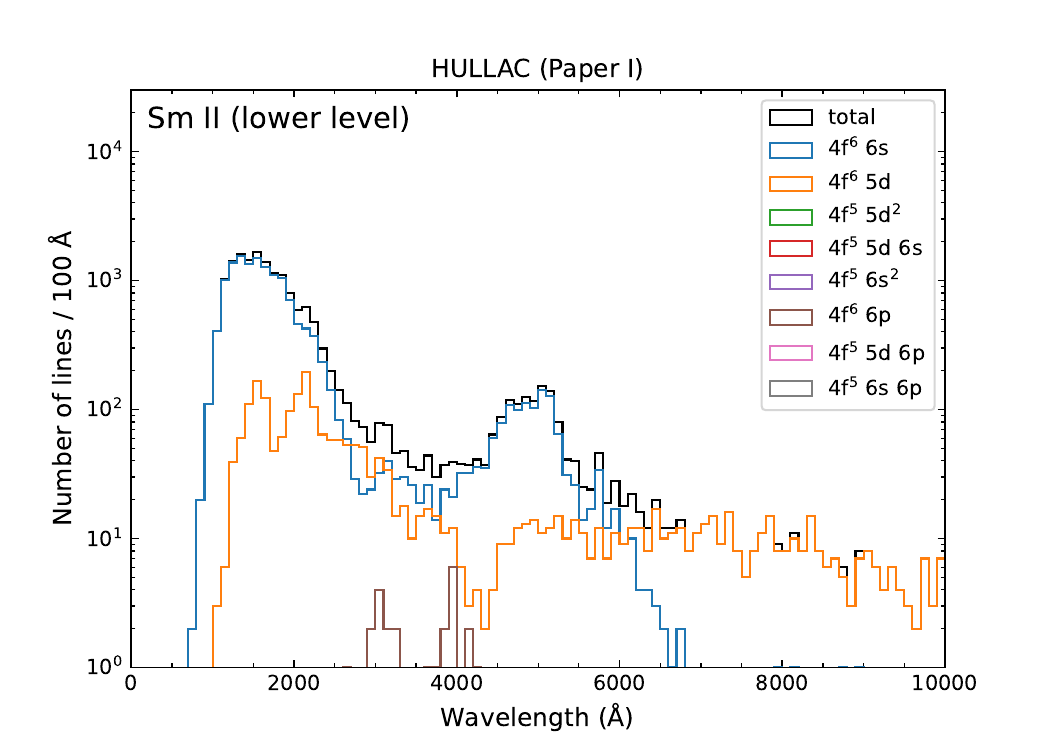}&
  \includegraphics[width=8.0cm]{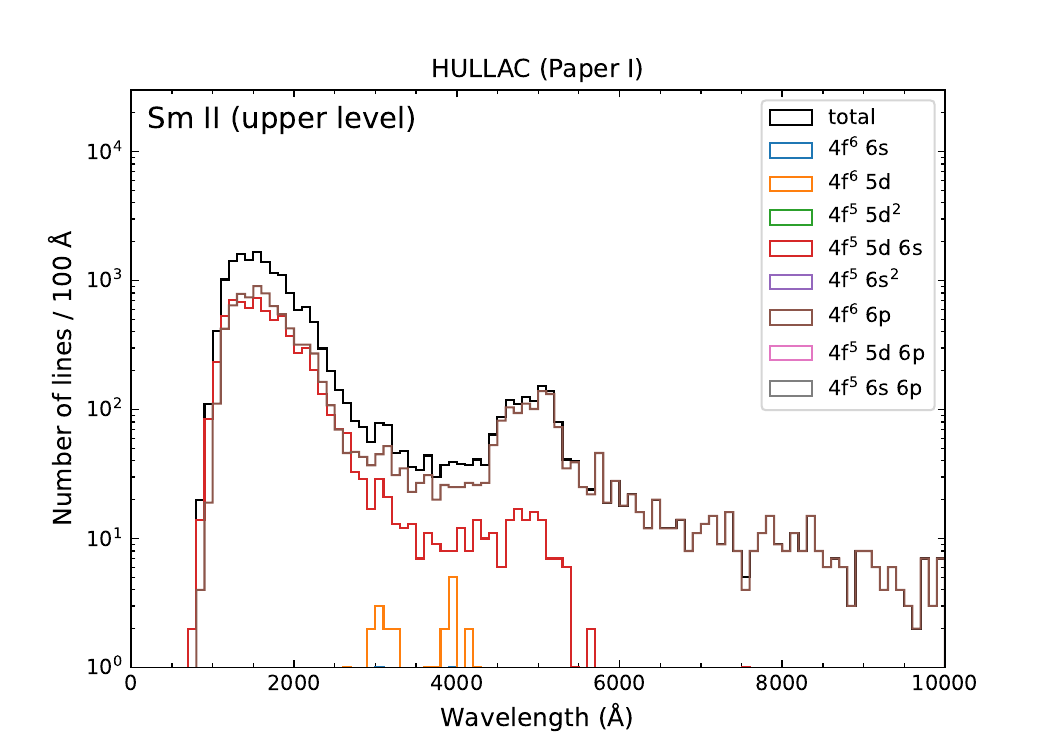}\\
  \includegraphics[width=8.0cm]{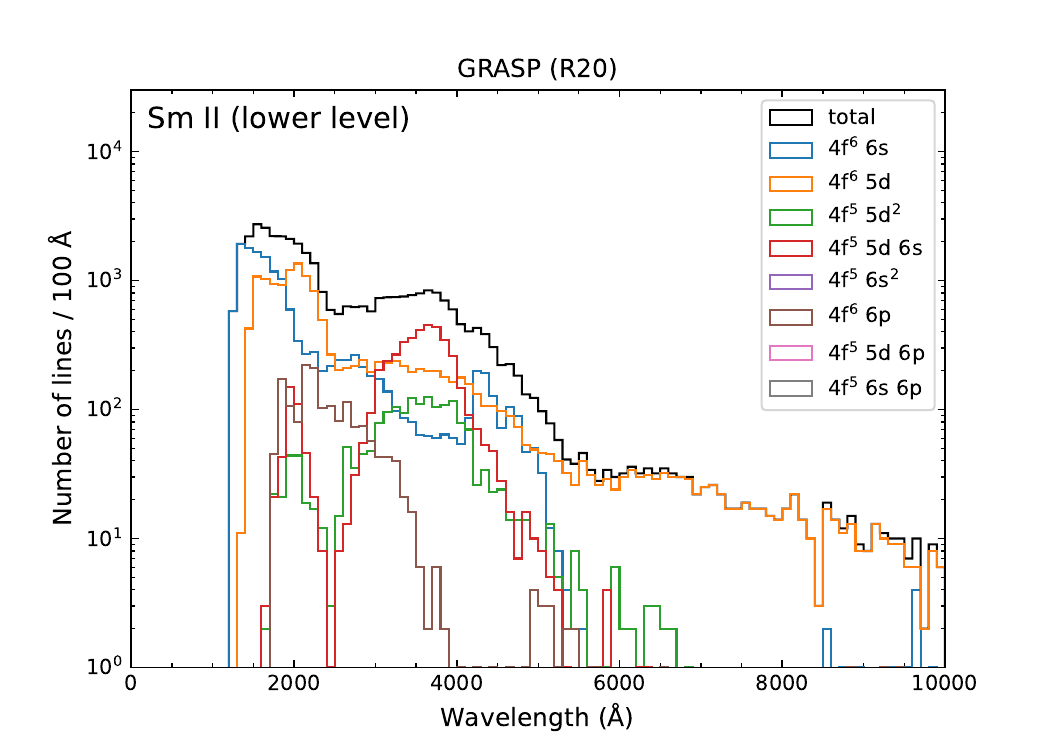}&
  \includegraphics[width=8.0cm]{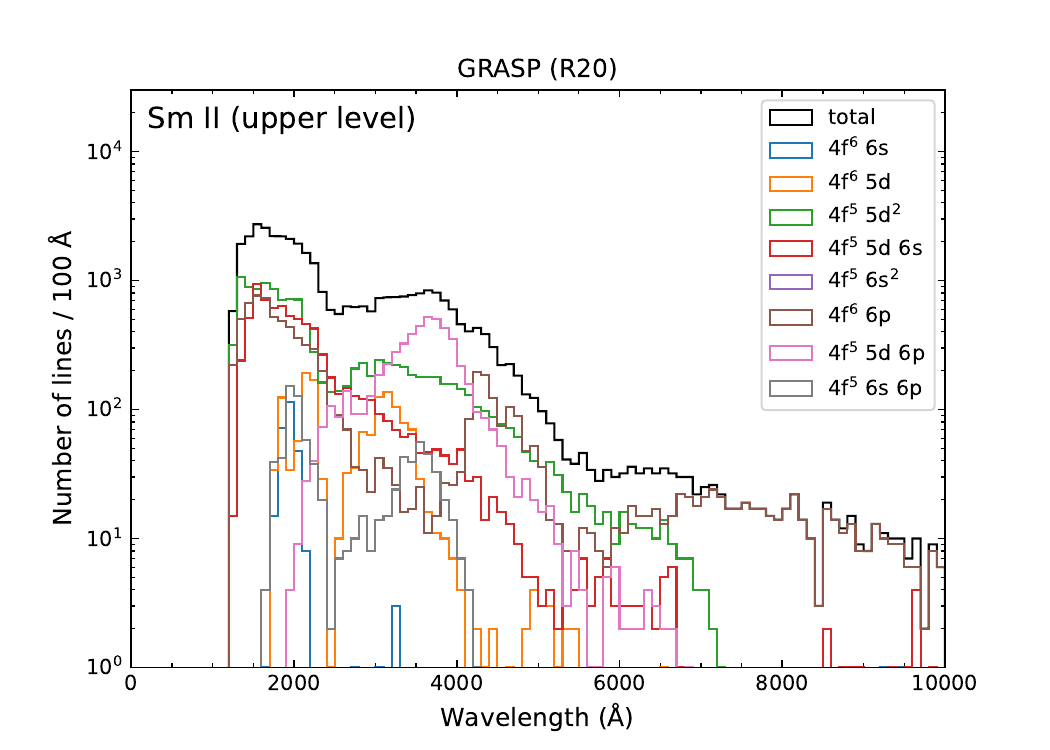}\\
  \includegraphics[width=8.0cm]{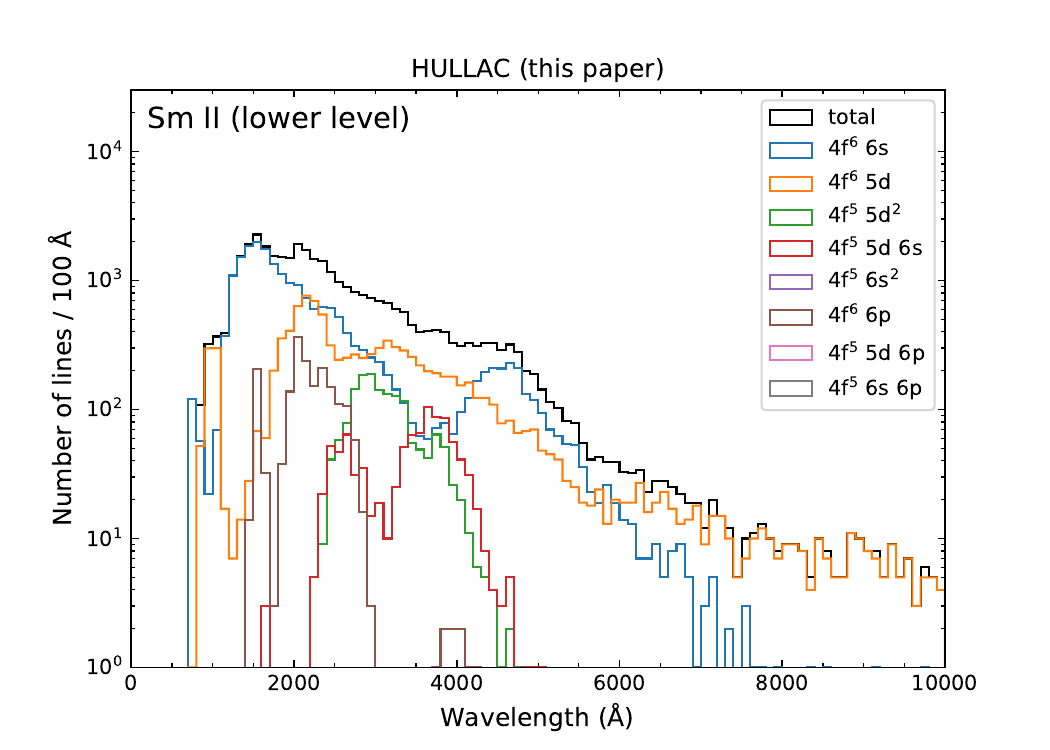}&
  \includegraphics[width=8.0cm]{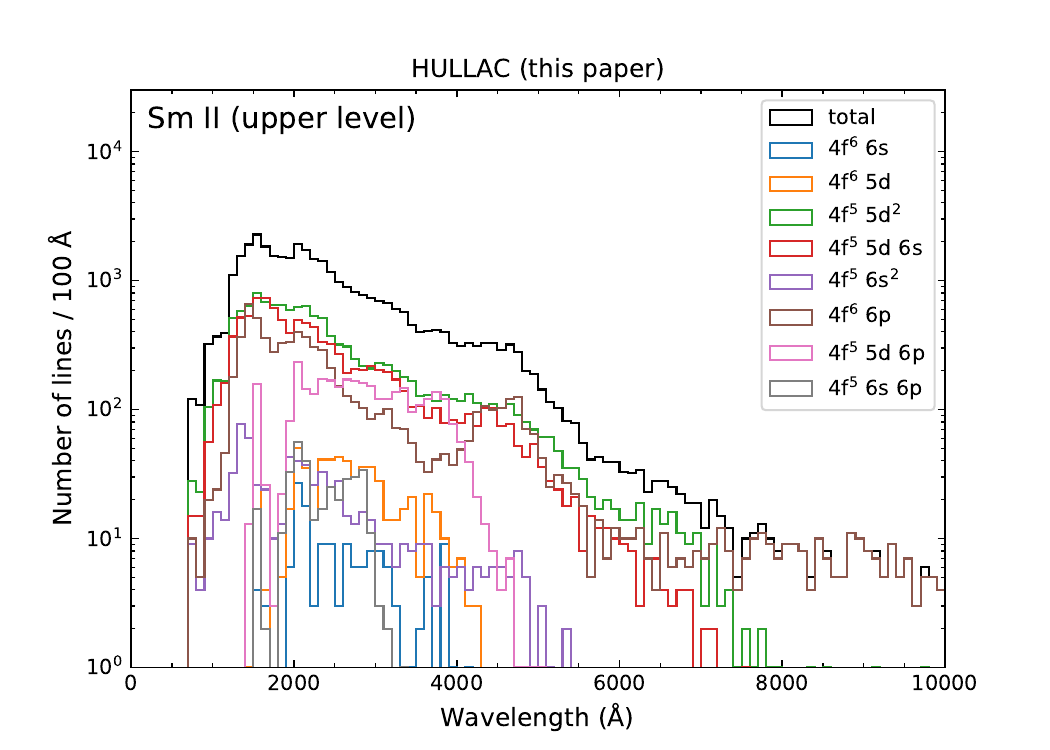}\\
\end{tabular}
\caption{The number of strong lines (per 100 \AA\ bin) for Sm II that satisfy $gf \exp(-E_l/kT) > 10^{-5}$ at $T = 5000$ K. Top, middle, and bottom panels show the cases using the {\sc Hullac} calculations in Paper I, {\sc Grasp} calculations (R20), and this paper, respectively. In the left and right panels, the number of the lines are shown according to their lower and upper configurations, respectively.}
\label{fig:nline_62}
\end{figure*}

\begin{figure*}
\centering
\begin{tabular}{cc}
  \includegraphics[width=8.0cm]{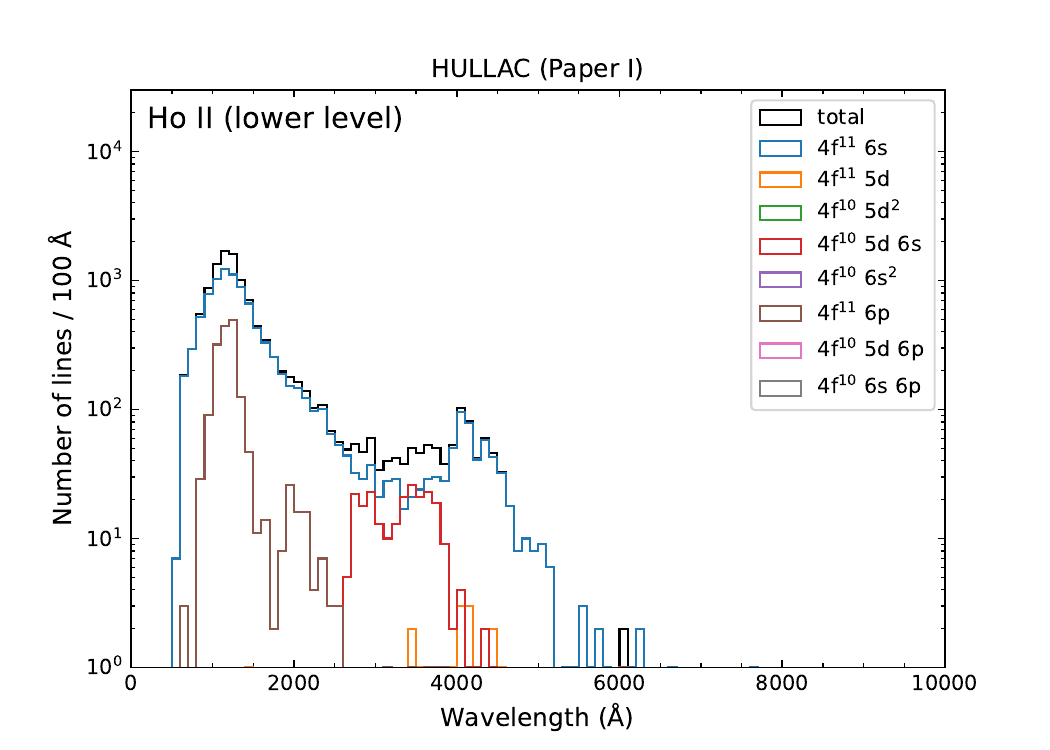}&
  \includegraphics[width=8.0cm]{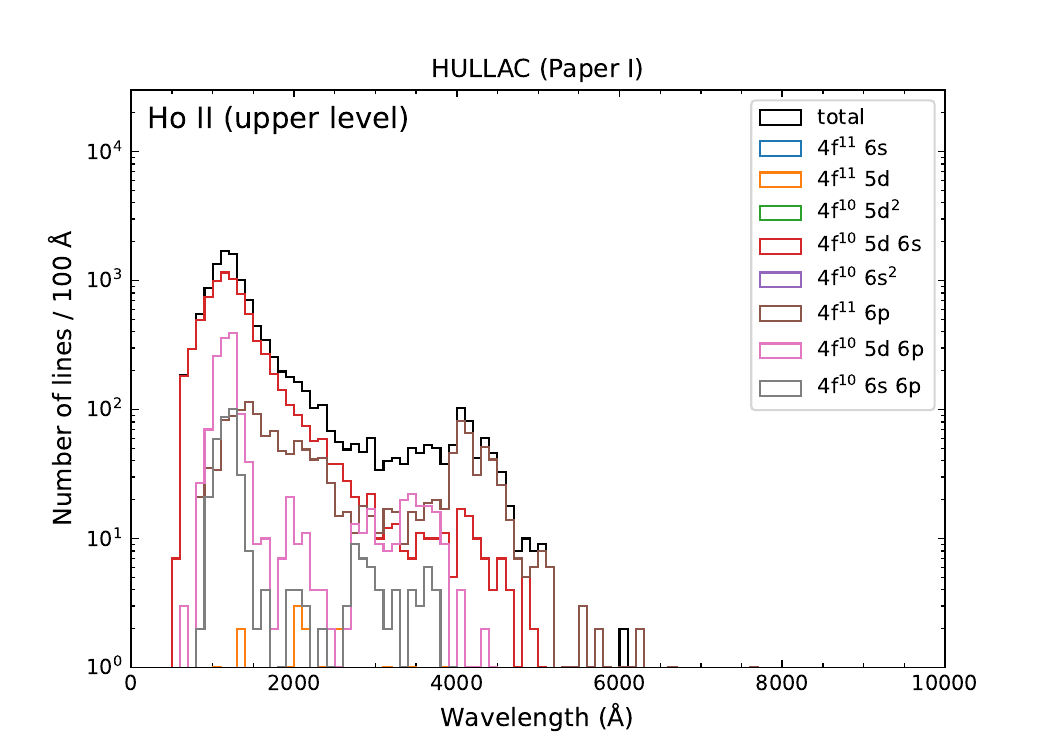}\\
  \includegraphics[width=8.0cm]{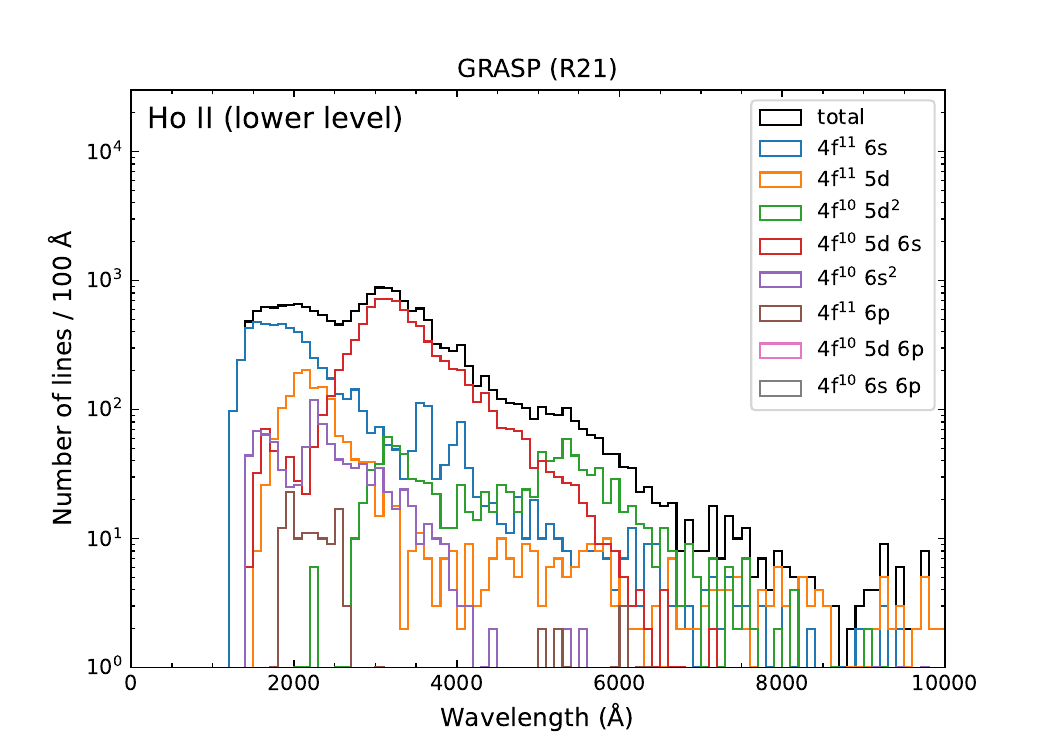}&
  \includegraphics[width=8.0cm]{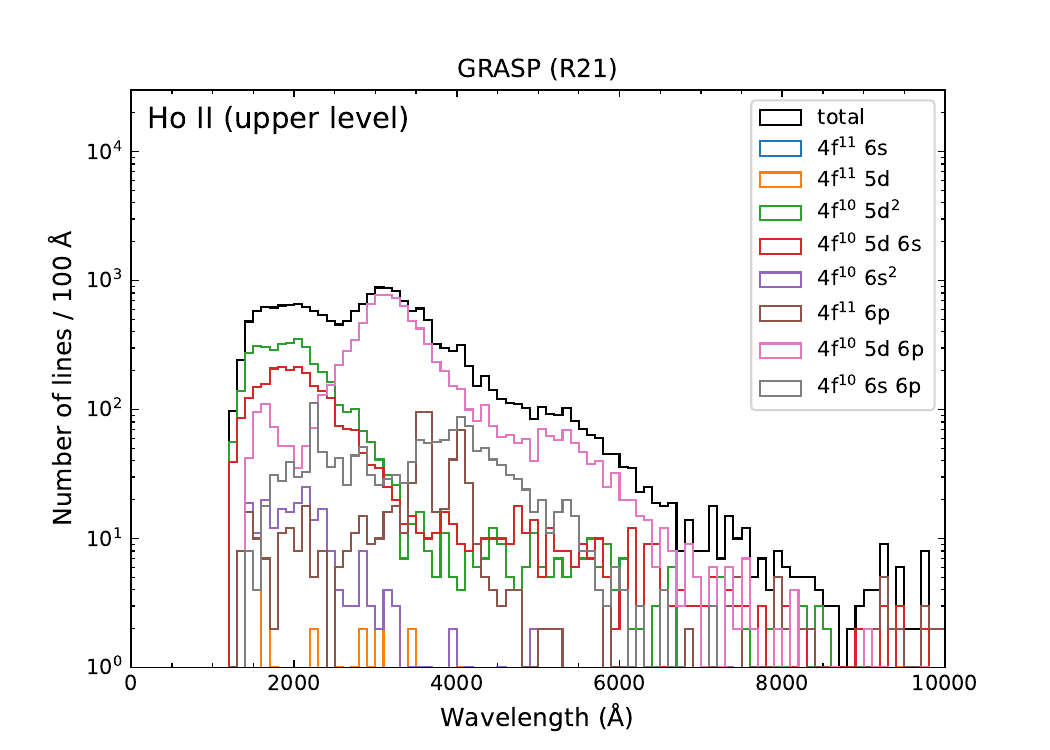}\\
  \includegraphics[width=8.0cm]{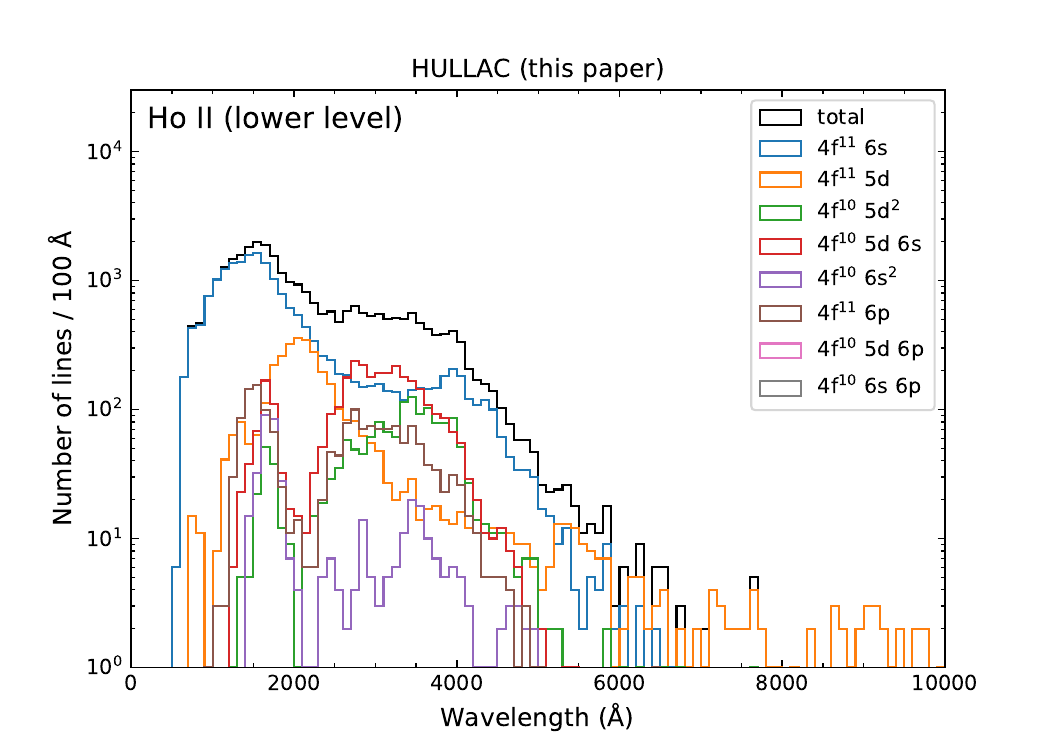}&
  \includegraphics[width=8.0cm]{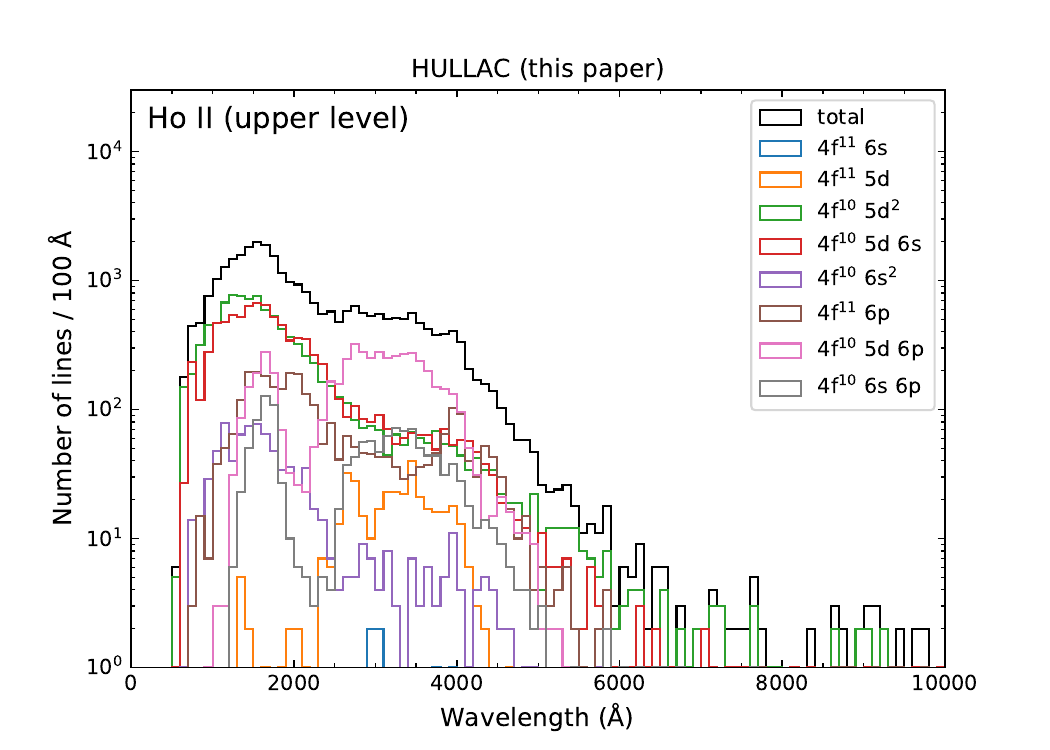}\\
\end{tabular}
\caption{Same as Figure \ref{fig:nline_62} but for Ho II. {\sc GRASP} results are from R21.}
\label{fig:nline_67}
\end{figure*}


\section{Discussions}
\label{sec:discussions}

\subsection{Properties of lanthanide opacities}

We have performed atomic calculations for singly ionized lanthanides
with {\sc Hullac} code with improved strategies.
We show that, compared with our previous calculations in Paper I,
the overall energy level distributions are shifted toward
lower energy.
This results in increase of the opacities through the higher
populations of excited levels for a given temperature.

To understand which configurations play important roles
in the lanthanide opacities, we here analyze the calculated opacities.
As demonstrated in \citet{gaigalas19},
the number of strong lines as a function of wavelength
gives a good measure of the opacity as the expansion opacity
is determined by the sum of $1-\exp(-\tau_l)$ for each wavelength bin.
Here, by following \citet{gaigalas19},
we select strong lines that satisfy
$gf \exp(-E_l/kT) > 10^{-5}$ at $T = 5000$ K.
Table \ref{tab:lines} summarizes the number of strong lines satisfying the condition above for each ion.

The results of the analysis are shown for the case of Sm II
in Figure \ref{fig:nline_62}.
The black lines show the number of strong lines for Sm II (black lines)
as a function of wavelength.
The left panels show that break down of the number of lines
according to the lower level configurations 
while the right panels show the same according to
the upper level configurations.
The same analysis are shown for our {\sc Hullac} calculations in Paper I (top),
{\sc Grasp} calculations from R20 (middle), and 
our new {\sc Hullac} calculations in this paper (bottom).
It is confirmed that the black line in each calculation
represent the characteristic features in the opacity,
as demonstrated by \citet{gaigalas19} for Nd.

At $\lambda < 6000$ \AA,
the strong lines are dominated by those from $4f^6 \ 6s$ as a lower configuration,
followed by those from $4f^6 \ 5d$, $4f^6 \ 6p$,  $4f^5 \ 5d \ 6s$,
and $4f^5 \ 5d^2$.
The corresponding upper configurations for these strong lines are either $4f^5 \ 5d^2$, $4f^5 \ 5d \ 6s$,
$4f^6 \ 6p$ or $4f^5 \ 5d \ 6p$.
At $\lambda > 6000$ \AA, the lower configuration of strong lines is
almost entirely  $4f^6 \ 5d$.
The corresponding upper configurations are either
$4f^6 \ 6p$ and $4f^5 \ 5d^2$.

From this analysis, 
we can understand the reason why the opacity of Sm II in Paper I is smaller than that from {\sc GRASP} calculations (R20) and our {\sc HULLAC} calculations in this paper.
In Paper I, atomic calculations for Sm II did not include $4f^5 \ 5d^2$ and $4f^5 \ 5d \ 6p$ configurations.
As shown in the {\sc GRASP} calculations (R20) and the {\sc HULLAC} calculations in this paper, $4f^5 \ 5d^2$ configuration is important as a lower configuration, 
and both $4f^5 \ 5d^2$ and $4f^5 \ 5d \ 6p$ configurations are important as upper configurations.
Thus, lack of these two configurations causes a strong dip in the opacity around 3000-4000 \AA.

Overall, a similar trend can be seen for the case of Ho II
as shown in Figure \ref{fig:nline_67}.
In this case, the low opacity in Paper I is caused by
lack of $4f^{10} \ 5d^2$ as upper configurations and upward energy distribution of $4f^{10} \ 5d \ 6s$ configuration.
For Ho II, our {\sc Hullac} calculations and {\sc Grasp} calculations (R21)
still give a relatively large discrepancy in the opacities at $\lambda > 5000-10000$ \AA.
This is due to the higher energy levels of 
$4f^{10} \ 5d^2$ configuration in the {\sc Hullac} calculations (see even parity in Figure \ref{fig:Sm+Ho}).

\begin{figure*}
\centering
\includegraphics[width=13.5cm]{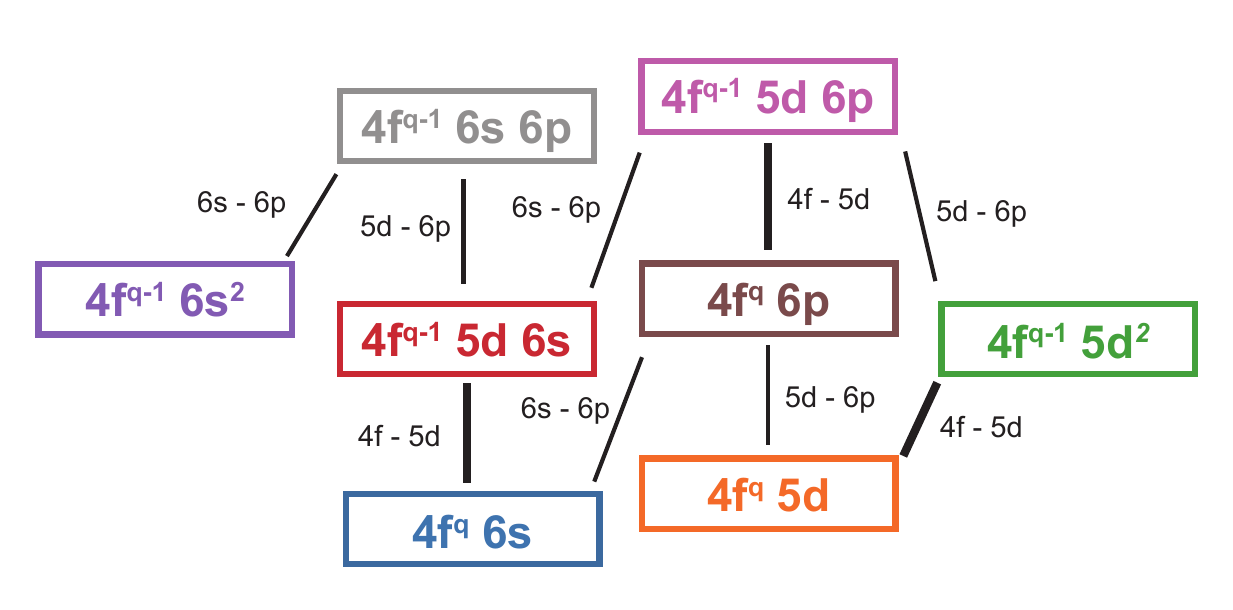}
\caption{Schematic summary of transition arrays for singly ionized lanthanides. Colors for configurations are according to the same color scheme in Figures \ref{fig:nline_62} and \ref{fig:nline_67}.
Since the energy levels of each configuration widely spread, each line shows the transition in either direction depending on the energy level ordering. For Gd II, the low-lying levels of $4f^7~5d~6s$ are largely overlapping with those of $4f^8~6s$ (see Figure~\ref{fig:elevel_config1}).
}
\label{fig:array}
\end{figure*}

It is interesting that transitions between the levels of
certain configurations are clustered in wavelength,
forming ``transition array'', as also discussed in
\citet{carvajalgallego24_TA}.
Since lanthanides have a large number of excited levels
with small energy separation,
many transitions can be clustered in a similar wavelength range.
The transition arrays for singly ionized lanthanides
is summarized in Figure \ref{fig:array}.
Our results demonstrate that it is important 
(1) to include these configurations
in the atomic structure calculations to
secure the completeness of the transitions,
and (2) to derive the accurate energy levels
for these configurations to obtain the reliable opacities.

\begin{figure*}
\centering
\begin{tabular}{cc}
\includegraphics[width=8.0cm]{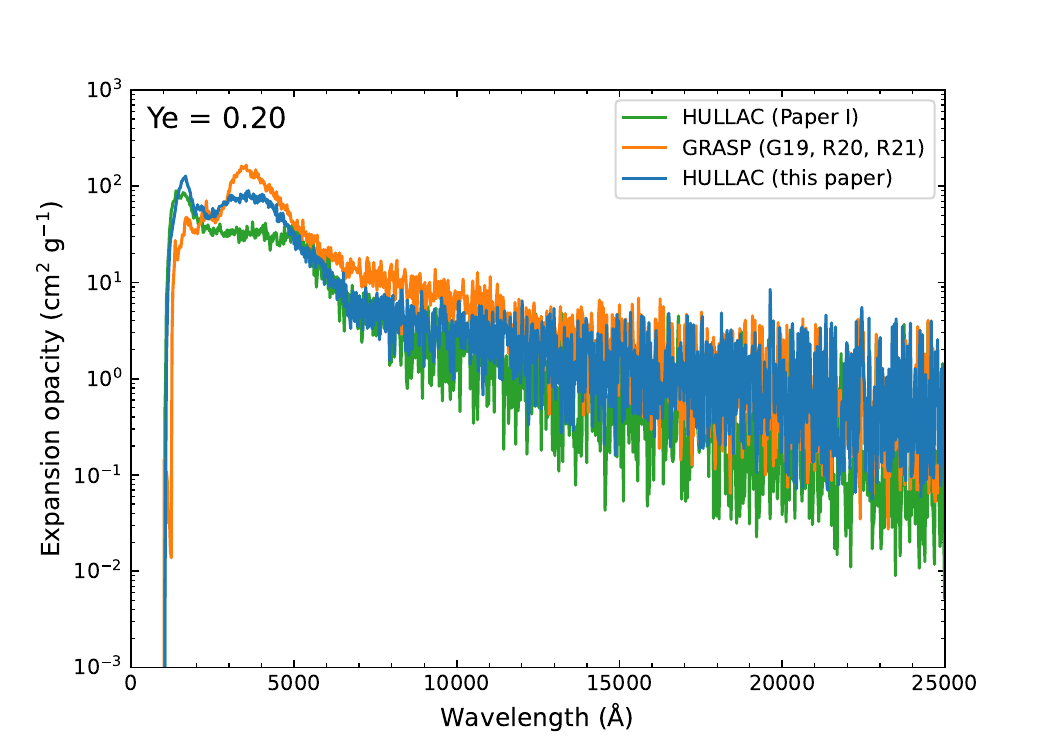}&
\includegraphics[width=8.0cm]{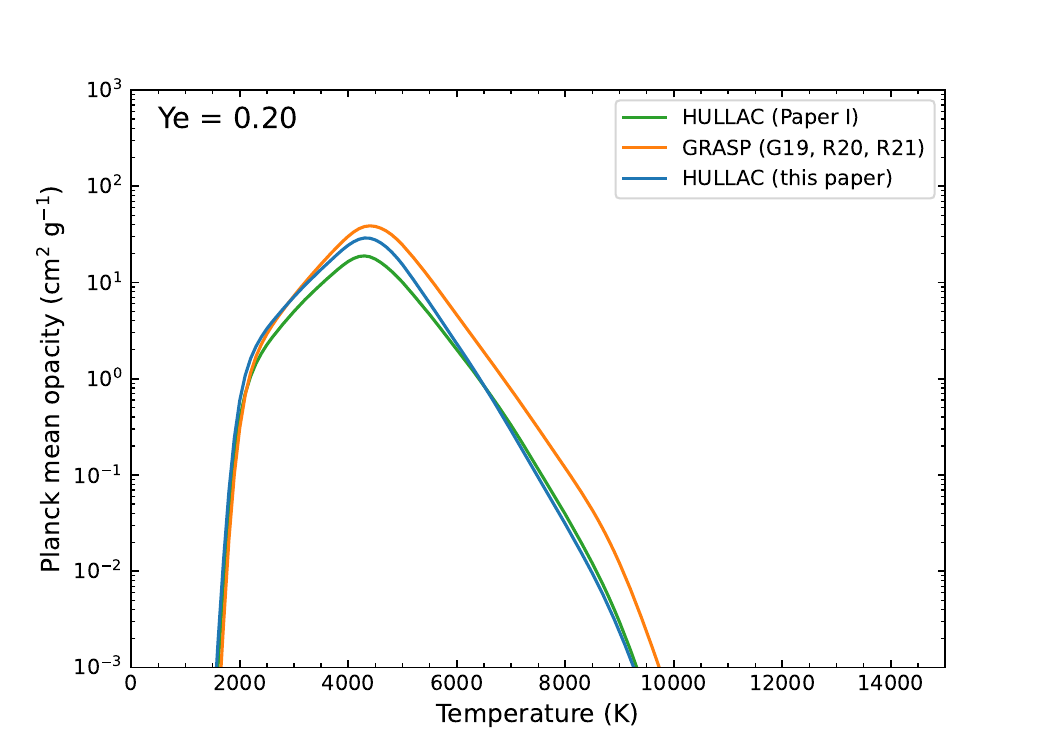}\\
\end{tabular}
\caption{Opacities of singly ionized lanthanides for the element mixture with $Y_{\rm e} = 0.2$. (Left) Expansion opacities for $\rho = 10^{-13} \ {\rm g\ cm^{-3}}$ and $T = 5000$ K at $t = 1$ day after the merger.
(Right) Planck mean opacities as a function of temperature. Note that the calculations include atomic data only for singly ionized ions, and actual opacities for the element mixture would be higher.}
\label{fig:Ye}
\end{figure*}

\subsection{Opacities of element mixture}

So far we have shown opacities for each element.
In realistic kilonova ejecta, however,
a variety of elements coexist in the plasma.
To demonstrate the impact of the improved atomic data,
we show the opacity for the element mixture in this section.
As a representative case, we use the abundance patterns
from a trajectory of $Y_{\rm e} = 0.20$ of \citet{wanajo14} as in Paper I.
The mass fraction of lanthanides is 11 \% in total.
Note that as our improved atomic data (as well as {\sc Grasp} data)
are available only for singly ionized lanthanides,  
we calculate the opacity by only including 
the atomic data of singly ionized lanthanides.
Thus, the actual opacities for the element mixture would be higher
than those given here.

The left panel of Figure \ref{fig:Ye} shows the expansion opacity
for singly ionized lanthanides calculated
for $\rho = 10^{-13} \ {\rm g\ cm^{-3}}$ and $T = 5000$ K at $t = 1$ day after the merger.
As expected from the opacities of individual elements,
our new opacity is higher than that in Paper I in particular at $\lambda < 5000$ \AA.
Overall, our new opacity show a sound agreement with 
that calculated with the {\sc Grasp} results.
At the wavelength ($\lambda = 5000-10000$ \AA), however,
the {\sc Hullac} opacity is lower than the {\sc Grasp} opacity by a factor of about 3.
This is mainly due to the difference the opacities of Tb II, Dy II, and Ho II (see Section \ref{sec:opacity}).

A similar trend is seen in the Planck mean opacities
(right panel of Figure \ref{fig:Ye}).
For the temperature range at which singly ionized states are dominant,
the Planck mean opacities of singly ionized lanthanides
from our new calculations are $\kappa = $ 24.4, 27.7, and 15.4 ${\rm cm^2 \ g^{-1}}$ at $T = $ 4000, 4500, and 5000 K, respectively.
There values are higher than those of Paper I by a factor of 1.5-1.6.
($\kappa = $ 16.6, 17.5, and 10.0 ${\rm cm^2 \ g^{-1}}$ at $T = $ 4000, 4500, and 5000 K, respectively).
The opacities from {\sc Grasp} results are $\kappa = $ 30.1, 38.4, and 24.6 at $T = $ 4000, 4500, and 5000 K, respectively.
These are higher than those from our new {\sc Hullac} calculations
by a factor of 1.2-1.6.

In fact, for the opacity of individual elements,
there are several cases showing larger discrepancy
between our new opacity and the {\sc Grasp} opacity
(see Figures \ref{fig:kappa1}, \ref{fig:kappa2}, \ref{fig:mean1},
and \ref{fig:mean2}).
But the difference in the opacity for the element mixture is rather moderate.
This is because the first few lanthanides, such as Pr ($Z=59$), Nd ($Z=60$), and Pm ($Z=61$), largely contribute to the opacities of the element mixture
and the agreement between two calculations are good for these elements.

With this degree of the difference, the impact to the kilonova light curve
is limited as singly ionized lanthanides are dominant opacity source
only around $T = 4000-5000$ K.
However, it is emphasized that our we perform intensive investigations only for
singly ionized states.
A similar level of investigation for other ionization states is necessary to fully understand the
impact of the accuracy in atomic calculations to kilonova light curves.
For such investigation, more bench mark calculations as well as
experimental measurements are important.


\section{Summary}
\label{sec:summary}

In this paper, we have performed {\sc Hullac} atomic calculations for
singly ionized lanthanides with improved strategies, aiming at understanding
the physics of the lanthanide opacities in kilonova ejecta
and necessary accuracy in atomic data.
Our results show the increased number of energy levels at low energies as compared with those in Paper I.
These are mainly due to choice of more appropriate
effective potentials and inclusion of more configurations in the calculations.

As a results of lower energy level distribution,
the opacities calculated with our new results are
higher than those by Paper I by a factor of up to $3- 10$,
depending on the elements and wavelength range.
We also present the opacities calculated by using the results of {\it ab-initio} {\sc Grasp} calculations (G19, R20 and R21).
Our new opacities show sound agreements with those with
{\sc Grasp} calculations.

Based on our results, we identify that structure of the opacities
are controlled by arrays of transitions.
At $\lambda < 6000$ \AA, transitions between
$4f^q \ 6s$ and $4f^{q-1} \ 5d \ 6s$ configurations
as well as those between $4f^q \ 5d$ and $4f^{q-1} \ 5d^2$ configurations
and $4f^{q-1} \ 5d \ 6s$ and $4f^{N-1} \ 5d \ 6p$ configurations
give dominant contributions.
At $\lambda > 6000$ \AA, transitions between
$4f^q \ 5d$ and $4f^q \ 6p$ configurations
and 
those between $4f^{q-1} \ 5d^2$ and  $4f^{q-1} \ 5d \ 6p$
give dominant contributions.
It is thus important to derive accurate energy distribution
for these configurations.

For a lanthanide-rich element mixture with $Y_{\rm e} = 0.20$,
our results give a higher opacity than that by Paper I
by a factor of about 1.5.
This is moderate as compared to the difference seen in the individual elements.
This is because the largest contribution comes from the first few lanthanides,
for which the differences between our new calculations are moderate.
To fully understand the impacts to kilonova light curves,
systematic investigation as done in this paper has to be performed
for other ionization states.

\section*{Acknowledgements}

This project has received funding from the 
Research Council of Lithuania (LMTLT), agreement No S-LJB-23-1
and JSPS Bilateral Joint Research Project (JPJSBP120234201).


\section*{Data Availability}

Energy level and transition data presented in this paper
are available at
{\sc Japan-Lithuania Opacity Database for Kilonova}:
\url{http://dpc.nifs.ac.jp/DB/Opacity-Database/}.



\bibliographystyle{mnras}

\begin{thebibliography}{}
\makeatletter
\relax
\def\mn@urlcharsother{\let\do\@makeother \do\$\do\&\do\#\do\^\do\_\do\%\do\~}
\def\mn@doi{\begingroup\mn@urlcharsother \@ifnextchar [ {\mn@doi@}
  {\mn@doi@[]}}
\def\mn@doi@[#1]#2{\def\@tempa{#1}\ifx\@tempa\@empty \href
  {http://dx.doi.org/#2} {doi:#2}\else \href {http://dx.doi.org/#2} {#1}\fi
  \endgroup}
\def\mn@eprint#1#2{\mn@eprint@#1:#2::\@nil}
\def\mn@eprint@arXiv#1{\href {http://arxiv.org/abs/#1} {{\tt arXiv:#1}}}
\def\mn@eprint@dblp#1{\href {http://dblp.uni-trier.de/rec/bibtex/#1.xml}
  {dblp:#1}}
\def\mn@eprint@#1:#2:#3:#4\@nil{\def\@tempa {#1}\def\@tempb {#2}\def\@tempc
  {#3}\ifx \@tempc \@empty \let \@tempc \@tempb \let \@tempb \@tempa \fi \ifx
  \@tempb \@empty \def\@tempb {arXiv}\fi \@ifundefined
  {mn@eprint@\@tempb}{\@tempb:\@tempc}{\expandafter \expandafter \csname
  mn@eprint@\@tempb\endcsname \expandafter{\@tempc}}}

\bibitem[\protect\citeauthoryear{{Abbott} et~al.,}{{Abbott}
  et~al.}{2017a}]{abbott17}
{Abbott} B.~P.,  et~al., 2017a, \mn@doi [Physical Review Letters]
  {10.1103/PhysRevLett.119.161101}, \href
  {http://adsabs.harvard.edu/abs/2017PhRvL.119p1101A} {119, 161101}

\bibitem[\protect\citeauthoryear{{Abbott} et~al.,}{{Abbott}
  et~al.}{2017b}]{abbott17MMA}
{Abbott} B.~P.,  et~al., 2017b, \mn@doi [\apjl] {10.3847/2041-8213/aa91c9},
  \href {http://adsabs.harvard.edu/abs/2017ApJ...848L..12A} {848, L12}

\bibitem[\protect\citeauthoryear{{Banerjee}, {Tanaka}, {Kawaguchi}, {Kato}  \&
  {Gaigalas}}{{Banerjee} et~al.}{2020}]{banerjee20}
{Banerjee} S.,  {Tanaka} M.,  {Kawaguchi} K.,  {Kato} D.,   {Gaigalas} G.,
  2020, \mn@doi [\apj] {10.3847/1538-4357/abae61}, \href
  {https://ui.adsabs.harvard.edu/abs/2020ApJ...901...29B} {901, 29}

\bibitem[\protect\citeauthoryear{{Banerjee}, {Tanaka}, {Kato}  \&
  {Gaigalas}}{{Banerjee} et~al.}{2024}]{banerjee24}
{Banerjee} S.,  {Tanaka} M.,  {Kato} D.,   {Gaigalas} G.,  2024, \mn@doi [\apj]
  {10.3847/1538-4357/ad4029}, \href
  {https://ui.adsabs.harvard.edu/abs/2024ApJ...968...64B} {968, 64}

\bibitem[\protect\citeauthoryear{{Bar-Shalom}, {Klapisch}  \&
  {Oreg}}{{Bar-Shalom} et~al.}{2001}]{bar-shalom01}
{Bar-Shalom} A.,  {Klapisch} M.,   {Oreg} J.,  2001, \mn@doi [\jqsrt]
  {10.1016/S0022-4073(01)00066-8}, \href
  {http://adsabs.harvard.edu/abs/2001JQSRT..71..169B} {71, 169}

\bibitem[\protect\citeauthoryear{{Barnes} \& {Kasen}}{{Barnes} \&
  {Kasen}}{2013}]{barnes13}
{Barnes} J.,  {Kasen} D.,  2013, \mn@doi [\apj] {10.1088/0004-637X/775/1/18},
  \href {http://adsabs.harvard.edu/abs/2013ApJ...775...18B} {775, 18}

\bibitem[\protect\citeauthoryear{{Bauswein}, {Goriely}  \& {Janka}}{{Bauswein}
  et~al.}{2013}]{bauswein13}
{Bauswein} A.,  {Goriely} S.,   {Janka} H.-T.,  2013, \mn@doi [\apj]
  {10.1088/0004-637X/773/1/78}, \href
  {http://adsabs.harvard.edu/abs/2013ApJ...773...78B} {773, 78}

\bibitem[\protect\citeauthoryear{{Carvajal Gallego}, {Deprince}, {Berengut},
  {Palmeri}  \& {Quinet}}{{Carvajal Gallego} et~al.}{2023}]{carvajalgallego23}
{Carvajal Gallego} H.,  {Deprince} J.,  {Berengut} J.~C.,  {Palmeri} P.,
  {Quinet} P.,  2023, \mn@doi [\mnras] {10.1093/mnras/stac3129}, \href
  {https://ui.adsabs.harvard.edu/abs/2023MNRAS.518..332C} {518, 332}

\bibitem[\protect\citeauthoryear{{Carvajal Gallego}, {Pain}, {Godefroid},
  {Palmeri}  \& {Quinet}}{{Carvajal Gallego}
  et~al.}{2024a}]{carvajalgallego24_TA}
{Carvajal Gallego} H.,  {Pain} J.~C.,  {Godefroid} M.,  {Palmeri} P.,
  {Quinet} P.,  2024a, \mn@doi [Journal of Physics B Atomic Molecular Physics]
  {10.1088/1361-6455/ad2182}, \href
  {https://ui.adsabs.harvard.edu/abs/2024JPhB...57c5001C} {57, 035001}

\bibitem[\protect\citeauthoryear{{Carvajal Gallego}, {Deprince}, {Maison},
  {Palmeri}  \& {Quinet}}{{Carvajal Gallego} et~al.}{2024b}]{carvajalgallego24}
{Carvajal Gallego} H.,  {Deprince} J.,  {Maison} L.,  {Palmeri} P.,   {Quinet}
  P.,  2024b, \mn@doi [\aap] {10.1051/0004-6361/202347723}, \href
  {https://ui.adsabs.harvard.edu/abs/2024A&A...685A..91C} {685, A91}

\bibitem[\protect\citeauthoryear{{Eastman} \& {Pinto}}{{Eastman} \&
  {Pinto}}{1993}]{eastman93}
{Eastman} R.~G.,  {Pinto} P.~A.,  1993, \mn@doi [\apj] {10.1086/172957}, \href
  {http://ads.nao.ac.jp/abs/1993ApJ...412..731E} {412, 731}

\bibitem[\protect\citeauthoryear{{Eichler}, {Livio}, {Piran}  \&
  {Schramm}}{{Eichler} et~al.}{1989}]{eichler89}
{Eichler} D.,  {Livio} M.,  {Piran} T.,   {Schramm} D.~N.,  1989, \mn@doi
  [\nat] {10.1038/340126a0}, \href
  {http://adsabs.harvard.edu/abs/1989Natur.340..126E} {340, 126}

\bibitem[\protect\citeauthoryear{{Fl{\"o}rs} et~al.,}{{Fl{\"o}rs}
  et~al.}{2023}]{floers23}
{Fl{\"o}rs} A.,  et~al., 2023, \mn@doi [\mnras] {10.1093/mnras/stad2053}, \href
  {https://ui.adsabs.harvard.edu/abs/2023MNRAS.524.3083F} {524, 3083}

\bibitem[\protect\citeauthoryear{{Fontes}, {Fryer}, {Hungerford}, {Wollaeger}
  \& {Korobkin}}{{Fontes} et~al.}{2020}]{fontes20}
{Fontes} C.~J.,  {Fryer} C.~L.,  {Hungerford} A.~L.,  {Wollaeger} R.~T.,
  {Korobkin} O.,  2020, \mn@doi [\mnras] {10.1093/mnras/staa485}, \href
  {https://ui.adsabs.harvard.edu/abs/2020MNRAS.493.4143F} {493, 4143}

\bibitem[\protect\citeauthoryear{{Fontes}, {Fryer}, {Wollaeger}, {Mumpower}  \&
  {Sprouse}}{{Fontes} et~al.}{2023}]{fontes23}
{Fontes} C.~J.,  {Fryer} C.~L.,  {Wollaeger} R.~T.,  {Mumpower} M.~R.,
  {Sprouse} T.~M.,  2023, \mn@doi [\mnras] {10.1093/mnras/stac2792}, \href
  {https://ui.adsabs.harvard.edu/abs/2023MNRAS.519.2862F} {519, 2862}

\bibitem[\protect\citeauthoryear{{Gaigalas}, {Kato}, {Rynkun},
  {Rad{\v{z}}i{\={u}}t{\.{e}}}  \& {Tanaka}}{{Gaigalas}
  et~al.}{2019}]{gaigalas19}
{Gaigalas} G.,  {Kato} D.,  {Rynkun} P.,  {Rad{\v{z}}i{\={u}}t{\.{e}}} L.,
  {Tanaka} M.,  2019, \mn@doi [\apjs] {10.3847/1538-4365/aaf9b8}, \href
  {https://ui.adsabs.harvard.edu/abs/2019ApJS..240...29G} {240, 29}

\bibitem[\protect\citeauthoryear{{Goriely}, {Bauswein}  \& {Janka}}{{Goriely}
  et~al.}{2011}]{goriely11}
{Goriely} S.,  {Bauswein} A.,   {Janka} H.-T.,  2011, \mn@doi [\apjl]
  {10.1088/2041-8205/738/2/L32}, \href
  {http://adsabs.harvard.edu/abs/2011ApJ...738L..32G} {738, L32}

\bibitem[\protect\citeauthoryear{{Hotokezaka}, {Tanaka}, {Kato}  \&
  {Gaigalas}}{{Hotokezaka} et~al.}{2021}]{hotokezaka21}
{Hotokezaka} K.,  {Tanaka} M.,  {Kato} D.,   {Gaigalas} G.,  2021, \mn@doi
  [\mnras] {10.1093/mnras/stab1975}, \href
  {https://ui.adsabs.harvard.edu/abs/2021MNRAS.506.5863H} {506, 5863}

\bibitem[\protect\citeauthoryear{{J{\"o}nsson}, {Gaigalas}, {Biero{\'n}},
  {Fischer}  \& {Grant}}{{J{\"o}nsson} et~al.}{2013}]{jonsson13}
{J{\"o}nsson} P.,  {Gaigalas} G.,  {Biero{\'n}} J.,  {Fischer} C.~F.,   {Grant}
  I.~P.,  2013, \mn@doi [Computer Physics Communications]
  {10.1016/j.cpc.2013.02.016}, \href
  {http://adsabs.harvard.edu/abs/2013CoPhC.184.2197J} {184, 2197}

\bibitem[\protect\citeauthoryear{{Karp}, {Lasher}, {Chan}  \&
  {Salpeter}}{{Karp} et~al.}{1977}]{karp77}
{Karp} A.~H.,  {Lasher} G.,  {Chan} K.~L.,   {Salpeter} E.~E.,  1977, \mn@doi
  [\apj] {10.1086/155241}, \href {http://ads.nao.ac.jp/abs/1977ApJ...214..161K}
  {214, 161}

\bibitem[\protect\citeauthoryear{{Kasen}, {Thomas}  \& {Nugent}}{{Kasen}
  et~al.}{2006}]{kasen06}
{Kasen} D.,  {Thomas} R.~C.,   {Nugent} P.,  2006, \mn@doi [\apj]
  {10.1086/506190}, \href {http://ads.nao.ac.jp/abs/2006ApJ...651..366K} {651,
  366}

\bibitem[\protect\citeauthoryear{{Kasen}, {Badnell}  \& {Barnes}}{{Kasen}
  et~al.}{2013}]{kasen13}
{Kasen} D.,  {Badnell} N.~R.,   {Barnes} J.,  2013, \mn@doi [\apj]
  {10.1088/0004-637X/774/1/25}, \href
  {http://adsabs.harvard.edu/abs/2013ApJ...774...25K} {774, 25}

\bibitem[\protect\citeauthoryear{{Kasen}, {Metzger}, {Barnes}, {Quataert}  \&
  {Ramirez-Ruiz}}{{Kasen} et~al.}{2017}]{kasen17}
{Kasen} D.,  {Metzger} B.,  {Barnes} J.,  {Quataert} E.,   {Ramirez-Ruiz} E.,
  2017, \mn@doi [\nat] {10.1038/nature24453}, \href
  {http://adsabs.harvard.edu/abs/2017Natur.551...80K} {551, 80}

\bibitem[\protect\citeauthoryear{{Kawaguchi}, {Shibata}  \&
  {Tanaka}}{{Kawaguchi} et~al.}{2018}]{kawaguchi18}
{Kawaguchi} K.,  {Shibata} M.,   {Tanaka} M.,  2018, \mn@doi [\apjl]
  {10.3847/2041-8213/aade02}, \href
  {http://adsabs.harvard.edu/abs/2018ApJ...865L..21K} {865, L21}

\bibitem[\protect\citeauthoryear{{Korobkin}, {Rosswog}, {Arcones}  \&
  {Winteler}}{{Korobkin} et~al.}{2012}]{korobkin12}
{Korobkin} O.,  {Rosswog} S.,  {Arcones} A.,   {Winteler} C.,  2012, \mn@doi
  [\mnras] {10.1111/j.1365-2966.2012.21859.x}, \href
  {http://adsabs.harvard.edu/abs/2012MNRAS.426.1940K} {426, 1940}

\bibitem[\protect\citeauthoryear{{Kramida}, {Ralchenko}, {Reader}  \& {NIST ASD
  Team}}{{Kramida} et~al.}{2018}]{kramida18}
{Kramida} A.,  {Ralchenko} Y.,  {Reader} J.,   {NIST ASD Team} 2018, NIST
  Atomic Spectra Database (version 5.6.1), {\tt https://physics.nist.gov/asd}.
  National Institute of Standards and Technology, Gaithersburg, MD.

\bibitem[\protect\citeauthoryear{{Lattimer} \& {Schramm}}{{Lattimer} \&
  {Schramm}}{1974}]{lattimer74}
{Lattimer} J.~M.,  {Schramm} D.~N.,  1974, \mn@doi [\apjl] {10.1086/181612},
  \href {http://adsabs.harvard.edu/abs/1974ApJ...192L.145L} {192, L145}

\bibitem[\protect\citeauthoryear{{Li} \& {Paczy{\'n}ski}}{{Li} \&
  {Paczy{\'n}ski}}{1998}]{li98}
{Li} L.-X.,  {Paczy{\'n}ski} B.,  1998, \mn@doi [\apjl] {10.1086/311680}, \href
  {http://adsabs.harvard.edu/abs/1998ApJ...507L..59L} {507, L59}

\bibitem[\protect\citeauthoryear{{Metzger} \& {Fern{\'a}ndez}}{{Metzger} \&
  {Fern{\'a}ndez}}{2014}]{metzger14}
{Metzger} B.~D.,  {Fern{\'a}ndez} R.,  2014, \mn@doi [\mnras]
  {10.1093/mnras/stu802}, \href
  {http://adsabs.harvard.edu/abs/2014MNRAS.441.3444M} {441, 3444}

\bibitem[\protect\citeauthoryear{{Metzger} et~al.,}{{Metzger}
  et~al.}{2010}]{metzger10}
{Metzger} B.~D.,  et~al., 2010, \mn@doi [\mnras]
  {10.1111/j.1365-2966.2010.16864.x}, \href
  {http://adsabs.harvard.edu/abs/2010MNRAS.406.2650M} {406, 2650}

\bibitem[\protect\citeauthoryear{{Perego}, {Radice}  \& {Bernuzzi}}{{Perego}
  et~al.}{2017}]{perego17}
{Perego} A.,  {Radice} D.,   {Bernuzzi} S.,  2017, \mn@doi [\apjl]
  {10.3847/2041-8213/aa9ab9}, \href
  {http://adsabs.harvard.edu/abs/2017ApJ...850L..37P} {850, L37}

\bibitem[\protect\citeauthoryear{{Pognan}, {Jerkstrand}  \& {Grumer}}{{Pognan}
  et~al.}{2022}]{pognan22}
{Pognan} Q.,  {Jerkstrand} A.,   {Grumer} J.,  2022, \mn@doi [\mnras]
  {10.1093/mnras/stac1253}, \href
  {https://ui.adsabs.harvard.edu/abs/2022MNRAS.513.5174P} {513, 5174}

\bibitem[\protect\citeauthoryear{{Rad{\v{z}}i{\={u}}t{\.{e}}}, {Gaigalas},
  {Kato}, {Rynkun}  \& {Tanaka}}{{Rad{\v{z}}i{\={u}}t{\.{e}}}
  et~al.}{2020}]{radziute20}
{Rad{\v{z}}i{\={u}}t{\.{e}}} L.,  {Gaigalas} G.,  {Kato} D.,  {Rynkun} P.,
  {Tanaka} M.,  2020, \mn@doi [\apjs] {10.3847/1538-4365/ab8312}, \href
  {https://ui.adsabs.harvard.edu/abs/2020ApJS..248...17R} {248, 17}

\bibitem[\protect\citeauthoryear{{Rad{\v{z}}i{\={u}}t{\.{e}}}, {Gaigalas},
  {Kato}, {Rynkun}  \& {Tanaka}}{{Rad{\v{z}}i{\={u}}t{\.{e}}}
  et~al.}{2021}]{radziute21}
{Rad{\v{z}}i{\={u}}t{\.{e}}} L.,  {Gaigalas} G.,  {Kato} D.,  {Rynkun} P.,
  {Tanaka} M.,  2021, \mn@doi [\apjs] {10.3847/1538-4365/ac1ad2}, \href
  {https://ui.adsabs.harvard.edu/abs/2021ApJS..257...29R} {257, 29}

\bibitem[\protect\citeauthoryear{{Rosswog}, {Sollerman}, {Feindt}, {Goobar},
  {Korobkin}, {Wollaeger}, {Fremling}  \& {Kasliwal}}{{Rosswog}
  et~al.}{2018}]{rosswog18}
{Rosswog} S.,  {Sollerman} J.,  {Feindt} U.,  {Goobar} A.,  {Korobkin} O.,
  {Wollaeger} R.,  {Fremling} C.,   {Kasliwal} M.~M.,  2018, \mn@doi [\aap]
  {10.1051/0004-6361/201732117}, \href
  {https://ui.adsabs.harvard.edu/abs/2018A&A...615A.132R} {615, A132}

\bibitem[\protect\citeauthoryear{{Tanaka} \& {Hotokezaka}}{{Tanaka} \&
  {Hotokezaka}}{2013}]{tanaka13}
{Tanaka} M.,  {Hotokezaka} K.,  2013, \mn@doi [\apj]
  {10.1088/0004-637X/775/2/113}, \href
  {http://adsabs.harvard.edu/abs/2013ApJ...775..113T} {775, 113}

\bibitem[\protect\citeauthoryear{{Tanaka} et~al.,}{{Tanaka}
  et~al.}{2017}]{tanaka17}
{Tanaka} M.,  et~al., 2017, \mn@doi [\pasj] {10.1093/pasj/psx121}, \href
  {http://adsabs.harvard.edu/abs/2017PASJ...69..102T} {69, 102}

\bibitem[\protect\citeauthoryear{{Tanaka} et~al.,}{{Tanaka}
  et~al.}{2018}]{tanaka18}
{Tanaka} M.,  et~al., 2018, \mn@doi [\apj] {10.3847/1538-4357/aaa0cb}, \href
  {http://adsabs.harvard.edu/abs/2018ApJ...852..109T} {852, 109}

\bibitem[\protect\citeauthoryear{{Tanaka}, {Kato}, {Gaigalas}  \&
  {Kawaguchi}}{{Tanaka} et~al.}{2020}]{tanaka20}
{Tanaka} M.,  {Kato} D.,  {Gaigalas} G.,   {Kawaguchi} K.,  2020, \mn@doi
  [\mnras] {10.1093/mnras/staa1576}, \href
  {https://ui.adsabs.harvard.edu/abs/2020MNRAS.496.1369T} {496, 1369}

\bibitem[\protect\citeauthoryear{{Wanajo}, {Sekiguchi}, {Nishimura}, {Kiuchi},
  {Kyutoku}  \& {Shibata}}{{Wanajo} et~al.}{2014}]{wanajo14}
{Wanajo} S.,  {Sekiguchi} Y.,  {Nishimura} N.,  {Kiuchi} K.,  {Kyutoku} K.,
  {Shibata} M.,  2014, \mn@doi [\apjl] {10.1088/2041-8205/789/2/L39}, \href
  {http://adsabs.harvard.edu/abs/2014ApJ...789L..39W} {789, L39}

\bibitem[\protect\citeauthoryear{{Wollaeger} et~al.,}{{Wollaeger}
  et~al.}{2018}]{wollaeger18}
{Wollaeger} R.~T.,  et~al., 2018, \mn@doi [\mnras] {10.1093/mnras/sty1018},
  \href {http://adsabs.harvard.edu/abs/2018MNRAS.478.3298W} {478, 3298}

\makeatother
\end{thebibliography}



\appendix
\section{Planck mean opacities}
\label{app:mean}

\begin{figure*}
  \centering
\begin{tabular}{cc}
  \includegraphics[width=8.0cm]{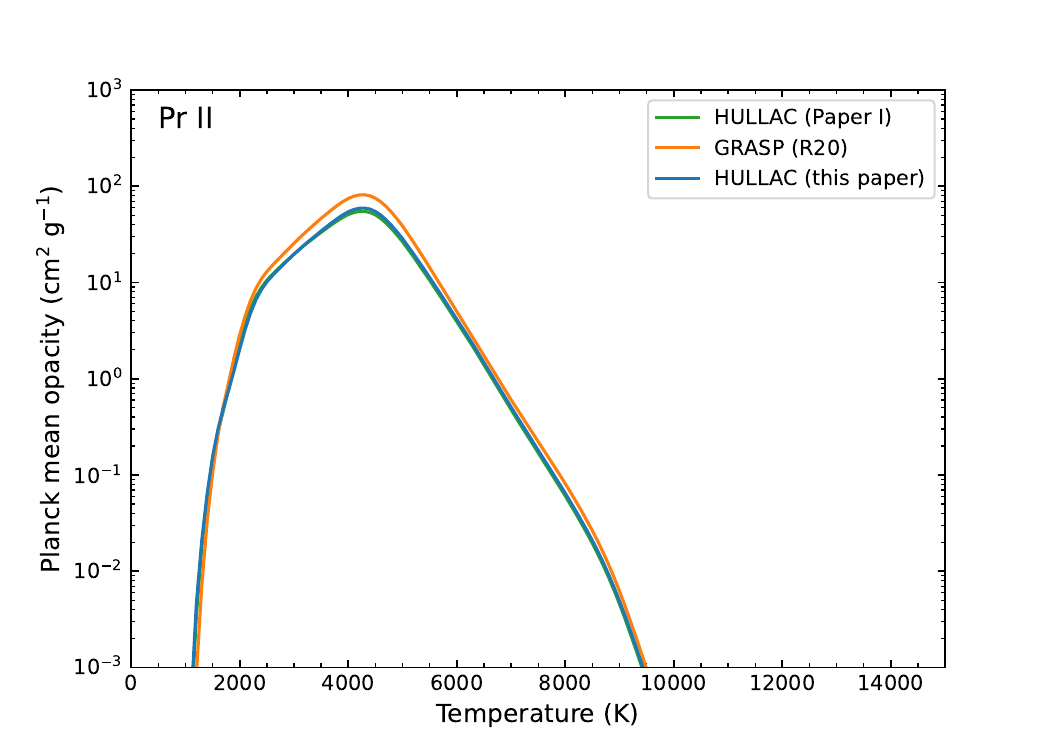}&
  \includegraphics[width=8.0cm]{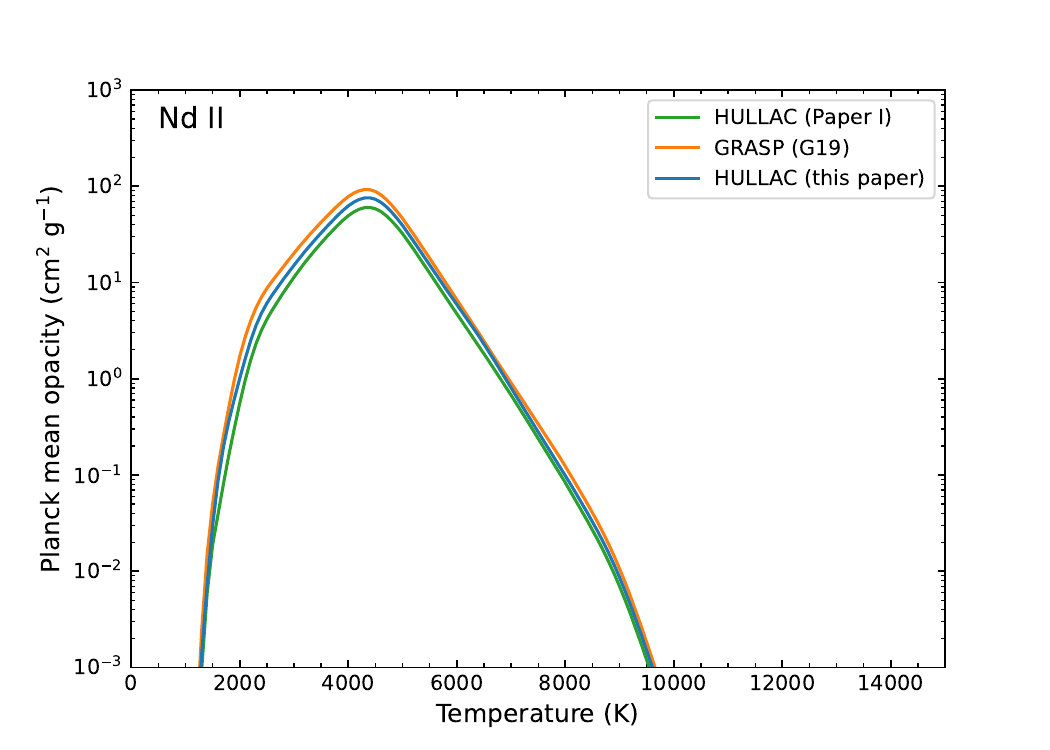}\\
  \includegraphics[width=8.0cm]{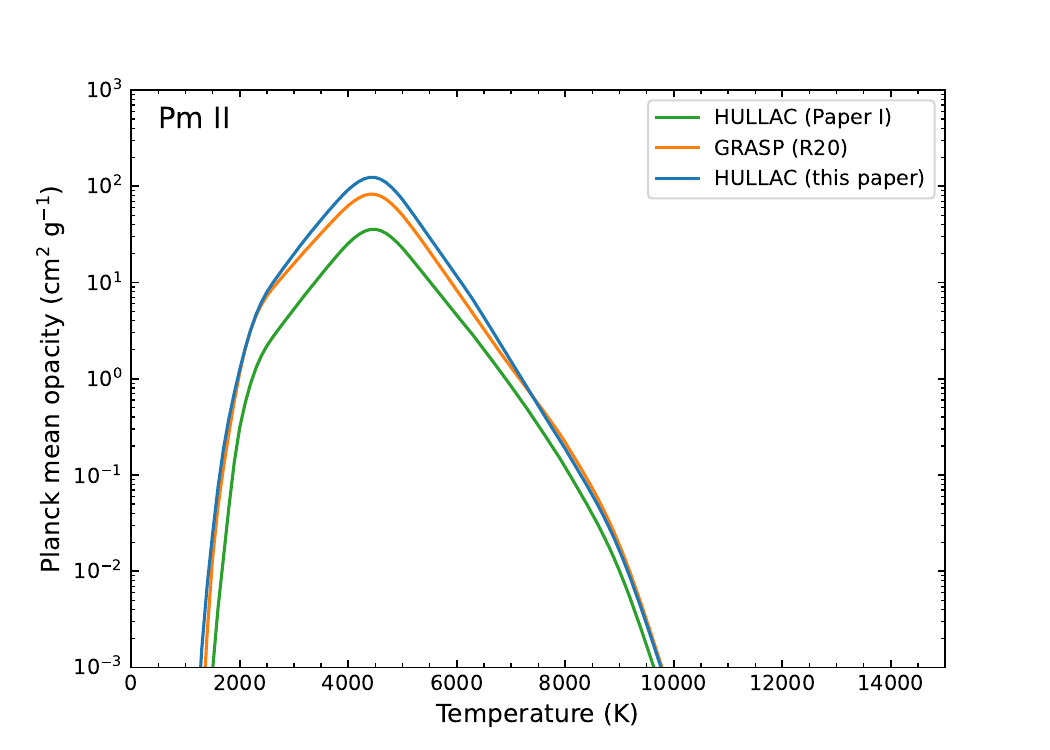}&
  \includegraphics[width=8.0cm]{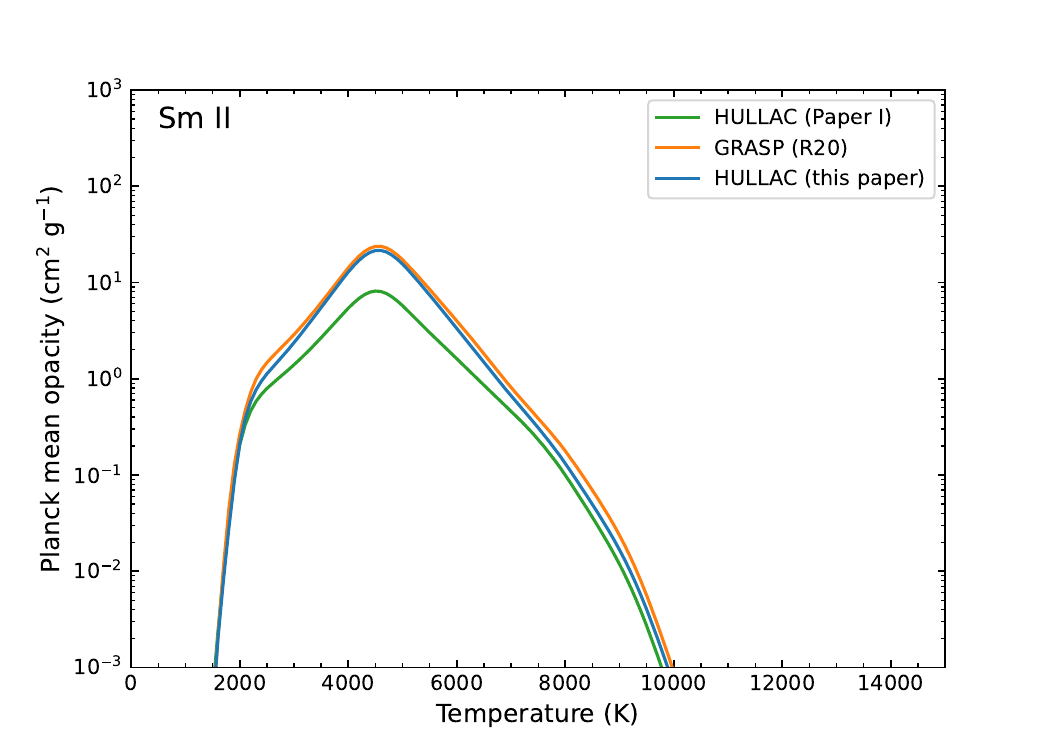}\\
  \includegraphics[width=8.0cm]{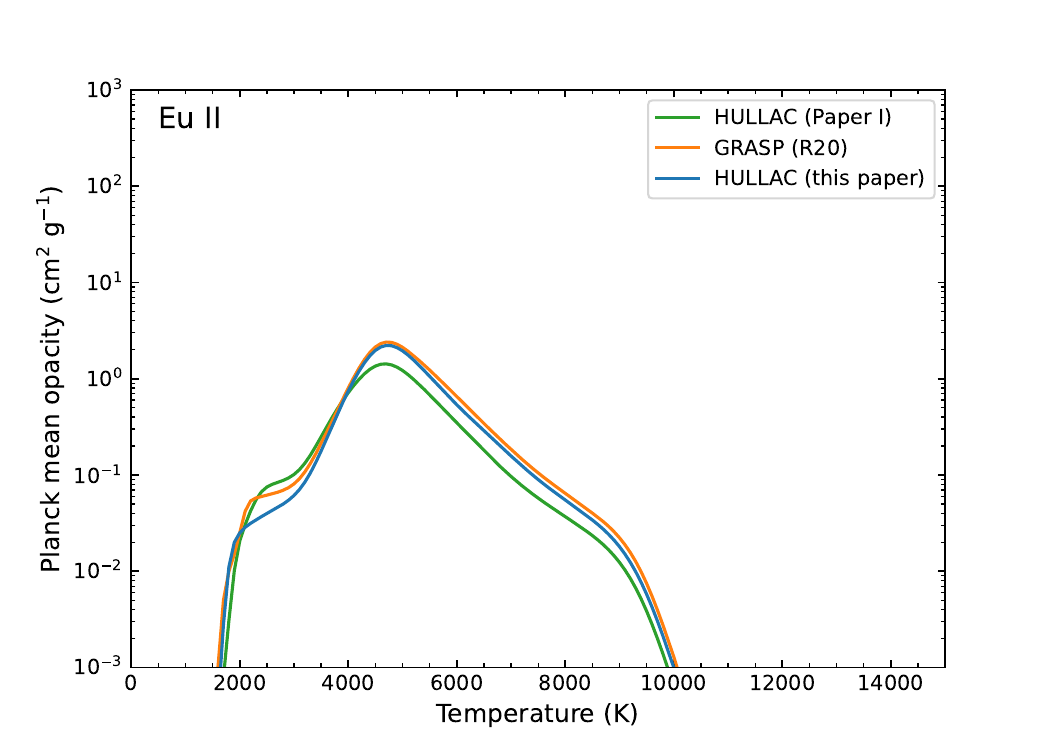}&
  \includegraphics[width=8.0cm]{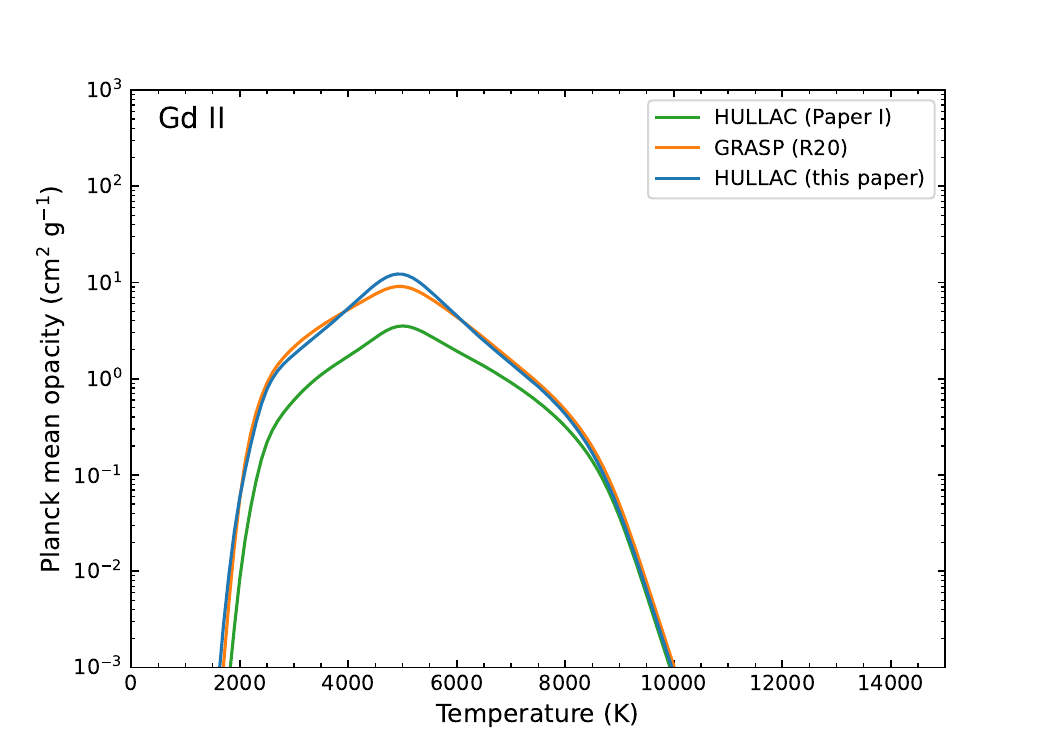}\\
\end{tabular}
\caption{Planck mean opacity for singly ionized lanthanides ($Z=59-64$) calculated
from {\sc Hullac} (Paper I and this paper) and {\sc Grasp} results (G19, R20, and R21)}.
\label{fig:mean1}
\end{figure*}

\begin{figure*}
  \centering
\begin{tabular}{cc}
  \includegraphics[width=8.0cm]{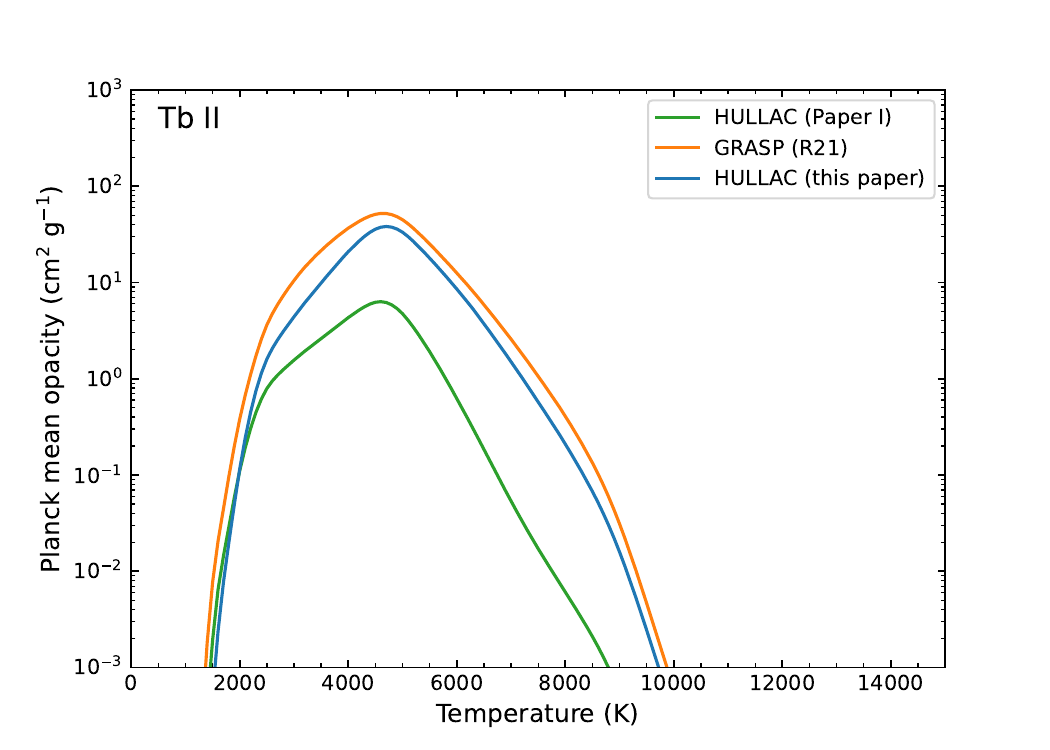}&
  \includegraphics[width=8.0cm]{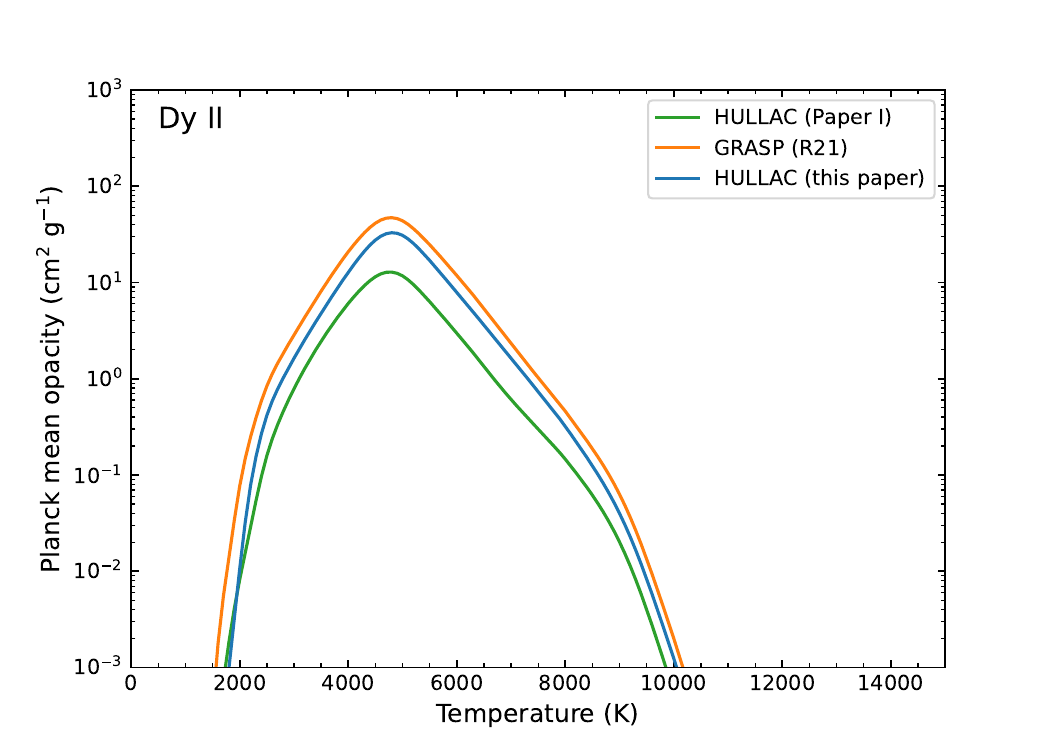}\\
  \includegraphics[width=8.0cm]{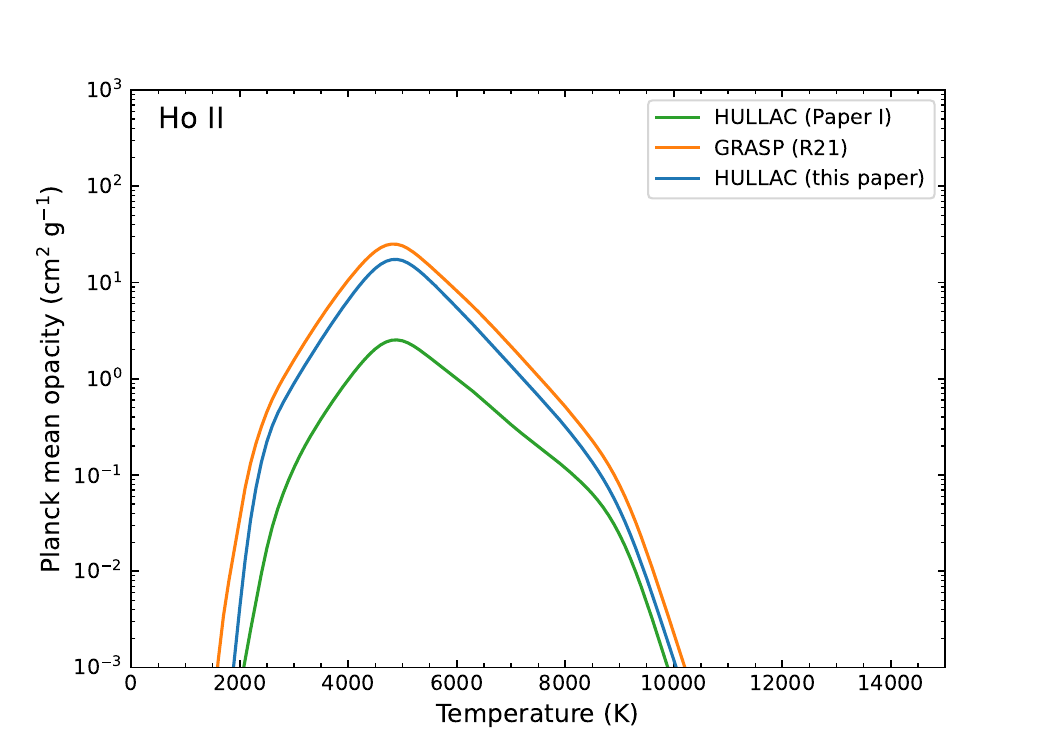}&
  \includegraphics[width=8.0cm]{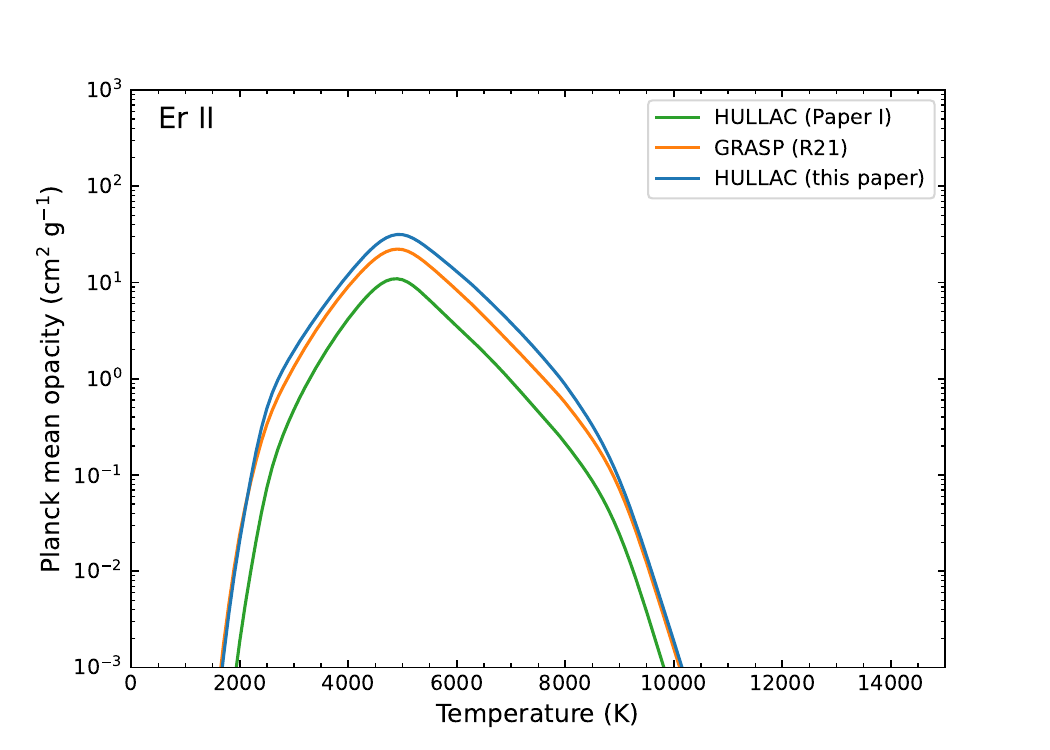}\\
  \includegraphics[width=8.0cm]{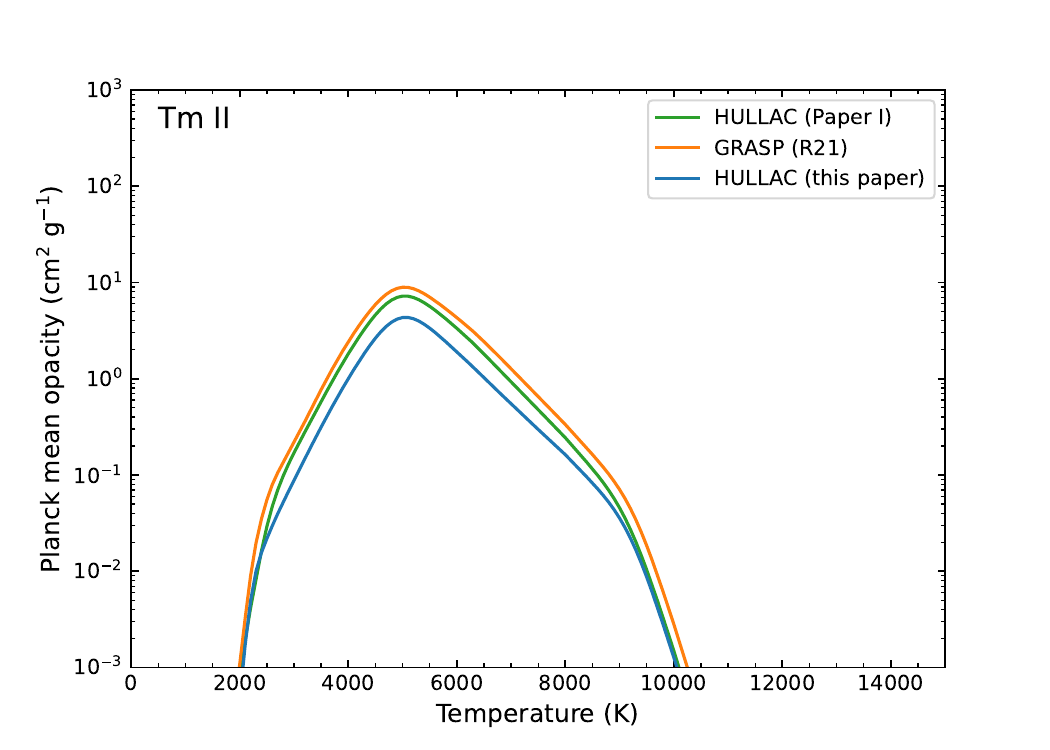}&
  \includegraphics[width=8.0cm]{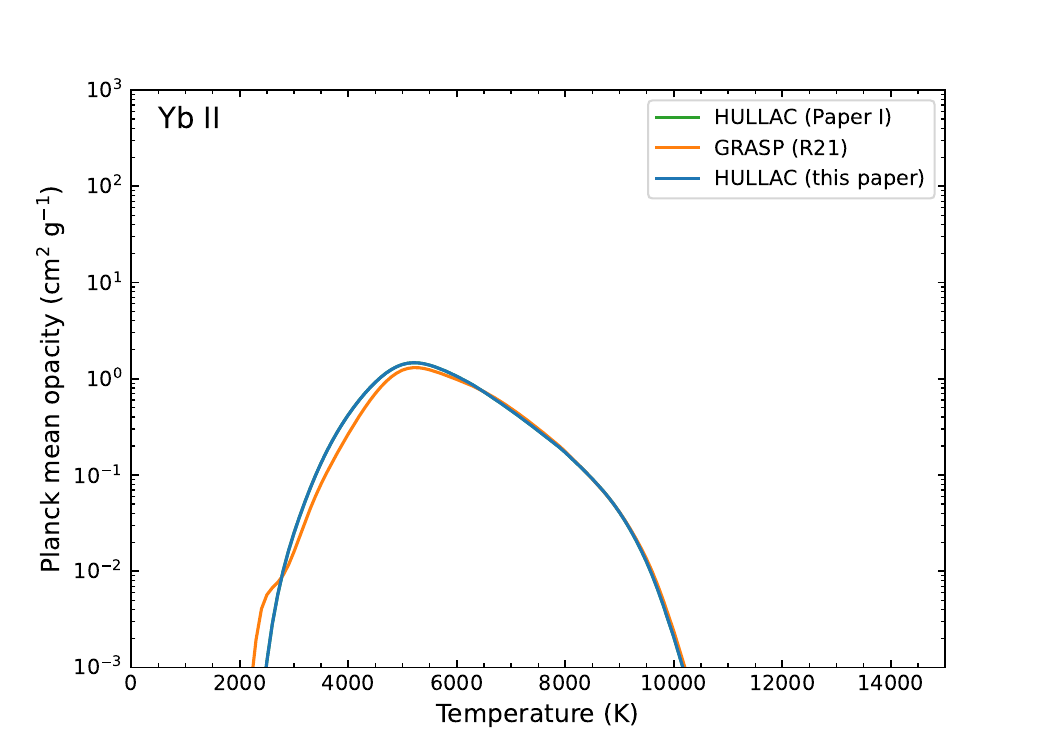}\\
\end{tabular}
\caption{Same as Figure \ref{fig:mean1} but for $Z=65-70$. Note that the {\sc HULLAC} results for Yb II in Paper I and this paper are identical.}
\label{fig:mean2}
\end{figure*}

Figures \ref{fig:mean1} and \ref{fig:mean2} show Planck mean opacities for each element. The opacities are calculated with $\rho = 10^{-13} \ {\rm g\ cm^{-3}}$ at $t = 1$ day after the merger.

.


\section{Nature of the opacities}
\label{app:nline}

\begin{figure*}
\centering
\begin{tabular}{cc}
  \includegraphics[width=8.cm]{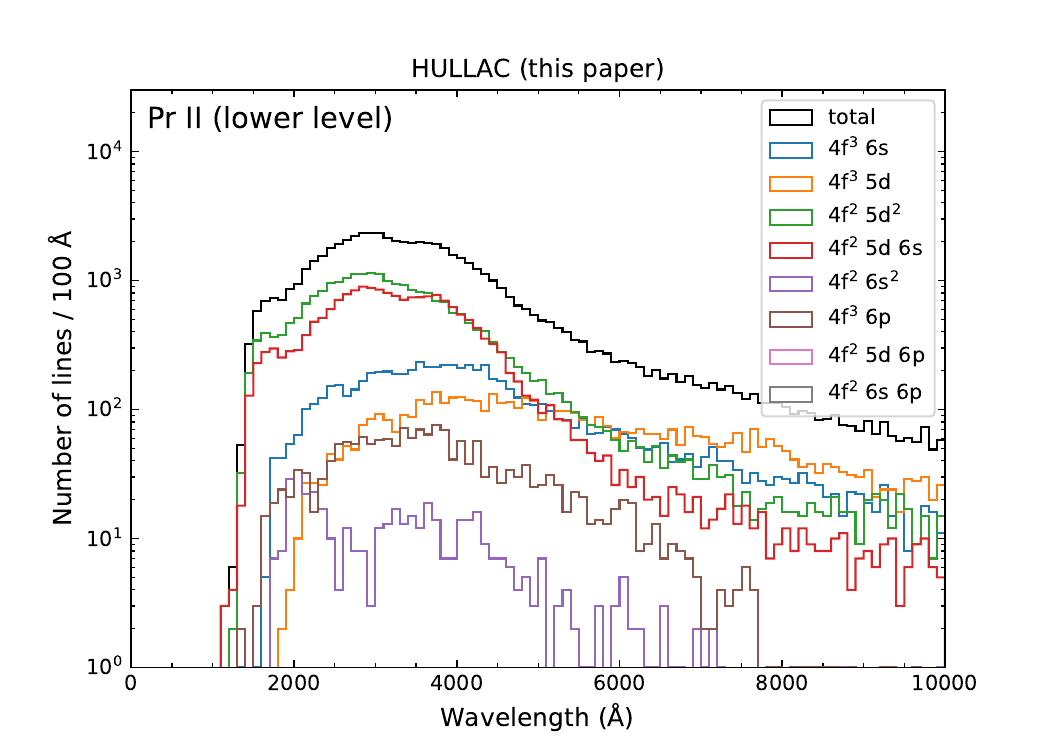}&
  \includegraphics[width=8.cm]{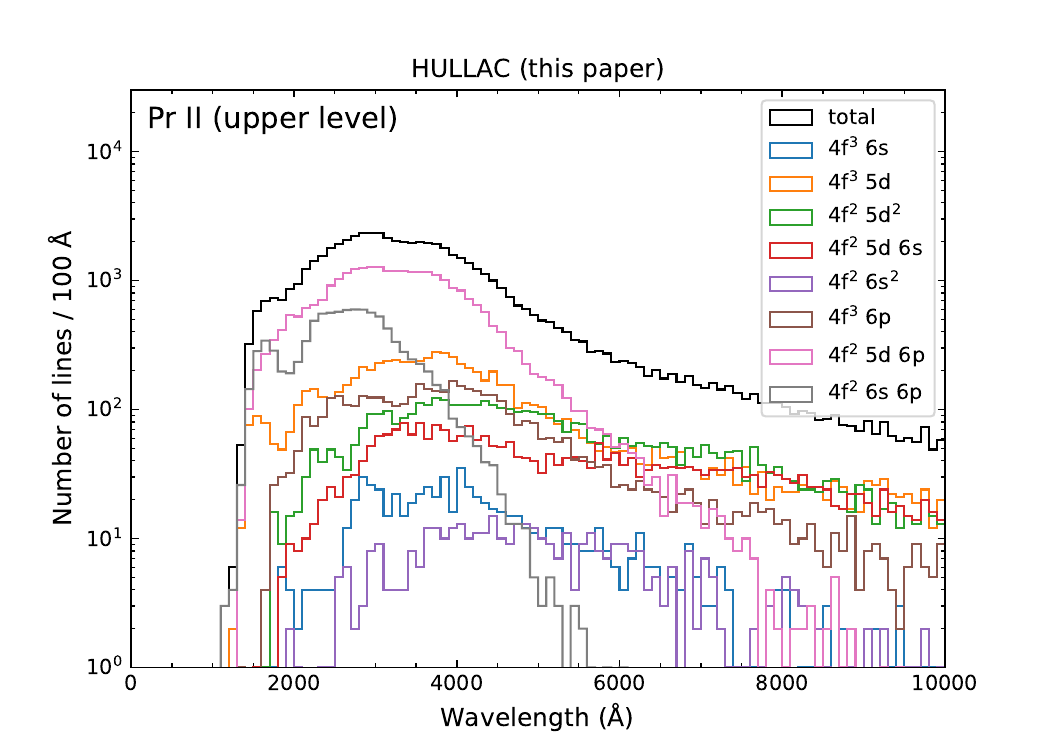}\\
\includegraphics[width=8.cm]{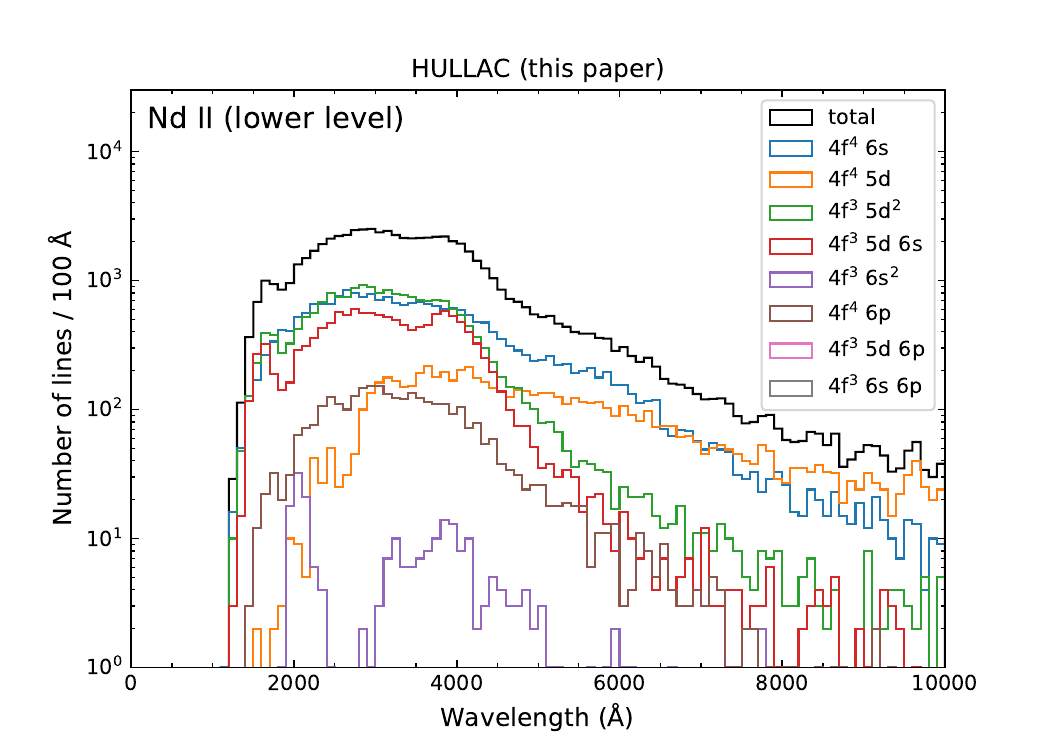}&
  \includegraphics[width=8.cm]{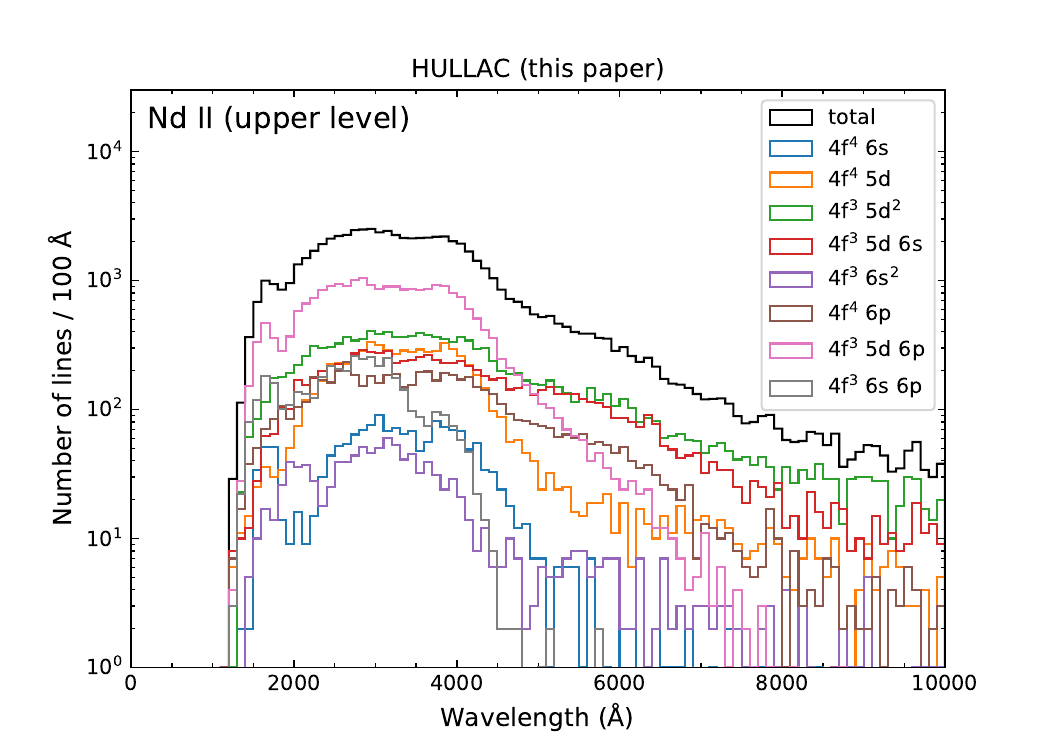}\\
  \includegraphics[width=8.cm]{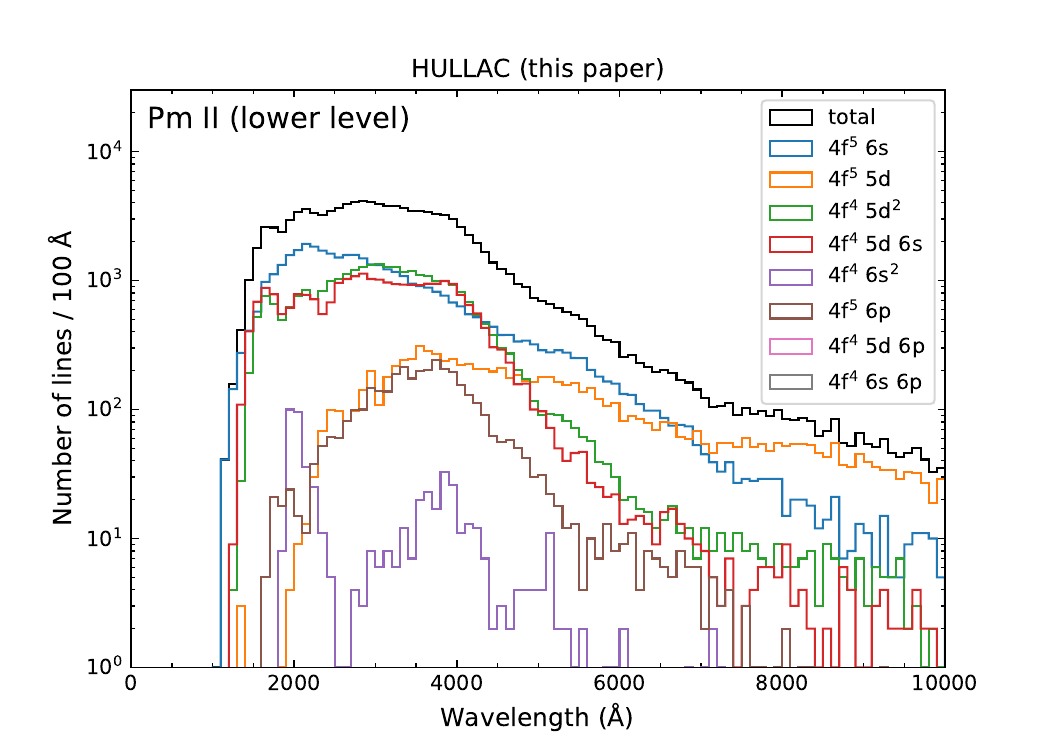}&
  \includegraphics[width=8.cm]{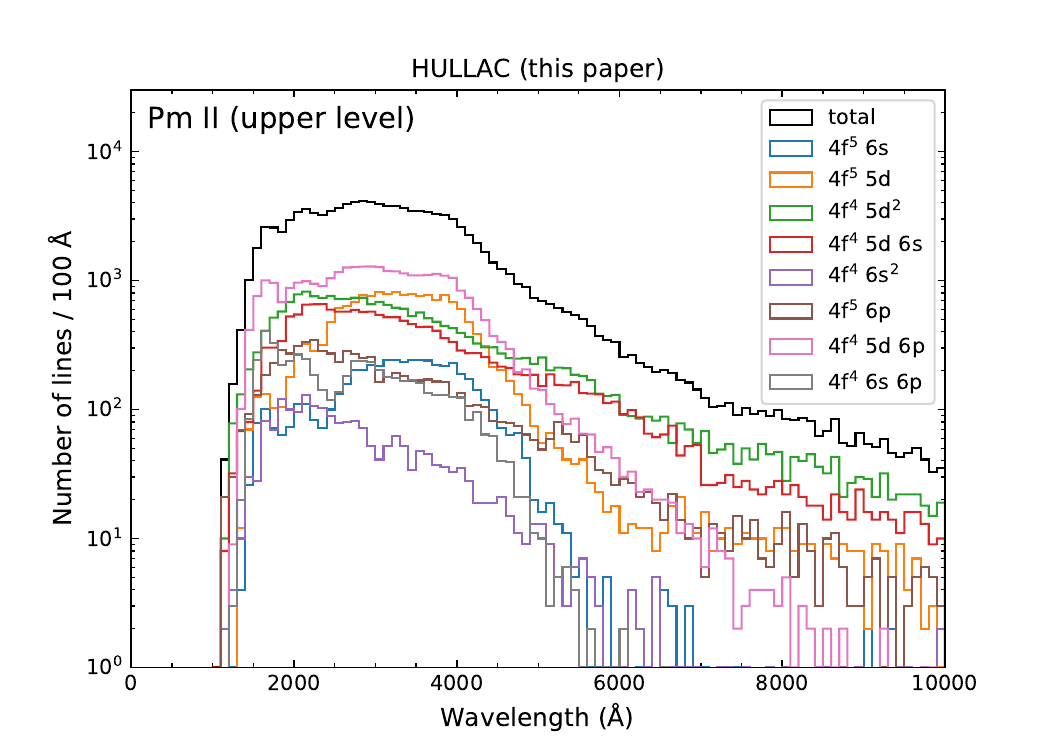}\\
  \includegraphics[width=8.cm]{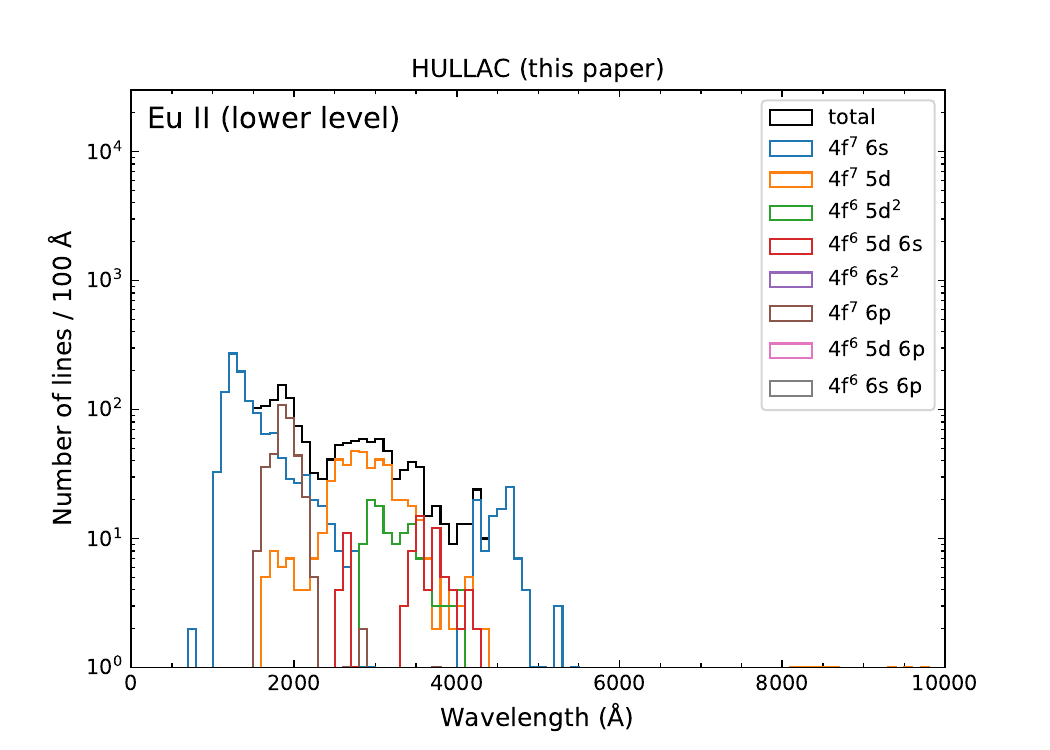}&
  \includegraphics[width=8.cm]{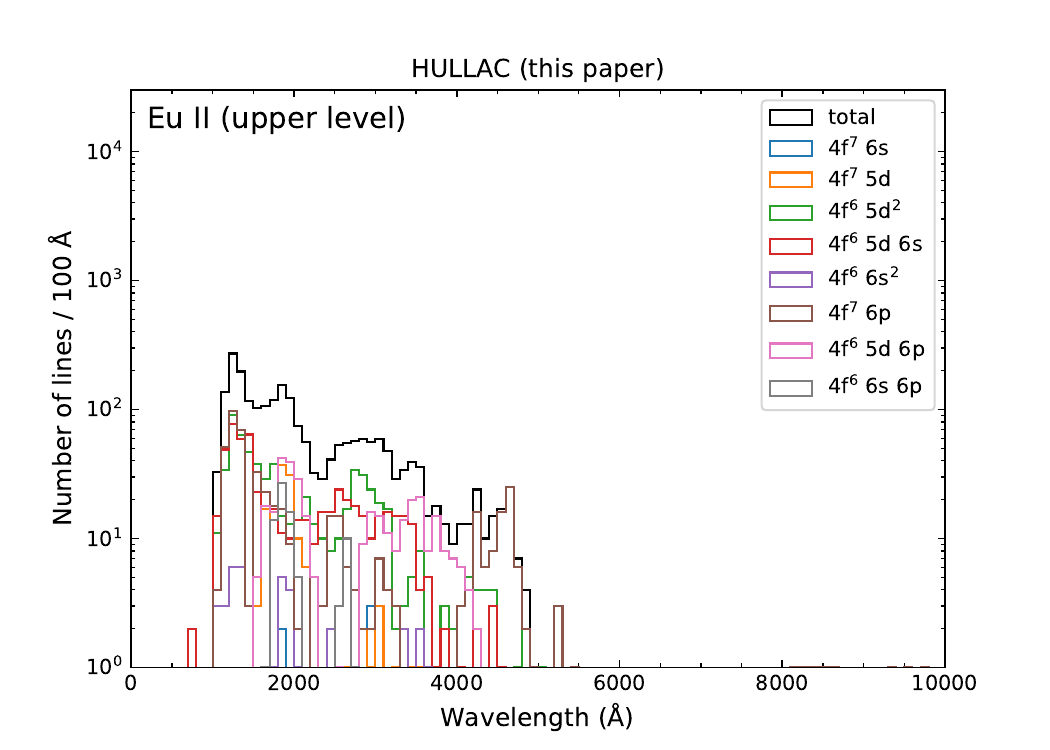}\\
\end{tabular}
\caption{The number of strong lines (per 100 \AA\ bin) for Pr II ($Z=59$), Nd II ($Z=60$), Pm II ($Z=61$), and Eu II ($Z=63$) that satisfy $gf \exp(-E_l/kT) > 10^{-5}$ at $T = 5000$ K
(see Figure \ref{fig:nline_62} for Sm II ($Z=62$)). 
The results calculated with {\sc Hullac} in this paper are shown.
In the left and right panels, the number of the lines are shown according to their lower and upper configurations, respectively.}
\label{fig:nline_59_63}
\end{figure*}

\begin{figure*}
\centering
\begin{tabular}{cc}
\includegraphics[width=8.cm]{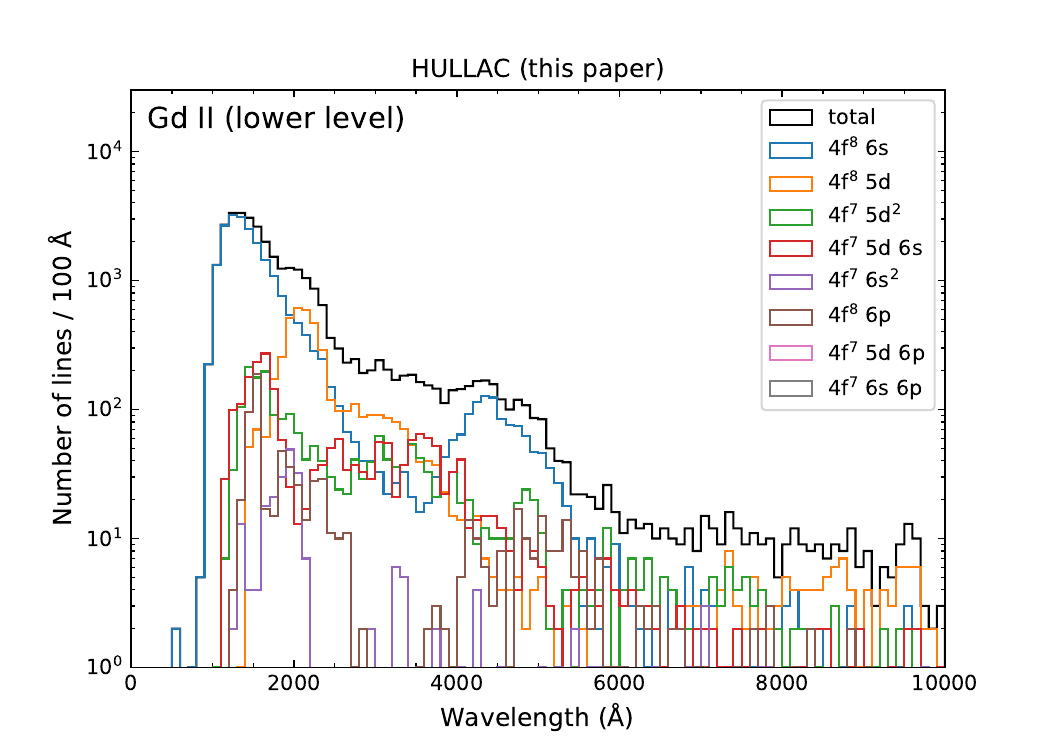}&
  \includegraphics[width=8.cm]{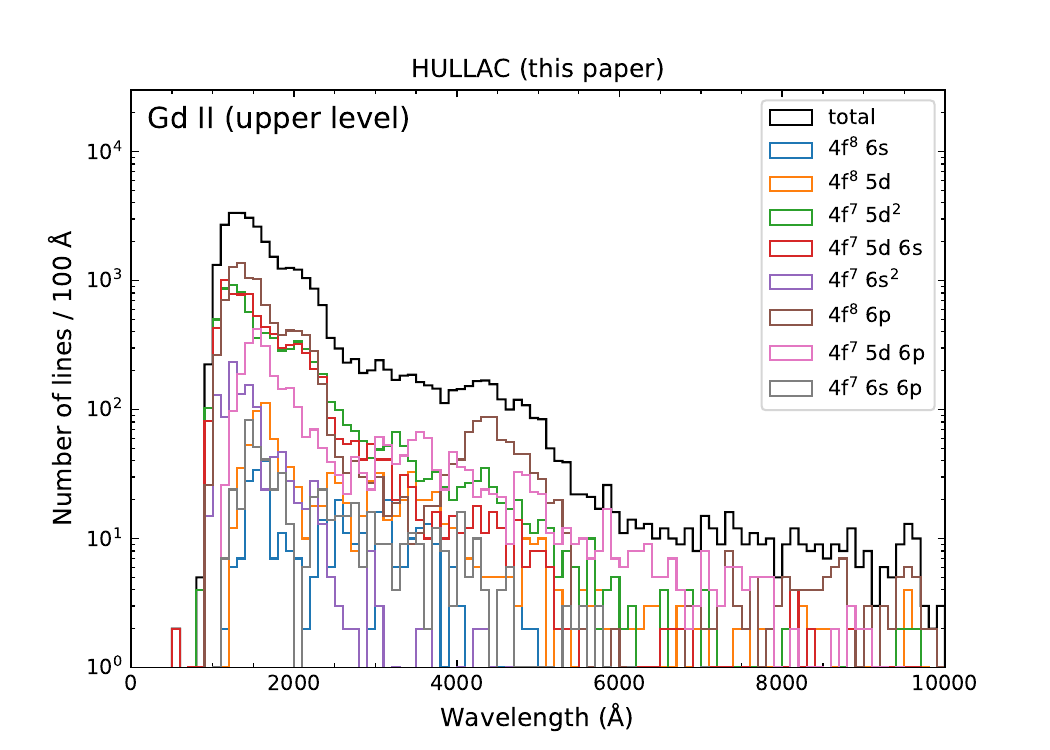}\\
\includegraphics[width=8.cm]{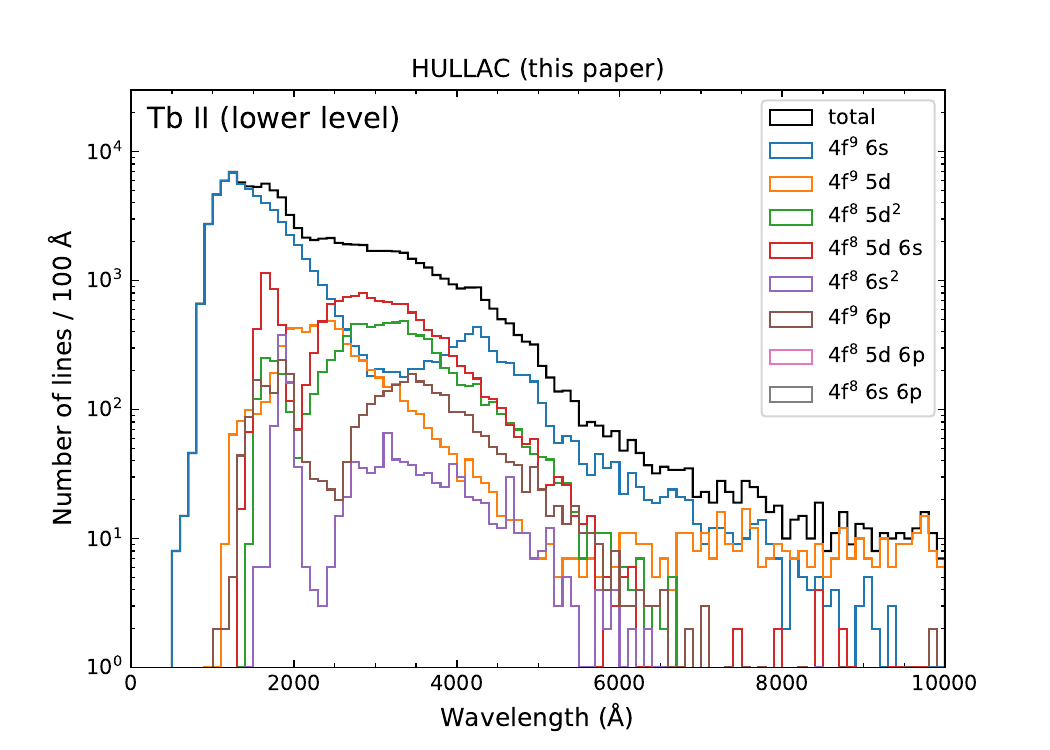}&
  \includegraphics[width=8.cm]{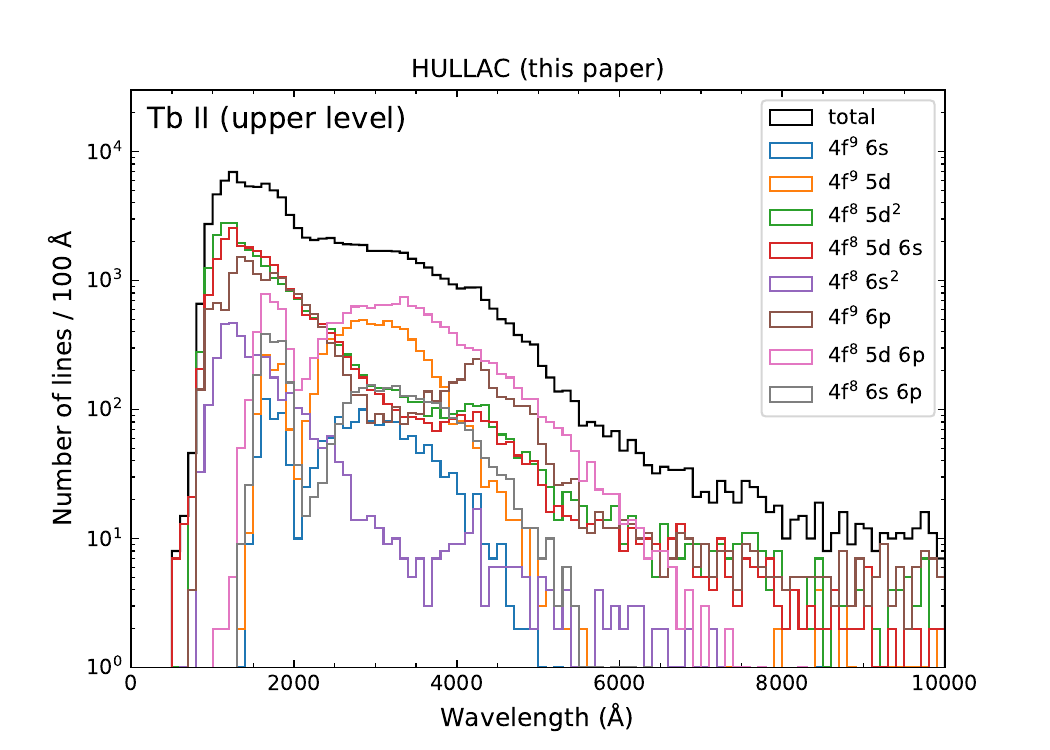}\\
  \includegraphics[width=8.cm]{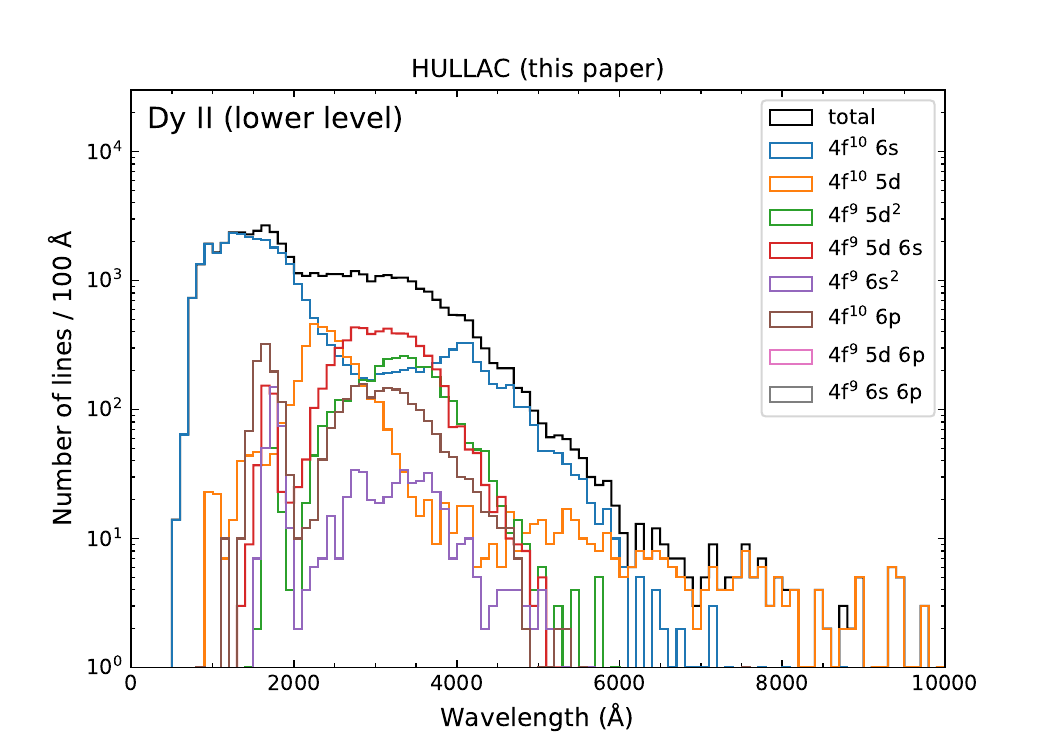}&
  \includegraphics[width=8.cm]{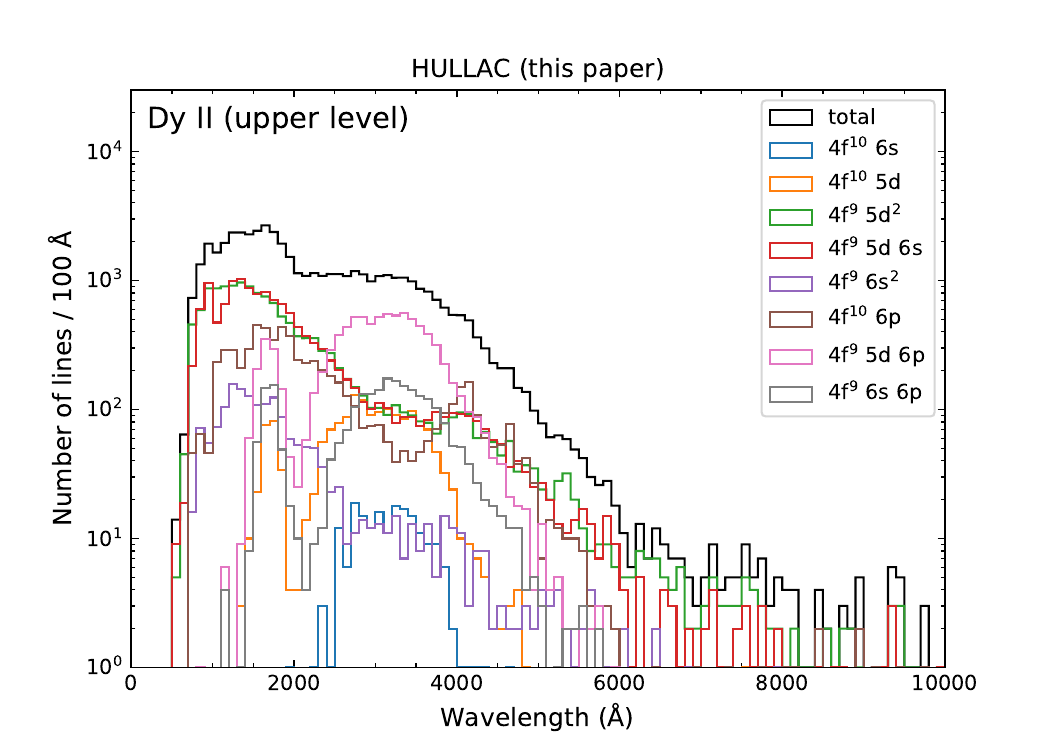}\\
\includegraphics[width=8.cm]{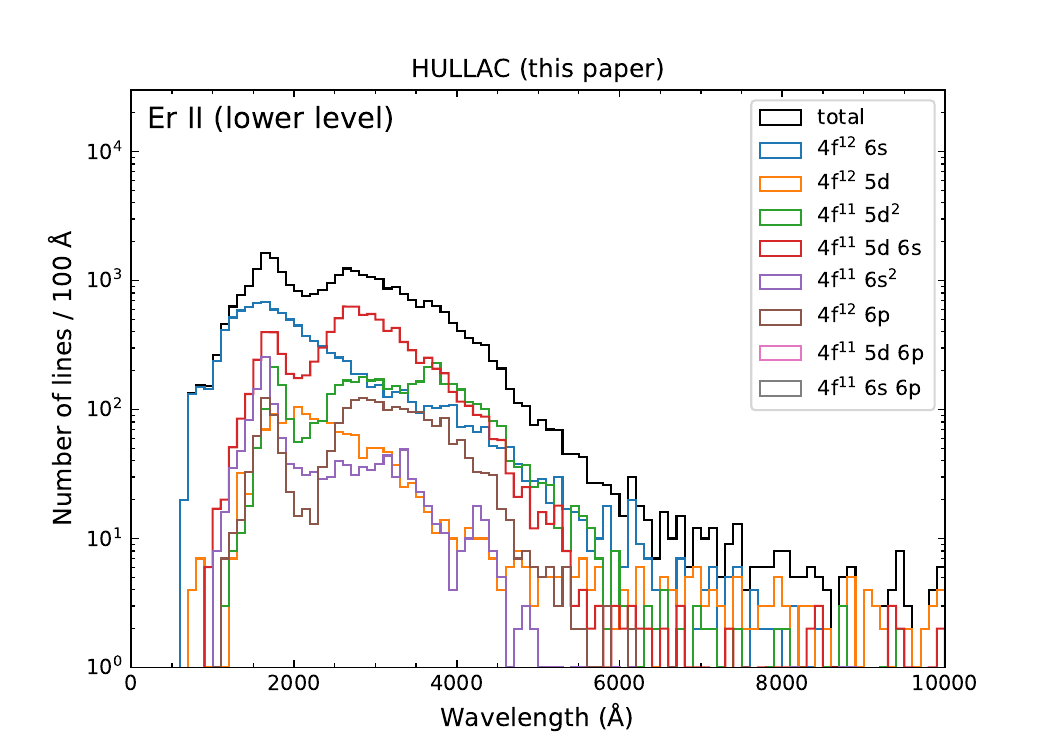}&
  \includegraphics[width=8.cm]{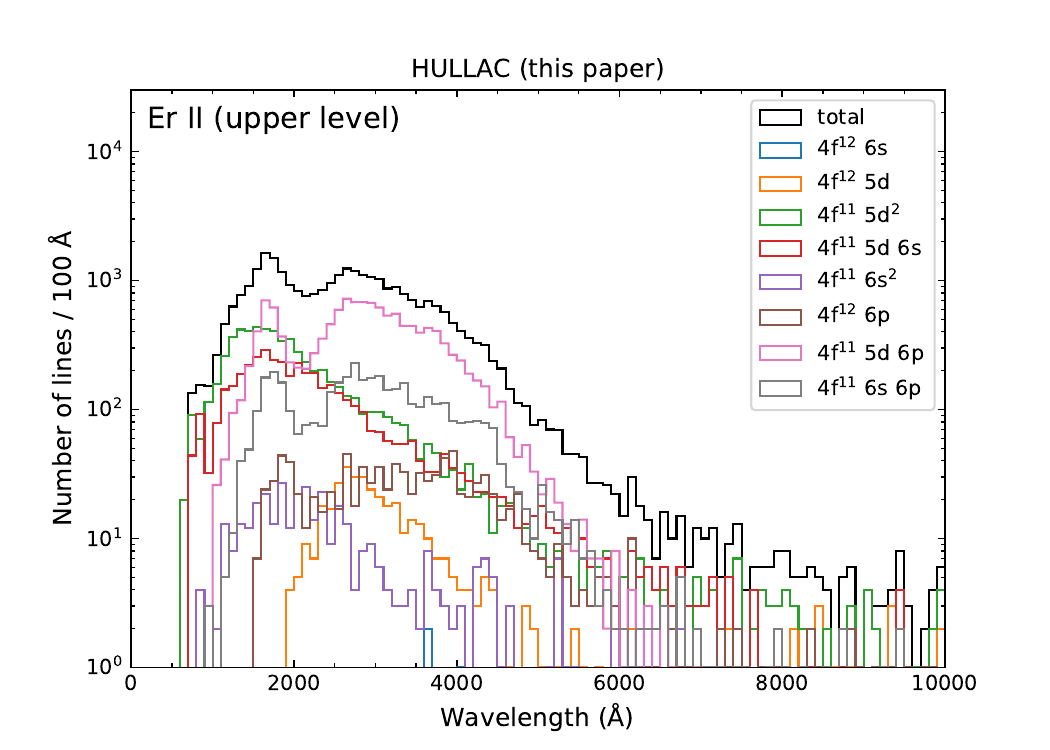}\\
\end{tabular}
\caption{Same as Figures \ref{fig:nline_59_63} but for Gd II ($Z=64$),  Tb II ($Z=65$), Dy II ($Z=66$), and Er II ($Z=68$) (see Figure \ref{fig:nline_67} for Ho II ($Z=67$)).}
\label{fig:nline_64_68}
\end{figure*}

\begin{figure*}
\centering
\begin{tabular}{cc}
  \includegraphics[width=8.cm]{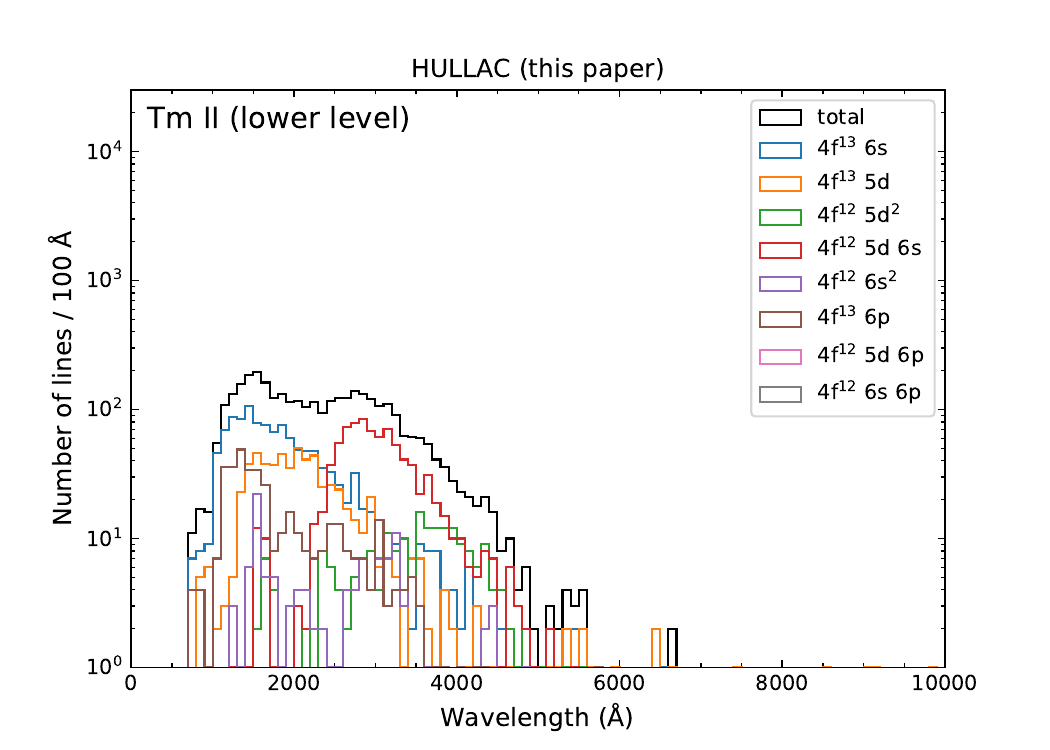}&
  \includegraphics[width=8.cm]{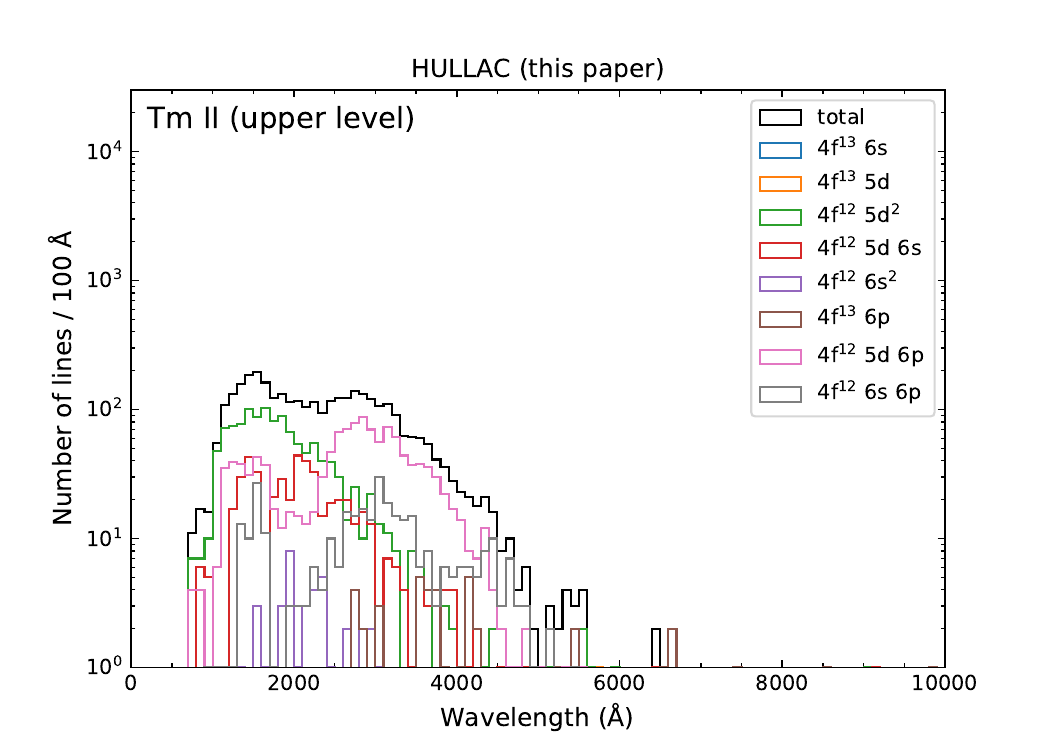}\\
\includegraphics[width=8.cm]{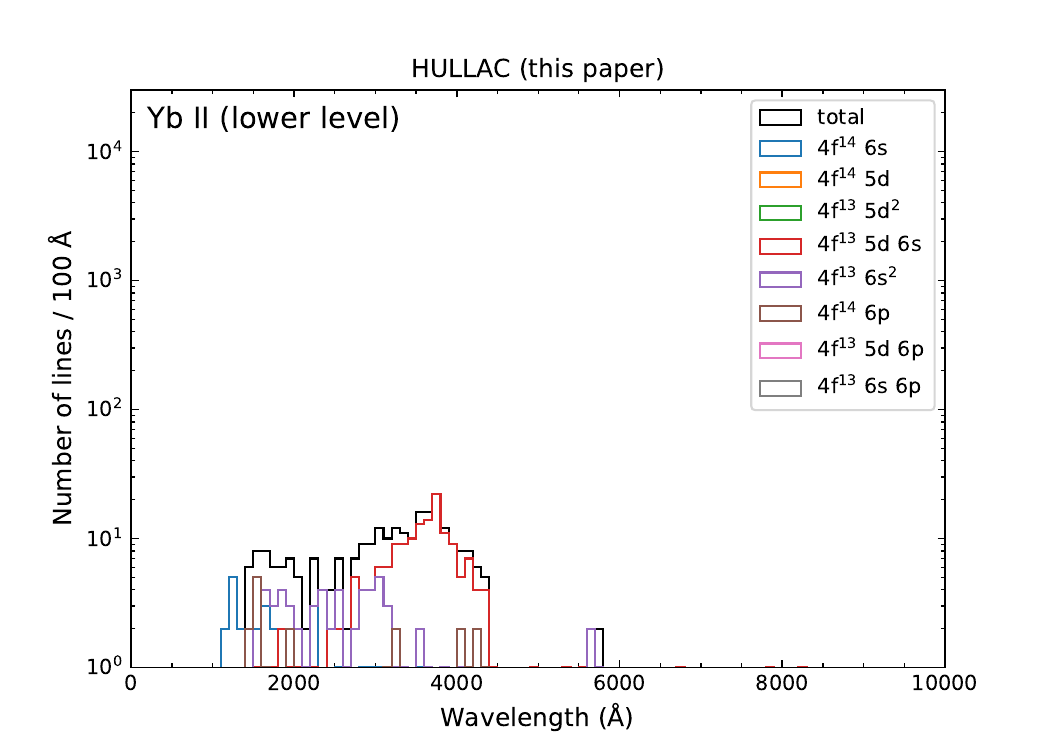}&
  \includegraphics[width=8.cm]{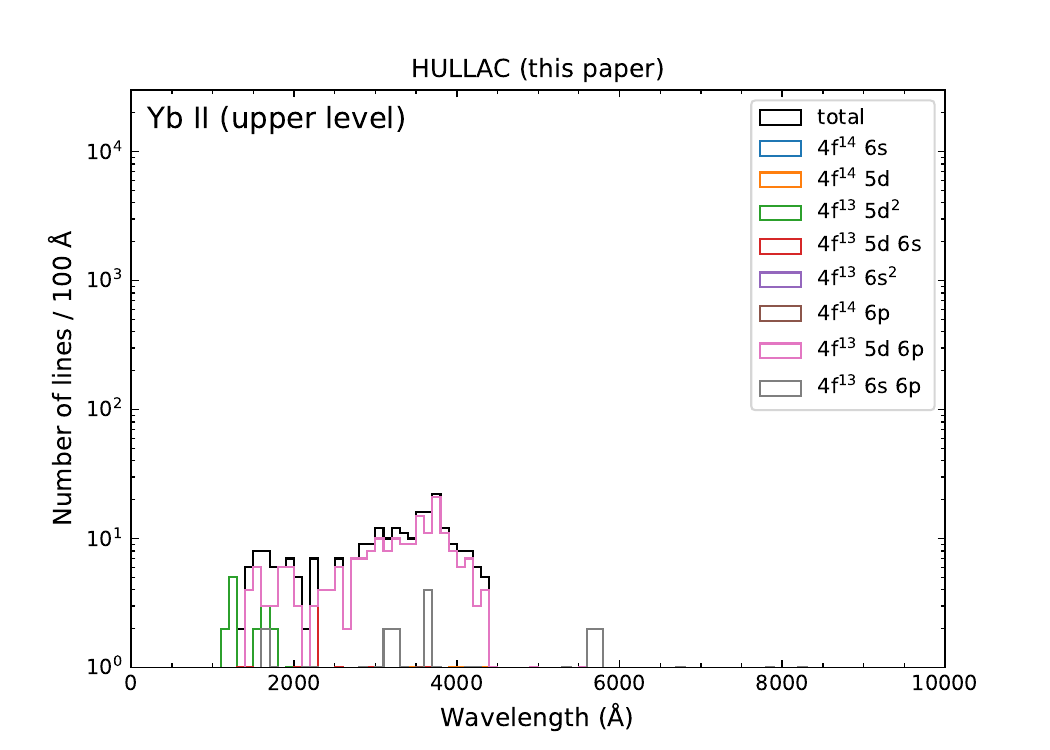}\\
\end{tabular}
\caption{Same as Figures \ref{fig:nline_59_63} but for Tm II ($Z=69$) and  Yb II ($Z=70$).}
\label{fig:nline_69_70}
\end{figure*}

Figures \ref{fig:nline_59_63}, \ref{fig:nline_64_68} and \ref{fig:nline_69_70} show the number of strong lines that satisfy $gf \exp(-E_l/kT) > 10^{-5}$ at $T = 5000$ K from our {\sc Hullac} calculations in this paper. 
In the left and right panels, the number of the lines are shown according to their lower and upper configurations, respectively

\bsp	
\label{lastpage}
\end{document}